\documentclass[graybox]{svmult} 

\usepackage{mathptmx}         
\usepackage{helvet}           
\usepackage{courier}          
\usepackage{makeidx}          
\usepackage{graphicx}         
\usepackage{multicol}         
\usepackage[bottom]{footmisc} 
\makeindex             
\definecolor{orange}{cmyk}{0.,0.353,1.,0.}    
\definecolor{dgreen}{cmyk}{1.,0.,1.,0.4}	

\newcommand{\jpsi} {\mbox{J\kern-0.05em /\kern-0.05em$\psi$} }

\newcommand{\K}{\mbox{$\mathrm {K}$}}
\newcommand{\Kzs}{\mbox{$\mathrm {K^0_S}$}}

\newcommand{\Jpsi} {\mbox{J\kern-0.05em /\kern-0.05em$\psi$}\xspace}

\def \llam {$\Lambda + \overline{\Lambda}$}

\newcommand{ \be }{\begin{eqnarray}}
\newcommand{ \ee }{\end{eqnarray}}
\newcommand{ \ben }{\begin{enumerate}}
\newcommand{ \een }{\end{enumerate}}
\newcommand{ \la }{\langle}
\newcommand{ \ra }{\rangle}

\newcommand{ \psirp }{\Psi_{\,\mathrm{RP}}}

\newcommand{ \eps }{\varepsilon}
\newcommand{ \BG}{\mathrm{BG}}
\def \vt{$v_{2}$}
\def \vtt{v_{2}\{2\}}
\def \vtf{v_{2}\{4\}}
\def \vtpp{v_{2,\mathrm{PP}}}
\def \vtrp{v_{2,\mathrm{RP}}}

\newcommand{\sqrtsNN}{\mbox{$\sqrt{\mathrm{s}_{_{\mathrm{NN}}}}$}}
\newcommand{\pt}{\ensuremath{p_T} }

\newcommand{\mean}[1]{\left\langle #1 \right\rangle}

\def \lt {\mbox{$\ <\ $}}
\def \gt {\mbox{$\ >\ $}}

\def \res {{\mathcal{R}}}
\def \sigdyn {\ensuremath{\sigma_{\mathrm{tot}}^2}}
\def \epsstd {\ensuremath{\eps_{\mathrm{std}}}}
\def \epspart {\ensuremath{\eps_{\mathrm{part}}}}
\def \ket {\ensuremath{\mathrm{KE}_T }}
\def \vtep {\ensuremath{v_2\{\mathrm{EP}\} }}

\newcommand{\dd}{\mathrm{d}}

\begin{document}

\title*{Collective phenomena in non-central nuclear collisions 
{\newline \small \hspace{3mm} \today \ Draft}}

\author{Sergei A. Voloshin, Arthur M. Poskanzer, and Raimond Snellings}
\authorrunning{S.A.~Voloshin, A.M.~Poskanzer, and R.~Snellings}

\institute{
Sergei A. Voloshin
\at Department of Physics and Astronomy, 
Wayne State University, 666~W.~Hancock, Detroit, Michigan 48201,
\email{voloshin@wayne.edu}
\and
Arthur M. Poskanzer 
\at MS70R319, LBNL, 1 Cyclotron Rd., Berkeley, California,
\email{AMPoskanzer@lbl.gov}
\and
Raimond Snellings 
\at NIKHEF, Kruislaan 409, 1098 SJ Amsterdam, The Netherlands 
\email{Raimond.Snellings@nikhef.nl}
}
\maketitle
  
\vspace*{-2.5cm}
\includegraphics[width=.165\textwidth]{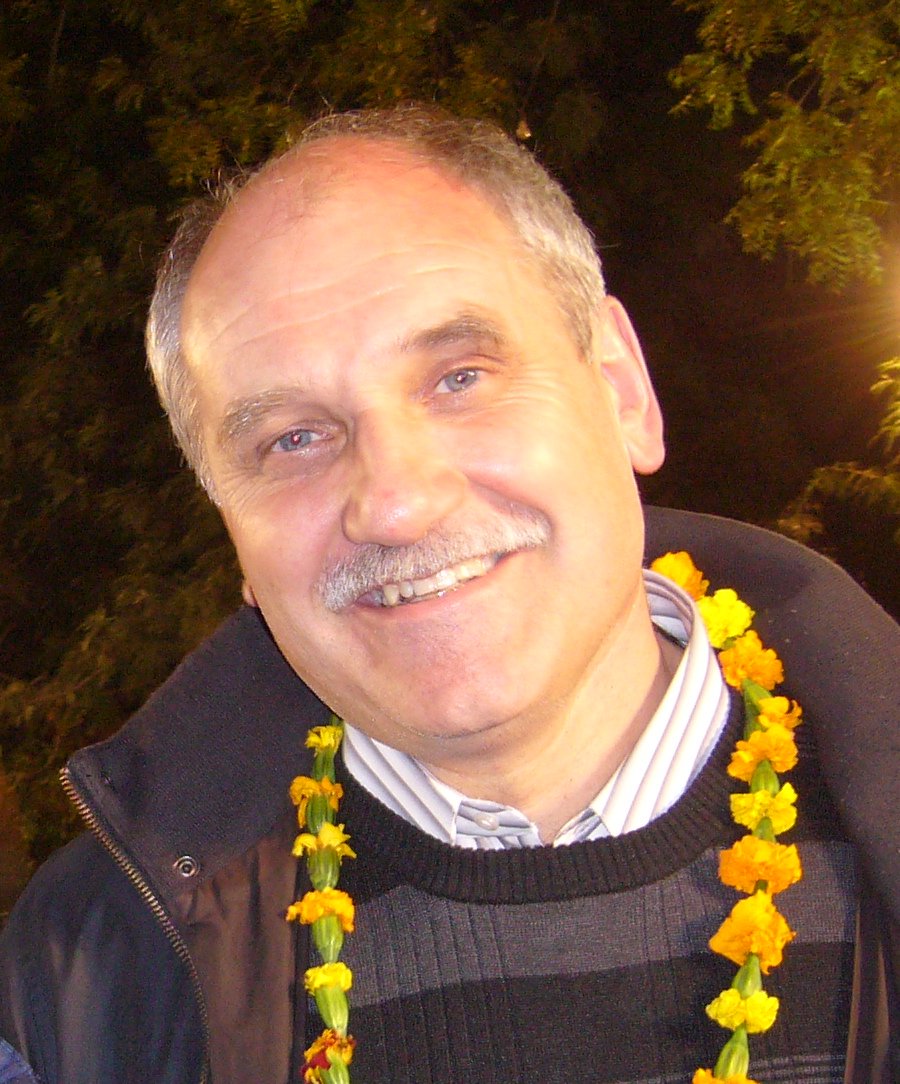} 
\hspace{1.7cm}
\includegraphics[width=.15\textwidth]{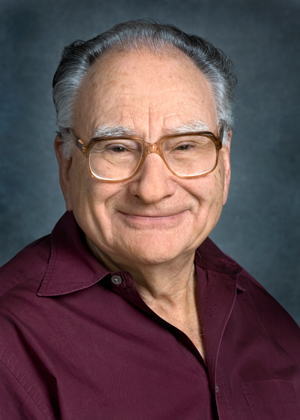} 
\hspace{1.7cm}
\includegraphics[width=.135\textwidth]{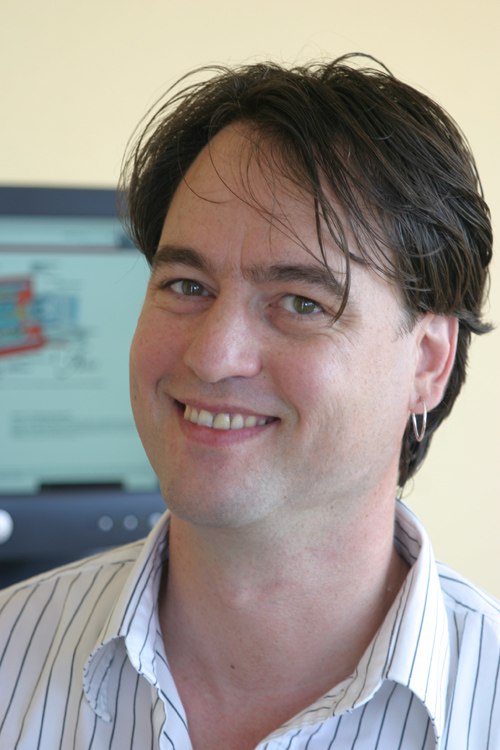} 
\vspace*{+2.5cm}

\abstract{
Recent developments in the field of anisotropic flow in nuclear
collision are reviewed. The results from the top AGS energy to the top
RHIC energy are discussed with emphasis on techniques, interpretation,
and uncertainties in the measurements.  
}


\setcounter{tocdepth}{4}
\newpage
\tableofcontents
\newpage
\section{Introduction}

\subsection{Unique observable}

Analysis of the azimuthal anisotropy resulting from non-central nuclear collisions appears to be one of the
most informative directions in studying the nature and properties of
matter created in high energy nuclear collisions. 
Anisotropies in particle momentum distributions 
relative to the reaction plane,
often referred to as anisotropic collective flow or event
anisotropies, have been in use for a few decades, starting from
the first Berkeley Bevalac experiments. 
Azimuthal anisotropies attracted even more attention 
when the so called in-plane elliptic flow, first suggested as a 
signature of collective flow in relativistic nuclear 
collisions by Ollitrault~\cite{Ollitrault:1992bk},  
was experimentally observed at the Brookhaven Alternate Gradient Synchrotron
(AGS)~\cite{Barrette:1994xr,Barrette:1996rs}, and later at 
the CERN Super Proton Synchrotron (SPS)~\cite{Alt:2003ab}. 
At the Brookhaven Relativistic Heavy Ion Collider (RHIC) the
observation of large elliptic 
flow~\cite{Ackermann:2000tr} is considered one of the
most important discoveries which lead to the concept of the strongly
interacting Quark Gluon Plasma (sQGP).
Anisotropic flow will be among the first results at the CERN 
Large Hadron Collider (LHC) heavy-ion program. 

The main interest in anisotropic flow is due to its sensitivity to the
system properties very early in its evolution. 
The origin of anisotropies in the particle
momentum distributions lies in the initial asymmetries in the geometry of
the system. Because the spatial asymmetries rapidly decrease with time,
anisotropic flow can develop only in the first fm/$c$. 
Based on this, one can conclude that anisotropic flow must
be sensitive to the particle interactions very early in the system
evolution, information usually available only via weakly interacting probes.  
In this sense, anisotropic flow  is
a unique hadronic observable providing direct information about the
stage where the QGP may be the main player. 
Constituent rescattering is by far the most common explanation 
of anisotropic flow. Although possibilities of a different origin of elliptic flow have been discussed
 (e.g. partonic structure of the nuclei~\cite{Boreskov:2008uy}, color dipole orientation~\cite{Kopeliovich:2008nx}, or direct anisotropy
in particle emission from the Color Glass Condensate (CGC) ~\cite{Teaney:2002kn,Krasnitz:2002ng})
we do not consider them here. 

Previous review papers~\cite{Reisdorf:1997fx, Herrmann:1999wu}
on collective flow in heavy-ion collisions have presented results 
from accelerators at lower energy than RHIC. 
In this review we concentrate on results obtained in recent years, with the data coming mostly from RHIC.
The field of anisotropic flow is growing with new data and
theoretical results appearing rapidly. In this review we concentrate on the general results and
interpretation, the status of the field, and major unresolved issues. 
The idea is that an interested reader, and we expect many graduate
students and young researches to be among them, could not only
appreciate the achievements of this field, but also identify interesting
problems and be ready to start working in those directions.  
For that reason we also give a rather detailed presentation of the
``technical'' side of flow measurements, discussing advantages and
disadvantages of different methods and associated systematic
uncertainties in the results. 
Though for the real details we refer to the original papers, we hope
that the information presented here could provide a good basis to
get involved. 
One will find that, unfortunately, the systematic uncertainties in
flow measurements are still rather large, up to 10--15\%, and often more. We identify two directions for future flow measurements, one
being large statistics to try to better understand
systematics, and the other being measurement of flow of rare particles.

\subsection{Definitions: flow and nonflow, the reaction and participant planes.}
\label{sec:defs}

\begin{figure}[htb]
\begin{minipage}[b]{0.48\textwidth} 
\begin{center}
  \includegraphics[width=.95\textwidth]{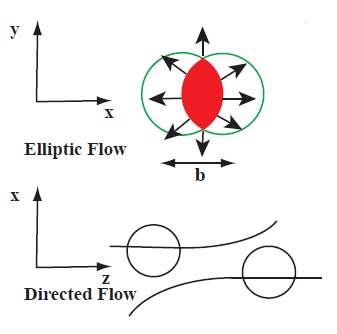} 
  \caption{Diagrams of elliptic and directed flow.}
  \label{fig:DirectedElliptic}
\end{center}
\end{minipage}
\hspace{\fill}
\begin{minipage}[b]{0.48\textwidth}
\begin{center}
  \includegraphics[width=.95\textwidth]{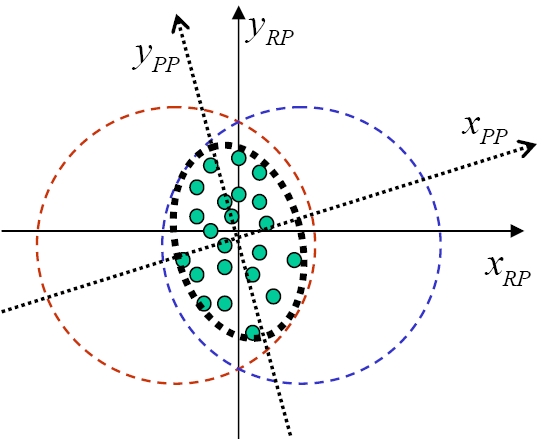}
  \caption{The definitions of the Reaction Plane and Participant Plane 
    coordinate systems.}
  \label{fig:planes}
\end{center}
\end{minipage}
\end{figure}

The {\em reaction plane} is spanned by the vector of the impact parameter
and the beam direction. Its azimuth is given by $\psirp$.  
The particle azimuthal distribution measured with respect to the
reaction plane is not isotropic; so it is customary to expand it in a Fourier
series~\cite{Voloshin:1994mz}:
\begin{equation}
  E \frac{d^3 N}{d^3 p} = \frac{1}{2\pi} 
\frac {d^2N}{\pt d \pt dy}(1+ \sum_{n=1}^\infty 2 v_n \cos(n(\phi-\psirp))),
\label{eq:ed3t}
\end{equation}
where the $v_n= \mean{\cos[n(\phi_i - \psirp)]}$ 
coefficients are used for a quantitative
characterization of the event anisotropy, and the angle brackets 
mean an average over all particles in all events. The sine terms are not present because of symmetry with respect to the reaction plane.
$v_1$ is referred to as directed flow, and $v_2$ as
elliptic flow (see Fig.~\ref{fig:DirectedElliptic}).
Radial flow in this paper refers to radial in the transverse plane.
The $v_n$ coefficients are functions of rapidity and
transverse momentum, and as such they are often referred to 
as $n^{th}$ harmonic {\em  differential} flow.
By {\em integrated} flow we mean the values of the $v_n$
coefficients averaged over transverse momentum and rapidity. 

The reaction plane angle can not be directly measured in high energy
nuclear collisions, but can be estimated from the particle azimuthal 
distribution event-by-event.
Then the different harmonic flow coefficients are reconstructed 
from two or many particle azimuthal correlations. 
This introduces uncertainty in the analysis, 
discussed in more detail in the methods section,
as the azimuthal correlations are not determined solely by anisotropic
flow but have other contributions, usually referred to as
{\em nonflow} and in case of two-particle correlations
quantified by parameters $\delta_n$:
\be
	\mean{\cos[n(\phi_i -\phi_j)]} = \mean{v_n^2} + \delta_n.
\label{eq:delta}
\ee
 
Anisotropic flow can fluctuate event to event, both in
magnitude and direction even at fixed impact parameter. 
We describe {\it flow fluctuations} 
by \be
\sigma_{vn}^2 = \mean{v_n^2}-\mean{v_n}^2.
\ee 
One of the important sources of flow fluctuations are fluctuations in
the initial geometry of the overlapping region due to the random nature of the interaction between constituents of the two nuclei.
The participants are those constituents which partake 
in the primary interaction.
The principal axis of the {\em participant} zone can deviate 
from the reaction plane. 
Fig.~\ref{fig:planes} shows the axes in the participant 
coordinate system, compared to the reaction plane system.
It is important to distinguish between flow values measured in
these two systems; the values in the reaction plane system being always
smaller than in the participant plane system: $\vtpp > \vtrp$. 
We discuss flow
fluctuations due to fluctuations in the initial participant zone
geometry in more detail in section~\ref{sec:v2_fluc}.

\section{Experimental methods}
  \subsection{Event plane method}

In the {\it standard event plane 
method}~\cite{Barrette:1996rs, Poskanzer:1998yz} one estimates 
the azimuthal angle of the {\em reaction plane} from the 
observed {\em event plane} angle determined from the anisotropic 
flow itself. This is done for each harmonic, $n$, of the Fourier
expansion. The event flow vector $\bf Q_n$ is a 2d
vector in the transverse plane:
\be
	Q_{n,x} \ &=& \ \sum_{i} w_i \cos(n \phi_i) \ 
	           =\  {\bf Q}_n \cos(n \Psi_n) ,\nonumber \\
	Q_{n,y} \ &=& \ \sum_{i} w_i \sin(n \phi_i)  \ =\ {\bf Q}_n \sin(n \Psi_n),
\label{eq:Qn}
\ee
where the sum goes over all particles $i$ used in the event plane 
calculation. 
The quantities $\phi_i$ and $w_i$ are the lab azimuthal angle and
weight for particle $i$, where for odd harmonics $w_i(-y) = -w_i(y)$. 
The optimal choice for $w_i$ is to approximate $v_n(\pt, y)$. 
Since often $v_n(\pt, y)$ almost linearly increases with $\pt$,
the transverse momentum is a common choice as a weight.
The {\it event plane angle} is the azimuthal angle of ${\bf Q}_n$ calculated as
\begin{equation}
\label{eq:Psi}
	\Psi_n \ =\ \arctan\!2(Q_{n,y}, Q_{n,x}) /n ,
\end{equation}
where $\arctan\!2$ is a C language mathematical function.

The observed $v_{n}$ is the $n^{th}$ harmonic of the azimuthal 
distribution of particles with respect to this event plane:
\begin{equation}
\label{eq:vobs}     
	v_{n}^{\mathrm{obs}}(\pt, y) \ =\ \langle \cos[n(\phi_i - \Psi_n)] \rangle ,
\end{equation}
where angle brackets denote an average over all particles in all 
events with their azimuthal angle $\phi_i$ in a given rapidity 
and \pt momentum space bin at a fixed centrality. 
To remove auto-correlations one has to subtract the ${\bf Q}$-vector 
of the particle of interest from the total event ${\bf Q}$-vector,
obtaining a $\Psi_n$ to correlate with the particle. 
To avoid binning problems one should store the cosine directly 
in a profile histogram, rather than making a histogram 
of $\phi - \Psi_n$ and then obtaining the mean {\rm cos}.

Since finite multiplicity limits the estimation of the angle of 
the reaction plane, the $v_n$ have to be corrected for the {\it event 
plane resolution} for each  harmonic given by
\begin{equation}
\label{eq:resDef}
	\res_n \ = \langle \cos[n(\Psi_n - \psirp)] \rangle ,
\end{equation}
where angle brackets denote an average over a large event sample.
The final flow coefficients are
\begin{equation}
\label{eq:vEP}
	v_{n} \ =\ \frac{v_n^{\mathrm{obs}}}{\res_n}.
\end{equation}
This equation should be applied in a narrow centrality bin. 
For a wide centrality bin, one should average the results from 
the narrow bins weighted with the multiplicity of the bin, 
since $v_n$ is a particle-wise average. 

\begin{figure}[htb]
\begin{center}
\includegraphics[width=0.5\textwidth]{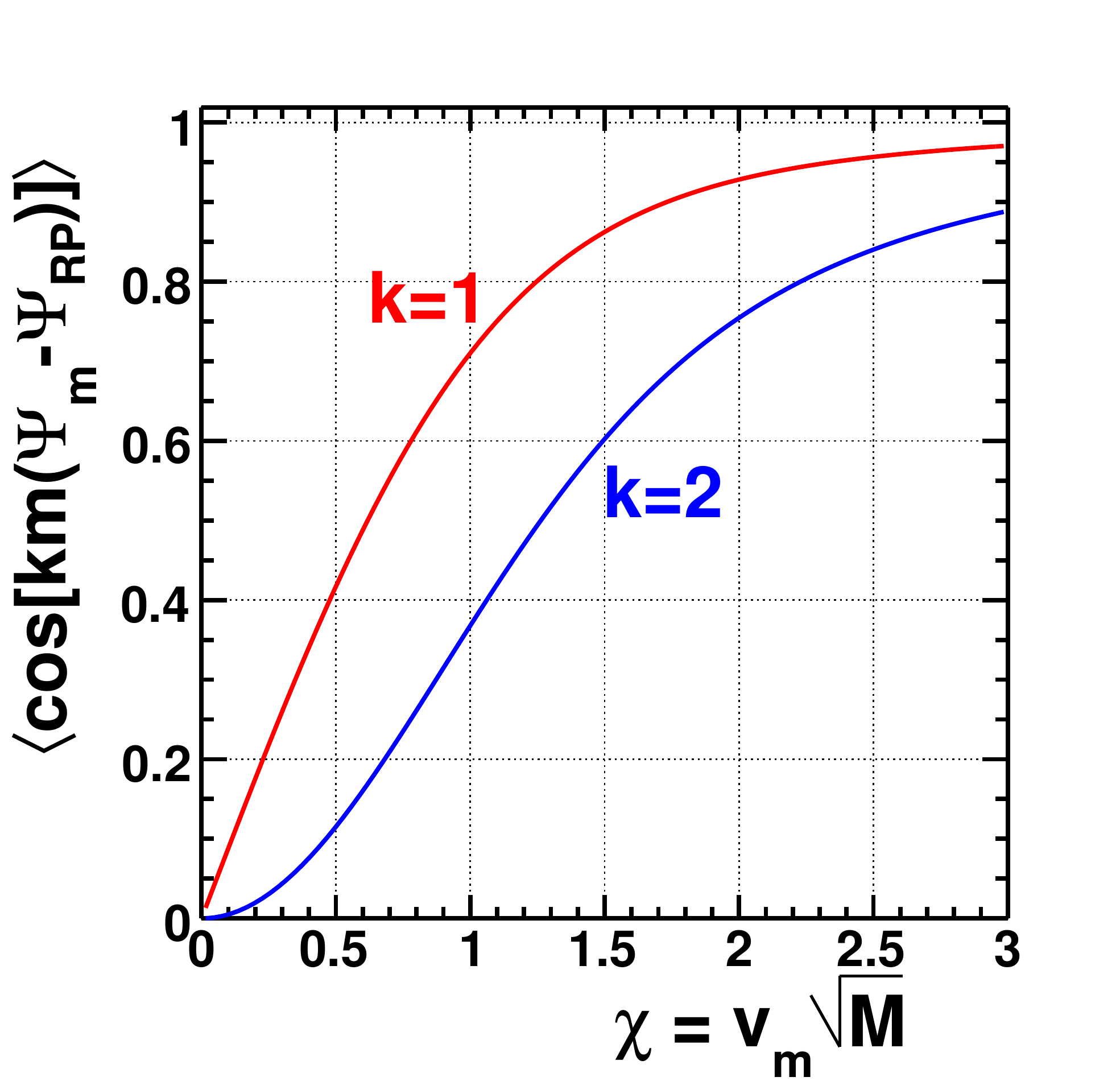}
\caption{The event plane resolution as a function of $v_m \sqrt{M}$. 
The harmonic number of the correlation $n$ is an integer $k$ times 
the harmonic number $m$ of the event plane. }
\label{fig:res}
\end{center}
\end{figure}

The reaction plane resolution depends on the multiplicity of particles
used to define the flow vector and the average flow of these
particles via the resolution parameter~\cite{Ollitrault:1997di,Ollitrault:1997vz,Poskanzer:1998yz}:
\be
\chi = v_n\ \sqrt{M}
\ee
\begin{equation}
 \res_k(\chi) = \sqrt{\pi}/2 \ \chi \exp(-\chi^2/2) \ (I_{(k-1)/2}(\chi^2/2) 
+ I_{(k+1)/2}(\chi^2/2)),
\label{eq:res}
\label{resk}
\end{equation}
where $I$ is the modified Bessel function.
(Note that the definition of parameter $\chi$ in
Ref.~\cite{Poskanzer:1998yz} was larger by $\sqrt{2}$.)
The dependence of Eq.~(\ref{eq:res}) on $\chi$
is shown for the case of $k=1$ in Fig.~\ref{fig:res}. 
To estimate the event plane resolution one divides the full event 
up into two independent sub-events~\cite{Danielewicz:1985hn,Ollitrault:1993ba} of equal multiplicity. 
Since the sub-events are positively correlated because each is 
correlated with the reaction plane, the event plane resolution 
for the sub-events is just the square-root of this correlation:
\begin{equation}
 \res_{n,\mathrm{sub}} 
	= \sqrt{\langle \cos[n(\Psi_{n}^{A} - \Psi_{n}^{B})] \rangle } ,
\label{eq:resSub}
\end{equation}
where A and B denote the two subgroups of particles.
Given $\res_{n,\mathrm{sub}}$, the solution for $\chi$ in Eq.~(\ref{resk}) is done by iteration.
The full event plane resolution is obtained using Eq.~(\ref{resk}) from
the resolution of the sub-events by
\begin{equation}
       \res_{\mathrm{full}} = \res(\sqrt{2}\ \chi_{\mathrm{sub}})
\label{eq:resFull}
\end{equation}
because $\chi \propto \sqrt{M}$ and the full event has twice as many
particles as the sub-events.
In the low resolution ($\!\! \lt 0.5$) linear region of the graph for $k = 1$, $\res_{\mathrm{full}} \approx \sqrt{2}  \res_{\mathrm{sub}}$.

There may be reasons not to use the ${\bf Q}$-vector of the full event, 
but to correlate particles from one sub-event with the ${\bf Q}$-vector 
of the other sub-event. 
In this case the resolution of the sub-event plane should be used, and the particle of interest is automatically not included in the ${\bf Q}$-vector.  
Criteria which have been used for dividing the event into sub-events 
are: random, pseudorapidity, charge, and combinations of these. 
Using sub-events separated in pseudorapidity is a good way of 
reducing contributions from short-range correlations, as flow 
is a large scale effect.

To remove {\it acceptance correlations} from an imperfect detector, 
one must first make the ${\bf Q}$-vector in Eqs.~(\ref{eq:Qn}) isotropic 
in the laboratory, both for the sub-events and the full event (if needed). 
Three methods have been used~\cite{Poskanzer:1998yz} for this 
flattening of the event plane azimuthal distribution:
\ben 
      	\item {\it Phi Weighting -}
	one weights each particle with the inverse of the azimuthal 
	distribution of the particles averaged over many events.
      	\item {\it Recentering -}
	one subtracts from the ${\bf Q}$-vector of each event, the
	${\bf Q}$-vector averaged over many events.
      	\item {\it Shifting -}
	one fits the non-flat distribution of $\Psi_{n}$ averaged over 
	many events with a Fourier expansion and calculates the shifts 
	for each event $\Psi_{n}$ necessary to force a flat
	distribution on average.
\een
By ``many events" we mean a large enough sample to obtain good 
averages, but small enough to avoid shifts in the beam position 
and/or detector response as a function of time. 
The first method is more intuitive, while the second is more practical because it guarantees zero average ${\bf Q}$-vector. 
It is also less sensitive to strong variations in acceptance and can deal 
with ``holes'' in the detector.  
If either of the first two methods is not sufficient, then the third 
method may also be used. 
However, it should be pointed out that only the $n^{th}$ harmonic 
of the flattened distribution needs to be small when $k=1$; that is when one is not dealing with mixed harmonics. 
Calculating the distribution of $\phi - \Psi_n$ and dividing by 
that for mixed events has no advantage over the phi-weight method. 
A complete rigorous treatment of acceptance effects can be achieved 
in the cumulant approach using generating functions~\cite{Borghini:2001vi} 
and similarly, but in a somewhat more transparent way, in the scalar product 
method~\cite{Selyuzhenkov:2007zi}.

An event plane determined from harmonic $m$ allows one to study the flow of harmonics $n=km$, where $k$ is an integer. In Eqs.~(\ref{eq:Qn}) and (\ref{eq:Psi}) 
for the event plane, $n$ is the harmonic number of the event plane, 
but the $n$ in Eq.~(\ref{eq:vobs}) is the harmonic number of 
the correlation, which must be an integral multiple $k$ of the event 
plane harmonic. The case of $k>1$ is called the {\it mixed harmonics method}. 
It was used widely at the AGS and SPS, where in the fixed target setting the detectors
usually cover well the region of rapidity where directed flow is large.
At RHIC it is mostly used to study higher ($n \ge 4$) harmonics relative to
elliptic flow, {\it e.g.} $v_4\{\mathrm{EP_2}\}$.
This is useful because elliptic flow at RHIC is very strong near midrapidity. 
However, it is also true that determining the event plane from directed flow of neutrons 
in a Zero Degree Calorimeter, allows one to greatly suppress nonflow effects in elliptic flow measurements at midrapidity.

The resolution from Eq.~(\ref{eq:res}) is lower
in the mixed harmonic method (see Fig.~\ref{fig:res}) for the case $k=2$,
because of the added difficulty of resolving higher harmonics.
However, the advantages are that nonflow correlations are greatly suppressed because one is correlating two different harmonics of the collective flow, 
which is essentially a three-particle correlation. Also, one can determine the sign of the correlation harmonic relative to the event plane harmonic. The sign of $v_1$ is defined to be positive for nucleons at large positive rapidity.

\begin{figure}[htb]
\begin{center}
\includegraphics[width=0.8\textwidth]{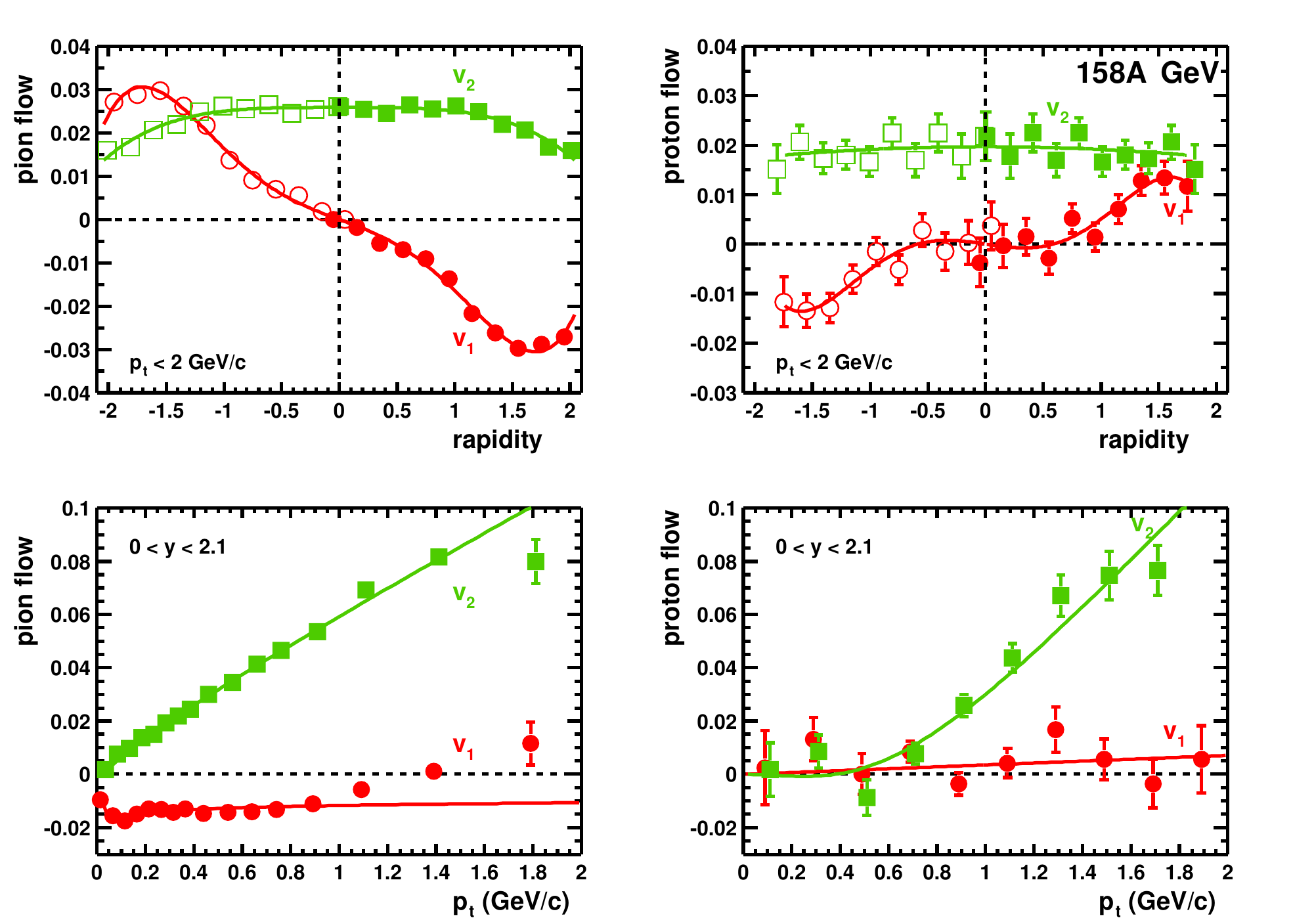}
\caption{Directed and elliptic flow as a function of rapidity 
and transverse momentum from minimum bias 158 A GeV Pb+Pb collisions~\cite{Alt:2003ab}.}
\label{fig:NA49}
\end{center}
\end{figure}

Examples of results from the event plane method are 
shown in Fig.~\ref{fig:NA49}.  
$v_1$ is an odd function of rapidity, with the proton flow in 
the opposite direction to the pion flow. 
$v_2$ is an even function of rapidity, peaking at midrapidity. 
As a function of \pt all curves go to zero at zero $\pt$.

Unless specified otherwise, all flow values presented in this paper will be from the full event plane method with the event plane determined near mid-rapidity.

\subsection{Two and many particle correlations}
\label{sec:two_particle}

The {\it pair-wise correlation method}~\cite{Wang:1991qh},
is based on the fit of the two-particle azimuthal distribution to
that expected from anisotropic flow:
\begin{equation}
  \frac{dN^{pairs}}{d\Delta \phi} 
  \propto (1 +  \sum_{n=1}^\infty 2 v_n^2 \cos(n \Delta \phi))
\label{eq:dPairs}
\end{equation}
where all pairs of particles in a given momentum region are correlated. 
No event plane is used. Acceptance correlations
are removed to first order by dividing by the mixed event distribution.
The harmonic coefficients are small 
because they are the squares of the flow coefficients.
This equation is only for integrated quantities, but normally 
the integrated values are obtained by averaging the differential quantities.

The {\it two-particle cumulant method} 
differs from the previous one only in that instead of the fit to the
two-particle distribution, it calculates the coefficients directly as
\begin{equation}       
  v_n\{2\}^2 = \mean{\cos[n(\phi_1 - \phi_2)]} = \la u_{n,1} u_{n,2}^* \ra 
\label{eq:cum2}     
\end{equation}
for all pairs of particles, where $u_n \equiv e^{in\phi}$ is 
a particle's unit flow vector. The differential flow in
the {\it scalar product method}~\cite{Adler:2002pu} differs from 
the event plane method of Eq.~(\ref{eq:vEP}) by using the magnitude 
of the flow vector as a weight
\begin{equation}       
  v_n(\pt, y) = \frac{ \mean{{\bf Q}_n   u_{n,i}^*(\pt, y)} }
                     {2 \sqrt{ \mean{Q^a_n  {Q^b_n}^*} } } , 
\label{eq:scalProd}     
\end{equation}       
where $u_{n,i}$ is the unit vector of the $i^{th}$-particle 
(which is not included in ${\bf Q}_n$) and $a$ and $b$ are two subevents. 
The resulting statistical errors are slightly smaller than 
with the standard event plane method. If ${\bf Q}_n$ is replaced by its unit vector, this reduces to the standard method.
The differential flow from the two-particle cumulant method is the same as Eq.~\ref{eq:scalProd} with each Q-vector weighted by one over its multiplicity.
The two-particle methods measure $\vtpp$ along the participant plane axis, except when the detector is sensitive to spectator neutrons in the reaction plane.

Since nonflow effects are mainly due to few particle correlations, 
estimates of flow coefficients based on {\it multi-particle correlations} have the distinct advantage of reducing nonflow correlations.
Higher order cumulants are multi-particle correlations where the contributions 
of lower order multiplets have been subtracted. 
In the {\it cumulant method} it can be 
shown~\cite{Borghini:2001vi, Adler:2002pu, Borghini:2001zr, :2008ed} 
that, for example, 
the four-particle correlation minus twice the square of the 
two-particle correlation eliminates two-particle nonflow effects
:
\be    
  \la \la  u_{n,1} u_{n,2} u_{n,3}^* u_{n,4}^* \ra \ra 
  \equiv \la u_{n,1} u_{n,2} u_{n,3}^* u_{n,4}^* \ra    
         -2 \la u_{n,1} u_{n,2}^* \ra ^2 = -v_n^4\{4\} \,,
\label{eq:cum4}    
\ee
where the double brackets indicate the cumulant and $u_{n,i}$ 
is again the $n^{th}$ harmonic unit vector of particle $i$.  
The fourth-root of this result is taken to get $v_n\{4\}$. 
The statistical errors are larger than with the event plane method.
A disadvantage is that $v_n^4$ can sometimes be negative, 
depending on the nature of flow fluctuations (see discussion of
sensitivity to flow fluctuations in Sec.~\ref{sec:sensitivity}).
An advantage is that the cumulant technique allows a consistent treatment of acceptance effects~\cite{Borghini:2001vi}.
Normally, generating functions are used to calculate the
cumulants~\cite{Borghini:2001zr}, but direct calculation is also
possible \cite{Adler:2002pu}. 

Multiparticle cumulants can also involve mixed 
harmonics~\cite{Adams:2003zg, Borghini:2002vp}. 
An important example here is the three-particle 
correlation
\be    
  \la u_{n,1} u_{n,2} u_{2n,3}^* \ra 
          = v_n^2 v_{2n} ,
\label{eq:cum3}    
\ee
which was successfully used at RHIC to suppress
nonflow in the study of $v_1$~\cite{Borghini:2002vp, Adams:2004bi} and $v_4$~\cite{Adams:2003zg}.

\subsection{$q$-distributions, Lee-Yang Zeros, Bessel and
Fourier transforms}

The flow vector ${\bf Q}$ involves all the particles. In the absence 
of correlations, its length would grow as the square-root of 
the multiplicity $M$. 
Thus, to remove most of the multiplicity dependence, a reduced 
flow vector was defined~\cite{Adler:2002pu} as
\begin{equation}
  {\bf q}_n = {\bf Q}_n / \sqrt{M} .
\label{eq:q}
\end{equation}
In the limit of $M \gg 1$
its magnitude is 
distributed~\cite{Voloshin:1994mz,Poskanzer:1998yz,Adler:2002pu,Ollitrault:1995dy}
as
\begin{equation}       
  \frac{dN}{d q_n} = \frac{q_n}{\sigma_n^2} 
                     e^{\displaystyle{-\frac{v_n^2\ M + q_n^2}{2\sigma_n^2}}}  
                     I_0\left(\frac{q_n v_n \sqrt{M}}{\sigma_n^2}\right) ,
\label{eq:dq}
\end{equation}       
where $I_0$ is a modified Bessel function. 
In this {\it $q$-distribution method} one looks at the length 
of the flow vector, not its angle~\cite{Barrette:1994xr}. 
The collective flow shifts the length distribution out 
by $v_n^2\ M$ and fluctuations broaden the distribution. 
Nonflow correlations reduce the effective multiplicity, 
thus also broadening the distribution. 
From just statistical effects, $\sigma_n^2$ in Eq.~\ref{eq:dq} would 
equal $1/2$, but broadening the distribution increases $\sigma_n$:
\begin{equation}       
  \sigma_n^2 = \frac{1}{2} (1 + M \: \sigdyn) ,
\label{eq:sig}
\end{equation}
where
\begin{equation}       
\sigdyn = \delta_n + 2 \sigma_{vn}^2.
\label{eq:sigdyn}
\end{equation}

The {\it Lee-Yang Zeros method} is an all-particle correlation 
designed to subtract nonflow effects to all 
orders~\cite{:2008ed, Bhalerao:2003xf, Borghini:2004ke}.  
It is based on a 1952 proposal of Lee and Yang to detect a liquid-gas 
phase transition. 
Using the second-harmonic flow vector ${\bf Q}_2$, the projection on to 
an arbitrary laboratory angle $\theta$ is
\begin{equation}
\label{eq:Qtheta}
  Q^\theta_2 = \sum_{i=1}^M w_i \cos[2(\phi_i - \theta)],
\end{equation}
where the sum is taken over all the particles $i$ with lab 
angles $\phi_i$ and weights $w_i$. 
Usually five equally spaced values of $\theta$ are used to average 
out detector acceptance effects. 
The essence of the method is to find a zero of a complex generating 
function, but in practice the first minimum of the modulus 
of the generating function along the imaginary axis is used. 
The sum generating function based on $Q_2^\theta$ (which is a sum) is given by
\begin{equation}
\label{eq:GthetaSum}
  G^\theta_2(ir) = \ \mid \langle e^{\mathit{i}rQ^\theta_2} \rangle \mid,
\end{equation}
where $r$ is a variable along the imaginary axis of the complex 
plane and the average is taken over all events. 
The square of the modulus is used to determine the first minimum. 
The position of the first minimum at the lab 
angle $\theta$ is $r_{0}^{\theta}$, and, for the case of unit weights, 
is related to the ``integrated'' flow by
\begin{equation}
\label{eq:Vtheta}
  V_2^\theta = \mathit{j}_{01} / r_0^\theta ,
\end{equation}
\begin{equation}
\label{eq:v}
  v_2 = \mean{V_2^\theta}_\theta / M,
\end{equation}
where $\mathit{j}_{01} = 2.405$ is the first root of the Bessel 
function $J_0$ and $M$ is the multiplicity. 
In Eq.~(\ref{eq:v}) the average is taken over the lab angles $\theta$. 
As in all multi-particle methods, this $v_2$ is $\vtrp$~\cite{Voloshin:2007pc, Bhalerao:2006tp}, along 
the reaction plane axis. This can be thought of in this way: $v_2$ is lower than $\mean{\vtpp}$ by the same amount as $\vtrp$ is lower than $\vtpp$.
A variant Lee-Yang Zeros sum generating function method has been 
devised~\cite{Bilandzic:2008nx} which produces an event 
plane ${\bf Q}$-vector calculated with weights designed to eliminate 
autocorrelations and nonflow effects.

The product generating function is
\begin{equation}
\label{eq:GthetaProd}
  G^\theta_2(ir) = \ \mid \langle \prod_{j=1}^M [ 1 + \mathit{i}r w_j \cos(2(\phi_j - \theta))] \rangle \mid .
\end{equation}
Calculation of this generating function requires more computer time because the product over all particles 
has to be calculated for each value of $r$. 
Although the sum generating function works fine for $v_2$, analyses 
for $v_4$ (and $v_1$~\cite{Borghini:2004ad}) relative to $v_2$ have 
to be based on the product generating function. 
This is because the product generating function is better 
at suppressing autocorrelation effects which are more important 
for mixed harmonics~\cite{Bhalerao:2003xf}.  
The Lee-Yang Zeros method only works for a sufficient 
signal-to-noise ratio. 
Since the signal is $v_2$ and the noise is proportional 
to $1/\sqrt{M}$, the parameter $\chi = v_2 \sqrt{M}$ determines 
the applicability of the method. 
It is found that the errors get large and the results scatter 
when $\chi \lt 0.8$, or the full event plane resolution is less than 0.6. 
When there is no flow the method will find a minimum from a fluctuation.
For STAR at $\sqrtsNN=200$~GeV Au+Au~\cite{:2008ed} the method fails for central collisions because $v_2$ is small, and for peripheral collisions because the multiplicity is small.

The method of {\em Fourier and Bessel transforms}~\cite{Voloshin:2006gz} 
of the flow vector distributions is intimately related to 
the Lee-Young Zeros method; it clearly
illustrates how the separation of nonflow effects happens.
Let $f_0(Q_{n,x})$ denote the distribution in the $x$ component
of the flow vector (Eq.~(\ref{eq:Qn})) for the case of zero flow, $v_n=0$.
Then, in the case of non-zero flow, and under the condition $\sqrt{M} \gg 1$, 
the corresponding distribution can be written as
a superposition of $f_{0}$ distributions ``shifted'' in the
direction of flow by an appropriate amount
depending on the reaction plane angle~\cite{Voloshin:1994mz}:
\be
f(Q_{n,x}) \equiv \frac{dP}{dQ_{n,x}} 
= \int \frac{d\Psi}{2\pi} f_{0}(Q_{n,x}-  v_n M\cos(n\Psi)).
\ee
The Fourier transform of this distribution is:
\be
\tilde{f}(k) &=& \la e^{ikQ_{n,x}} \ra =
 \int \frac{d\Psi}{2\pi} \int dQ_{n,x} e^{ikQ_{n,x}} f_{0}(Q_{n,x}
-v_n M\cos(n\Psi))   
\nonumber \\
&=& 
 \int \frac{d\Psi}{2\pi} e^{ikv_n M\cos(n\Psi)}   \int dt e^{ikt} 
f_{0}(t) 
 = J_0(k v_n M) \tilde{f}_{0}(k).
\label{eq:efQx}
\ee
Remarkably, the flow contribution is completely factored out, and
the zeros of the Fourier transform are determined by
the zeros of the Bessel function $J_0(kv_nM)$. One finds  
\be
\label{eq:vBT}
  v_n = j_{01} / (k_1 M),
\ee  
where $k_1$ is the first zero of the Fourier transform.
The above result is the same as that obtained applying the
Lee-Yang Zeros method using the sum generating
function. In fact this relation to
the Fourier transform was already pointed out in the original 
paper~\cite{Bhalerao:2003xf}.

The two-dimensional Fourier transform of
$d^2P/dQ_{n,x} dQ_{n,y}$
\be
\tilde{f}(k) &=& 
\int dQ_{n,x} e^{ik_x Q_{n,x}} dQ_{n,y} e^{ik_y Q_{n,y}}  
\frac{d^2P}{dQ_{n,x} dQ_{n,y}}
\nonumber \\
&=& 
\int dQ_n J_0(kQ_n) \frac{dP}{dQ_n}
~\sim~
 J_0(k v_n M), 
\ee
is reduced to the Bessel transform of the
distribution in the magnitude of the flow vector. 
Note that in this approach (valid in the limit of $\sqrt{M} \gg 1$) the
flow contribution is decoupled from all other
correlations, due to the collective nature of flow.
Note also that in the same limit one expects the distribution of flow
vectors to be Gaussian due to the Central Limit Theorem, thus
explaining why fitting the distribution to the form derived
in Ref.~\cite{Voloshin:1994mz} (such fits have been
used in Ref.~\cite{Barrette:1996rs,Adler:2001nb}) 
is also not sensitive to nonflow correlations. Thus
in this limit all three methods, the Bessel Transform, Lee-Yang
Zeros, and fitting the $q$-distribution, become very similar, 
if not equivalent.

\subsection{Methods comparison: sensitivity to nonflow and flow fluctuations}
\label{sec:sensitivity}

The results obtained with different methods discussed in the previous section are affected by nonflow and flow fluctuations in different ways. 
Also, in some methods the results are closer to flow values in the
participant plane and in others to the reaction plane. 
For example, correlating
particles in the same rapidity range with a single harmonic would
measure flow in the participant plane, but using a mixed harmonic
method, with the first harmonic determined from spectator 
neutrons, would provide elliptic flow in the reaction plane.

{\it Nonflow} $\delta_n$ is defined by Eq.~(\ref{eq:delta}). 
These are correlations not associated with the reaction plane. 
Included in nonflow effects are jets, resonance decay, 
short-range correlations such as the Hanbury-Brown Twis (HBT) effect,
and momentum conservation~\cite{Borghini:2000cm, Dinh:1999mn}. 
There exists several methods to evaluate and
suppress nonflow contributions, such as using
rapidity gaps between correlated particles, 
using different charge combinations
for correlated particles (to assess the contribution of resonances), etc.

Since nonflow correlations are mainly few-particle effects,
$\delta_n$ roughly scales as
the inverse of the multiplicity\footnote{
This is true under assumption that the relative strengths
of nonflow effects do not change with centrality.
In reality one should probably expect some increase in nonflow
effects in more central collisions due to larger relative
contribution of hard parton collisions, and/or tighter azimuthal correlations
from modification of the correlation due to stronger radial
flow~\cite{Voloshin:2003ud}.
}.
It leads to an almost constant contribution to the dependence
of the expression
$M\mean{u u^*}$ on centrality while flow has a maximum
for mid-central collisions, because in
peripheral collisions the multiplicity is small
and in central collision the anisotropic flow become small.
Similarly, one can ``subtract'' nonflow contribution in flow
measurement using the so called {\it $AA-pp$ method}, as the nonflow
contribution to the correlator $\mean{uQ^*}$, is constant.
For the scalar product method one can consider using
\be
\mean{uQ^*}_{AA,corrected}=\mean{uQ^*}_{AA}-\mean{uQ^*}_{pp} ,
\label{eq:QAApp}
\ee
where $AA$ refers to nucleus-nucleus and $pp$ refers to proton-proton.
One can also {\em assume} a particular shape of nonflow correlations
as a function of
the difference in particle rapidity and/or azimuthal angles.
Then one can estimate the flow value with the fit to the correlation function.
We denote the results obtained from the fit to a 2-dimensional
correlation function as $v_2\{2d\}$.
Note that although this notation is similar to
one employed in Ref.~\cite{Trainor:2007fu}, the meaning is different.

\begin{figure}[htb]
\begin{center}
\includegraphics[width=0.6\textwidth]{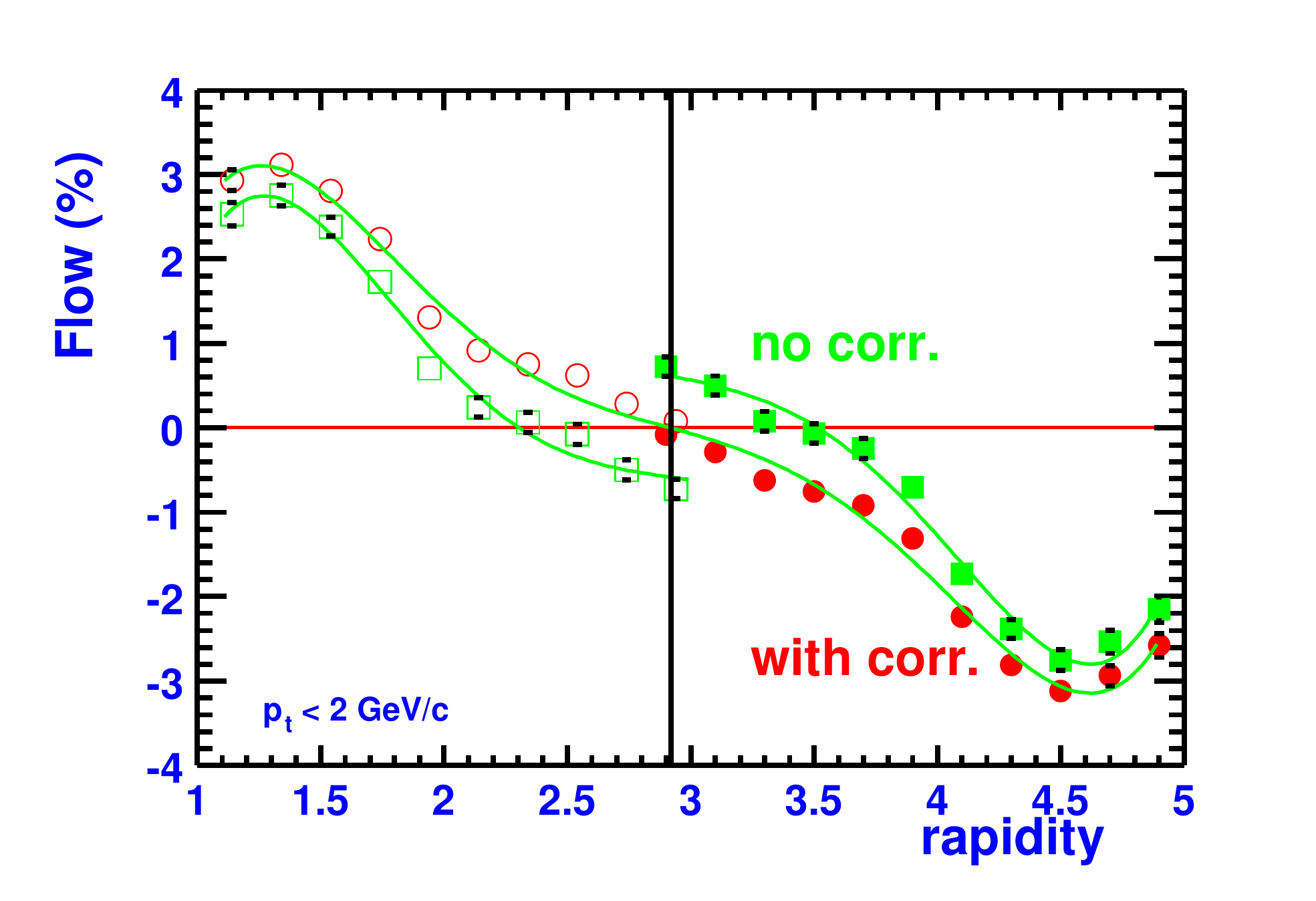}
\caption{Directed flow as a function of rapidity for charged pions from 
minimum bias 158 A GeV Pb+Pb collisions~\cite{Borghini:2002mv}. 
Shown are $v_1$ before (squares) and after (circles) correction 
for momentum conservation.}
\label{fig:pCons}
\end{center}
\end{figure}

Unfortunately, the above mentioned techniques do not allow real quantitative
estimates of the residual nonflow.
Using multi-particle methods is more attractive in this sense, as they suppress nonflow effects by $\sim 1/M$ for each extra particle in the correlator. 
Estimates show that measuring elliptic flow at RHIC using
4-particle correlations almost completely removes nonflow effects.
The largest remaining systematic uncertainty is due to contributions to higher order cumulants from correlations when two particles, which are daughters
of a resonance decay, are correlated with all other particles in the
multiplet via flow of the resonance. Unfortunately, this effect can not be suppressed by using higher order cumulants.  
Global {\it momentum conservation} can affect measurements of directed
flow, or elliptic flow measured with respect to the first harmonic
event plane, if the detector acceptance is not symmetric about
mid-rapidity. It is a long-range effect and not reduced by a gap in pseudorapidity. The effect causes a discontinuity in $v_1$ at mid-rapidity, 
as seen in Fig.~\ref{fig:pCons}. 
If one can estimate the fraction of all produced particles which are detected, 
then a correction can be made for this 
effect~\cite{Borghini:2002mv}, as is also shown in Fig.~\ref{fig:pCons}.
Momentum conservation is unimportant when the event plane is 
determined from an even Fourier harmonic, or for a detector having 
symmetric acceptance around mid-rapidity. 

{\em Flow fluctuations} also affect different methods differently.
The effect on cumulant results are the easiest to trace
(though it does not mean it is easy to measure the fluctuations). 
The dependence follows directly from the definitions, 
Eqs.~(\ref{eq:cum2}, \ref{eq:cum4}):
\be
v\{2\} &=& \sqrt{\mean{v^2}} = (\mean{v}^2 + \sigma_v^2)^{1/2} , \\
v\{4\} &=& (2\mean{v^2}^2-\mean{v^4})^{1/4}, \\
v\{6\} &=& ((1/4) (\left<v^6\right> - 9 \mean{v^4} \mean{v^2} + 12 \mean{v^2}^3))^{1/6},
\ee
etc. Note that although $v_2\{2\}$ can be written in terms of only
$\mean{v}$ and $\sigma_v^2$, $v\{4\}$ and higher cumulants 
in general require knowledge of higher order
moments of the distribution in $v$.
If needed, a  model has to be  used for the distribution of $v$
to relate the contribution of flow fluctuations to
different order cumulants. For example, for a Gaussian distribution~\cite{Voloshin:2007pc} in $v$, 
\be
v_2\{2\} &=& (\mean{v}^2+\sigma_v^2)^{1/2} 
\approx \mean{v} + \sigma_v^2/(2 \mean{v}), \\
\vtf &=&(\mean{v}^4-2\sigma_v^2 \mean{v}^2 -\sigma_v^4)^{1/4}
\approx \mean{v} - \sigma_v^2/(2 \mean{v}), \\
v_2\{6\} &=& (\mean{v}^6-3\sigma_v^2 \mean{v}^4)^{1/6} 
\approx \mean{v} - \sigma_v^2/(2 \mean{v}). 
\ee
Note that the above relations are also valid for any other
distribution in the limit $\sigma_v \ll \mean{v}$.   

\begin{figure}[htb]
\begin{minipage}[b]{0.48\textwidth}
\begin{center}
  \includegraphics[width=.95\textwidth]{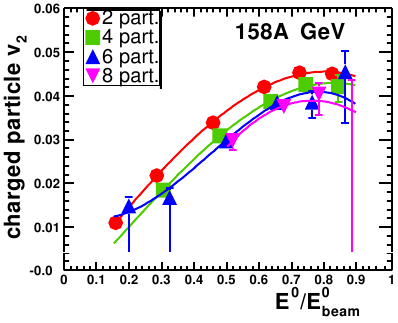}
  \caption{Many particle cumulant results for charged hadron $v_2$ from 
158A Pb+Pb as a function of centrality with the most central at the left. 
Lines are polynomial fits~\cite{Alt:2003ab}.}
  \label{fig:v2cumulantsNA49}
\end{center}
\end{minipage}
\hspace{\fill}
\begin{minipage}[b]{0.48\textwidth}
\begin{center}
  \includegraphics[width=.99\textwidth]{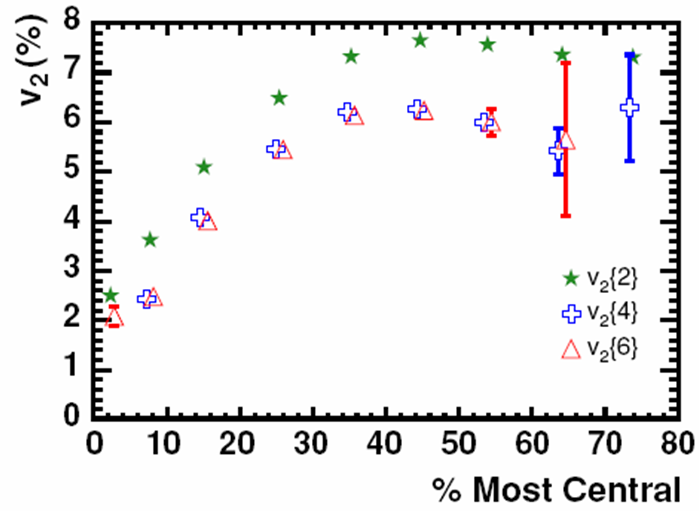}
  \caption{Many particle cumulant results for charged hadron $v_2$ 
  from $\sqrtsNN = 130$~GeV Au+Au as a function of centrality~\cite{Adler:2002pu}.}
  \label{fig:v2cumulantsSTAR}
\end{center}
\end{minipage}
\end{figure}

For $v_n$ fluctuations according to a Bessel-Gaussian distribution:
\begin{equation}       
  \frac{dN}{v_n d v_n} = \frac{1}{\sigma_n^2} 
                         e^{\displaystyle{-\frac{v_n^2 + v_0^2}{2\sigma_n^2}}}  
                         I_0\left(\frac{v_n v_0}{\sigma_n^2}\right)
	\equiv \BG(v_n;v_0,\sigma_n)
\label{eq:BG}
\end{equation}       
the cumulants are~\cite{Voloshin:2007pc}:
\be
v_2\{2\}^2 &=& v_0^2+2 \sigma^2_n , \\
v_2\{n\} &=& v_0,~n \ge 4 .
\ee
Note, that the Gaussian model of 
elliptic flow fluctuations due to eccentricity fluctuations
discussed in section~\ref{sec:v2_fluc}
results in a Bessel-Gaussian distribution in $v_2$ with the parameter $v_0$ 
being $v_{2,\mathrm{RP}}$~\cite{Voloshin:2007pc}, 
elliptic flow along the reaction plane axis.
We also note that 
Eq.~(\ref{eq:dq}) is a Bessel-Gaussian for $q$, while
Eq.~(\ref{eq:BG}) is the Bessel-Gaussian for $v_n$. 
Figures~\ref{fig:v2cumulantsNA49} and \ref{fig:v2cumulantsSTAR} are 
consistent with these equations based on the Bessel-Gaussian distribution.

The event plane method dependence on flow fluctuations is more
complicated and ranges from 
$\vtep=\vtt = \mean{v^2}^{1/2}$ to $\vtep =\mean{v}$
depending on the reaction plane resolution~\cite{Alver:2008zz}. 
Defining parameter $\alpha$ via  $\vtep=\mean{v_2^\alpha}^{1/\alpha}$
one finds that $\alpha \approx 2$ for small values of resolution and 
approaches  unity for large values. 
In Ref.~\cite{Alver:2008zz} this observation was made based on Monte-Carlo
simulations, but
this dependence also can be obtained analytically in the case
of small fluctuations, or numerically with direct integration over the
$v$ distribution in the numerator and 
denominator of~Eq.~(\ref{eq:vEP})~\cite{opv}. 

The dependence on fluctuations of the Lee-Yang Zeros  and other 
similar methods, such as fitting the $q$-distribution or using Fourier-Bessel
transforms, is also non-linear. As was shown in Ref.~\cite{Voloshin:2006gz} by
Monte-Carlo simulations, the dependence is close to that 
of higher ($n \gt 2$) cumulants. 
In the small fluctuation limit it also can be obtained analytically.
Note that for the Bessel-Gaussian distribution in $v$, these
methods yield the same results as higher cumulants, 
namely $\mean{v}=v_0$ of the Bessel-Gaussian. 
Thus, if the distribution of $v_2$ is Bessel-Gaussian, 
all the multi-particle methods should give the same result: 
$\mean{v} = v_0 = \vtrp$~\cite{Voloshin:2007pc}.

\begin{figure}[htb]
\begin{center}
\includegraphics[width=0.8\textwidth]{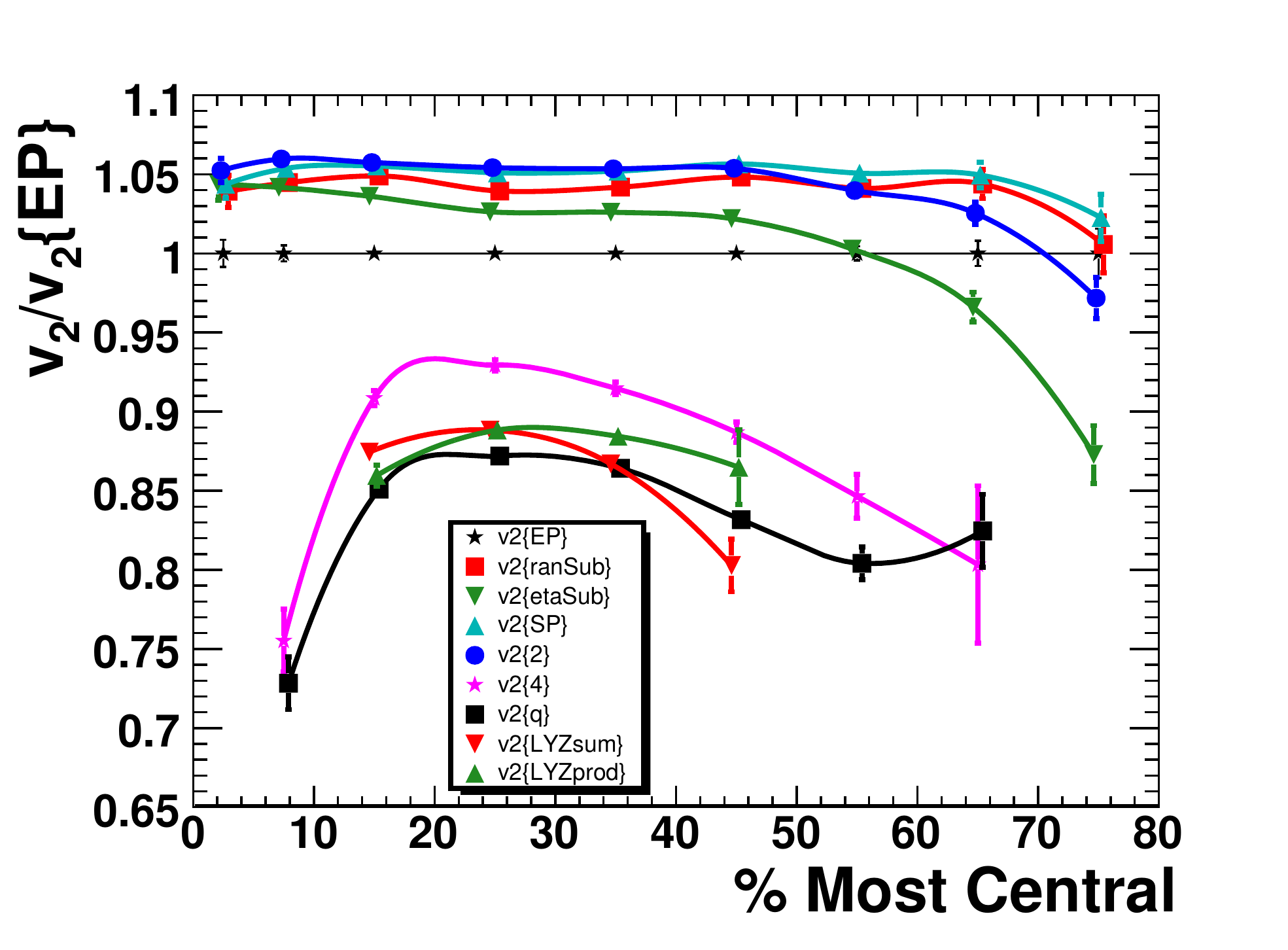}
\caption{Charged hadron $v_2$ divided by \vtep as a function of geometrical 
cross section for $\sqrtsNN = 200$~GeV Au+Au. The values are integrated 
over \pt for $\mid \eta \mid \ < 1.0$, except for the LYZ product generating function which 
had $\mid \eta \mid < \ 1.3$. Results are shown for the event plane method, 
random sub-events, pseudorapidity sub-events, scalar product, two-particle 
cumulants, four-particle cumulants, $q$-distribution, and Lee-Yang Zeros sum 
generating and product generating functions~\cite{:2008ed, Adams:2004bi}.}
\label{fig:comp}
\end{center}
\end{figure}

The above equations can be used for removing the ``trivial'' effect
of flow fluctuations due to variation of elliptic flow within
a wide centrality bin~\cite{Adler:2002pu}. As discussed below, real flow 
fluctuations~\cite{Adler:2002pu,Aguiar:2001ac}
are very difficult to separate from nonflow effects.

{\it Methods comparisons} 
are shown in Fig.~\ref{fig:comp} for charged hadron 200 GeV Au+Au results. Plotted is the integrated $v_2$ divided by the values for the standard event plane method $\vtep$. This is an update of Fig.~29b from 
Ref.~\cite{Adams:2004bi} by the addition of values for the Lee-Yang Zeros 
methods~\cite{:2008ed}. The results fall into two bands: the upper 
band for the two-particle correlation results and the lower band for 
the multi-particle results. The event plane values are about 5\% below 
most of the other two-particle results, and the pseudorapidity sub-event 
values for peripheral collisions drop below the other two-particle results.
The pseudorapidity sub-event method apparently succeeds in removing more 
nonflow, especially for peripheral collisions.

According to our current interpretation, the upper band represents averages 
along the participant plane with nonflow and fluctuation contributions,
and the lower band averages along the reaction
plane mostly free of nonflow and fluctuation contributions. 
The separation of the two bands is a function of \sigdyn:
\begin{equation}       
  v_2\{2\}^2 - v_2\{4\}^2 = \sigdyn = \delta_2 + 2 \sigma_{v2}^2 .
\label{eq:cumSigDyn}
\end{equation}      
The event plane method is a special 
case with results being somewhere between $\vtt$ and $\vtf$ depending on the
reaction plane resolution.
For the {\it higher order cumulants} it can be seen in 
Figs.~\ref{fig:v2cumulantsNA49} and \ref{fig:v2cumulantsSTAR} that although 
the two-particle cumulant values are above the four-particle cumulants, all 
the still higher order cumulants agree with $v_2\{4\}$.
\begin{figure}[htb]
\begin{center}
\includegraphics[width=0.48\textwidth]{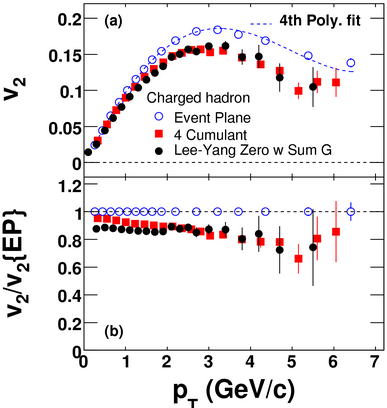}
\caption{A comparison of results for charged hadron four-particle cumulants 
and Lee-Yang Zeros with sum generating function, to the event plane method 
as a function of \pt for $\mid \eta \mid \lt 1.0$ in 10--40\% most central 
Au+Au collisions at $\sqrtsNN = 200$~GeV. The bottom panel shows the ratios 
of $v_2$ divided by \vtep~\cite{:2008ed}.}
\label{fig:nonflow}
\end{center}
\end{figure}
Figure~\ref{fig:nonflow} compares the four-particle cumulant, the Lee-Yang 
Zeros, and the event plane results as a function of $\pt$. The cumulant 
values decrease with \pt compared to \vtep as one would expect for a nonflow contribution from jets at high $\pt$, but the Lee-Yang Zeros values do not for some unknown reason.

           %

\section{Anisotropic flow: results and physics}

\subsection{General}

It has been found that many physical processes contribute to the
development of anisotropic flow during the evolution of the system
created in the collision. 
Though different flow harmonics often reflect different
physics, there are some common features, such as 
the mass dependence of the differential flow in the low transverse
momentum region, or the role of coalescence, both of which we discuss in more details later in this section.
The entire collision is viewed as going through several stages:  
the formation of the initial state (the result of this stage is
often referred as the ``initial conditions'') that takes time of the order
of the spatial dimensions of the Lorentz contracted nuclei, 
thermalization period, 
(viscous) hydrodynamic expansion, ``post-hydro'' expansion that is most
frequently simulated by a hadronic cascade, and finally chemical and kinetic
freeze-out. These stages might not have well identified boundaries.
The freeze-out stage is very likely continuous as a natural
evolution of the cascade phase and not occurring suddenly.
Formation of the anisotropic flow occurs during
each and every one of these stages, but depending on the collision energy,
rapidity and transverse momentum, the importance of different stages changes. From the theoretical point of view our knowledge of
different stages is also non uniform.
2d hydro is obviously much better studied than 3d. 
Though there has been significant progress in understanding the
initial condition in the Color Glass Condensate (CGC) model, one would find that probably the least is  known about the very first stages and in particular
about the rapidity dependence of the ``initial'' conditions.
Note that although several mechanisms
could lead to very fast thermalization (at the level of a fraction of a
fermi), the size of the nuclei after relativistic contraction $\sim R_A \, 2 m_N/\sqrtsNN$ is
not negligible, and as noticed by Stock~\cite{Stock:private} could
delay the start of elliptic flow development and consequently lower
$v_2$ values, except probably at the highest RHIC energies. 

Very little is known about
flow fields at the end of the thermalization stage; this question is just starting to be explored.
An example here would be the calculations of elliptic flow in the CGC
approach, where the entire result might be thought of as due to the
initial flow field. The results obtained in this model do not agree
with the data, but might be used as an initial condition for hydrodynamic calculations. Another interesting 
attempt~\cite{Broniowski:2008vp} (see also similar ideas in an earlier
paper~\cite{Gyulassy:2007zz}) to obtain the initial
flow field is via the so called ``Landau matching condition'' for the
energy momentum tensor, assuming initial free streaming of partons
with rapid thermalization at times of about one fermi.
The possibility for such initial flow in particular allows one to start
the hydrodynamic evolution at realistic times of about one fermi, and
not loose the ability to reproduce large elliptic flow.

\begin{figure}[htb]
\begin{minipage}[b]{0.48\textwidth}
  \centerline{\includegraphics[width=0.95\textwidth]{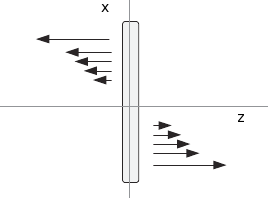}}
\caption{Initial longitudinal velocity profile in non-central nuclear 
collisions~\cite{Becattini:2007sr}.
\vspace{5mm}
}
\label{fig:becat}
\end{minipage}
\hspace{\fill}
\begin{minipage}[b]{0.48\textwidth}
\centerline{\includegraphics[width=0.9\textwidth]{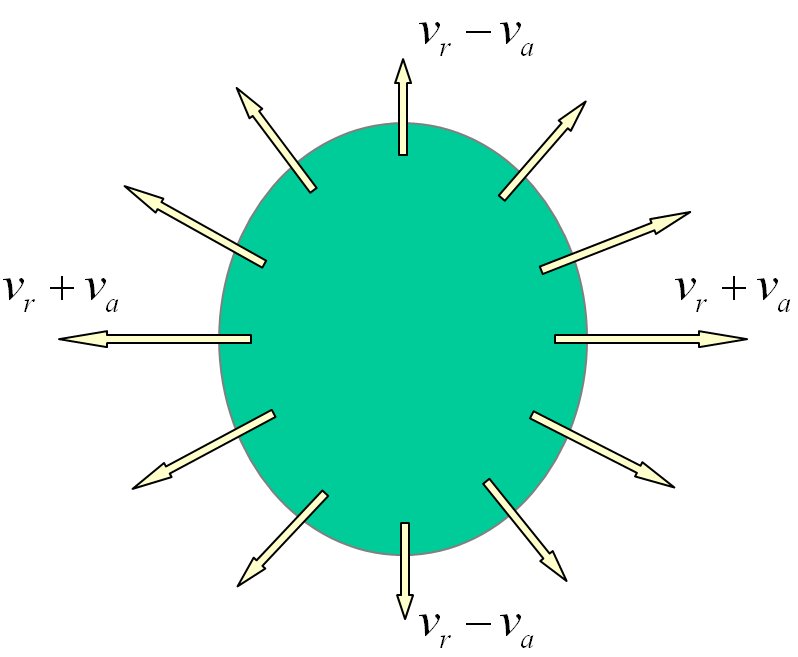}}
\caption{
Radial flow modulated by the elliptic component.
The solid ellipse shows the overlap region and the
arrows show the direction and magnitude of the expansion velocity.
}
\label{fig:radExp}
\end{minipage}
\end{figure}

In terms of initial longitudinal velocity fields, it might be
important to take into account the initial
velocity  gradient along the impact parameter as illustrated in Fig.~\ref{fig:becat}. Such a gradient directly contributes to the
in-plane expansion rate (see Eq.~23 in Ref.~\cite{Becattini:2007sr}). 
This effect naturally
also leads to {\em directed} flow (see the same Eq.~23), 
which is briefly addressed in~\cite{Troshin:2007cp}. 
It will be very interesting to compare the calculations in such 
a model to very precise data from STAR~\cite{Wang:2007kz}
on directed flow.
The relation to other models~\cite{Csernai:2006yk,Snellings:1999bt}
predicting non-trivial dependence of directed flow on rapidity would
be very interesting. 
Speculating on this subject one notices that viscous effects
must also play an important role in such a scenario. 
  
\subsubsection{Interplay of anisotropic and radial flow}
\label{sec:rad-ani}

In the collision scenario where the final particles are produced via freeze-out of the (locally) thermalized matter exhibiting collective flow, 
an interplay of radial expansion and anisotropic flow 
leads to a characteristic dependence 
of the differential flow $v_n(\pt)$ on the mass of 
the particle~\cite{Voloshin:1996nv,Huovinen:2001cy}.
At low transverse momenta heavier particles have lower $v_n(\pt)$;
for large mass and low temperature the flow $v_n(\pt)$  
can become even negative (opposite to the integrated value).
The mass ``splitting'' of the differential flow being dependent
simultaneously on all three velocities (radial, anisotropic and
thermal) can be a sensitive test of the theoretical models, in
particular to the presence of the phase transition during the
evolution of the system (see discussion below in section~\ref{EllipticFlow}). 

The physics of this phenomena is rather simple and depends on the relative
magnitudes of the three velocities: average radial expansion velocity,
anisotropic velocity 
(the amplitude of modulation in radial expansion velocity as
function of the relative angle to the reaction plane), 
and thermal velocity, which
depends on the temperature and the mass of the particle.
In the case of elliptic flow,
the anisotropic component is positive for the in-plane direction and negative
for the out-of-plane direction, see Fig.~\ref{fig:radExp}.
When thermal velocities are small, particles with low \pt are produced
mostly from regions of the source with low radial
velocity, where the anisotropic flow velocity component is 
negative in order to partially compensate the radial flow.
In the case when thermal velocities are large (light particles) compared to
anisotropic velocity modulations, the effect becomes small. 
The description of the effect is most transparent in the blast wave
model~\cite{Adler:2001nb,Voloshin:1996nv,Huovinen:2001cy,Retiere:2003kf}.

Blast wave fits to differential flow measurements of data have been first applied in Ref.~\cite{Barrette:1997pt}, fitting proton $v_1(\pt)$ in Au+Au
collisions at the AGS and 
reasonable values for anisotropic flow velocities were obtained.
Elliptic flow was very successfully fit in Ref.~\cite{Adler:2001nb} where
the model was further developed including a parameter responsible
for the spatial geometry of the system. It is further discussed
in section~\ref{sec:v2_lowpt}.
The effect is stronger for heavier particles. It is noteworthy that 
elliptic flow of $\jpsi$~\cite{Franz:2008ri} and deuterons~\cite{Liu:2007wu} 
become negative at low transverse momenta.

\subsubsection{Flow amplification by coalescence}
\label{sec:coalescence}

The distributions of particles produced via coalescence of some
primordial particles should reflect the original distributions of 
the primordial particles. If the coalescence process does not significantly affect the distributions of
the original particles, then the standard coalescence formula can be applied 
(written here for the case of nucleon coalescence into light nuclei
of atomic number $A$):
\be
 \frac{E_{A} d^3n_{A}}{d^3p_A} = B_A  \left( \frac{E_{p}
d^3n_{p}}{d^3p_p} \right)^A ,
\label{eq:Acoales}
\ee
where the coefficient $B_A$ reflects the coalescence probability;
in particular it includes the integration over the spatial distribution
of nucleons and is inversely proportional to the correlation volume. 
For simplicity in this equation it is assumed that
protons and neutrons have similar momentum distributions.
Neglecting for the moment possible non-uniformity in the spatial 
distribution and concentrating on the momentum distribution, one finds
that if nucleons are subject to anisotropic flow, the
distribution in Eq.~(\ref{eq:Acoales}) of nuclei $A$  becomes even more anisotropic. In this context coalescence leads to a simple scaling relation for flow:
\be
v_{n,A}(p_{T,A}) \approx A v_{n,p}(p_{T,A}/A).
\ee
The E877 Collaboration~\cite{Voloshin:1997rs,Barrette:1998bz} at the AGS 
observed that the directed flow 
of deuterons indeed followed this coalescence rule.
Scaling violations were observed at rapidity close to the beam
rapidity, and attributed to the change in spatial distribution of
nucleons in- and out-of-plane.
Elliptic flow of deuterons was studied at RHIC by the
PHENIX~\cite{Afanasiev:2007tv} and STAR~\cite{Liu:2007wu} Collaborations.  
Both collaborations found good consistency with predictions of a
coalescence model.

Anisotropic flow enhancement due to coalescence can be used also as a tool for the study of resonance regeneration, such as $\phi$ or $K^*$, during hadronic
evolution of the system. The idea is that if regeneration is important, and a
significant fraction of the resonances are produced via coalescence of other hadrons, then the elliptic flow of the resonances should be enhanced compared 
to that of direct production~\cite{Adams:2004ep}.

It appears that {\em constituent quark}
coalescence~\cite{Voloshin:2002wa} plays a very important role
in particle production in the intermediate \pt region; its relation
to the formation of elliptic flow is discussed in section
\ref{sec:v2_coalescence}.

\subsection{Directed flow}

\subsubsection{Physics of directed flow}

\begin{figure}[htb]
   \begin{center}
     \includegraphics[width=0.55\textwidth]{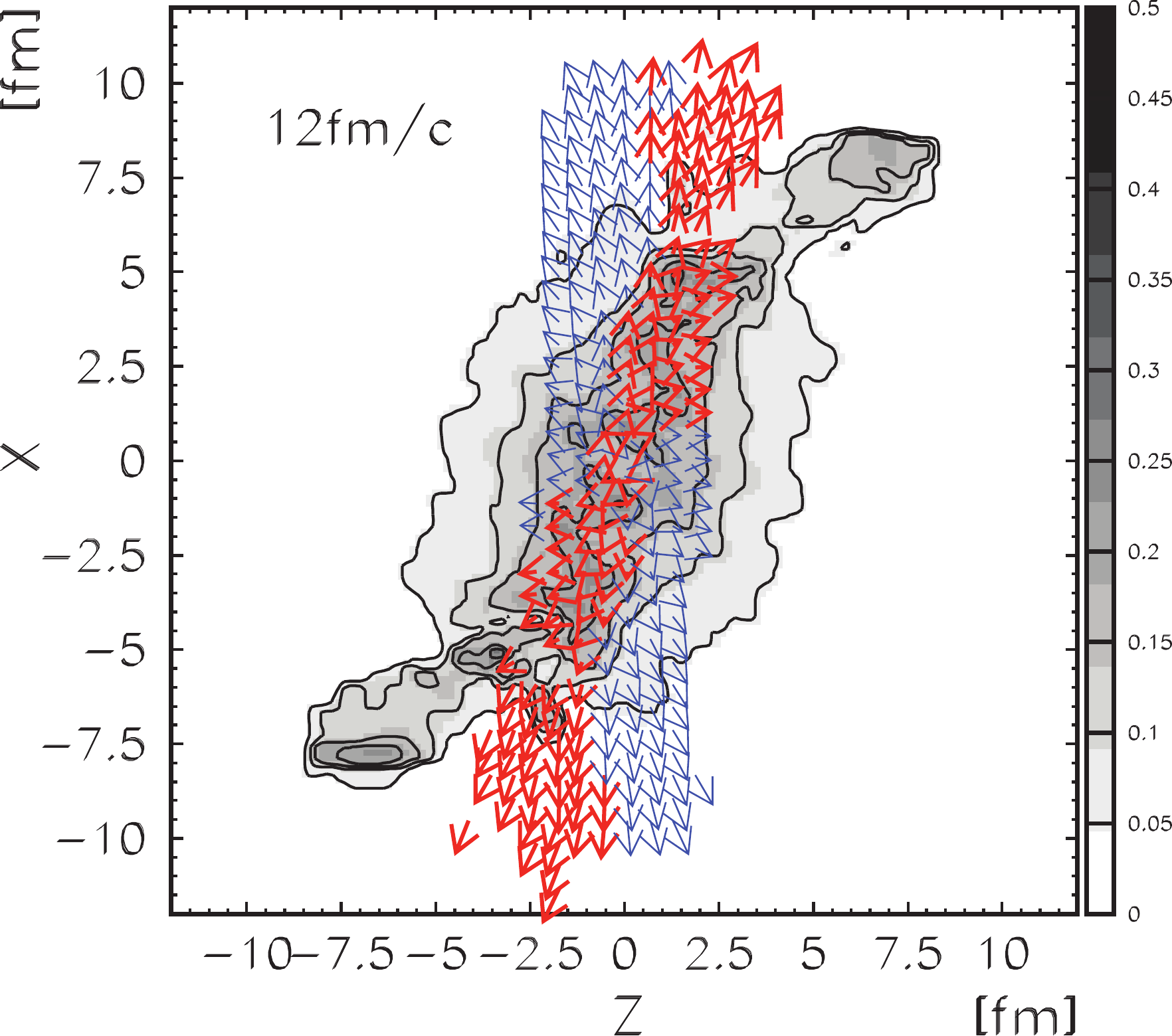}
     \caption{Net-baryon density at t = 12 fm/$c$ in the
	reaction plane with velocity arrows for midrapidity ($|y| < 0.5$) 
	fluid elements: Antiflow -
	thin arrows, Normal flow - bold arrows. 
	From Ref.~\cite{Brachmann:1999xt}. }
     \label{fig:antiflow}
   \end{center}
\end{figure}

\begin{figure}[htb]
   \begin{center}
     \includegraphics[width=0.75\textwidth]{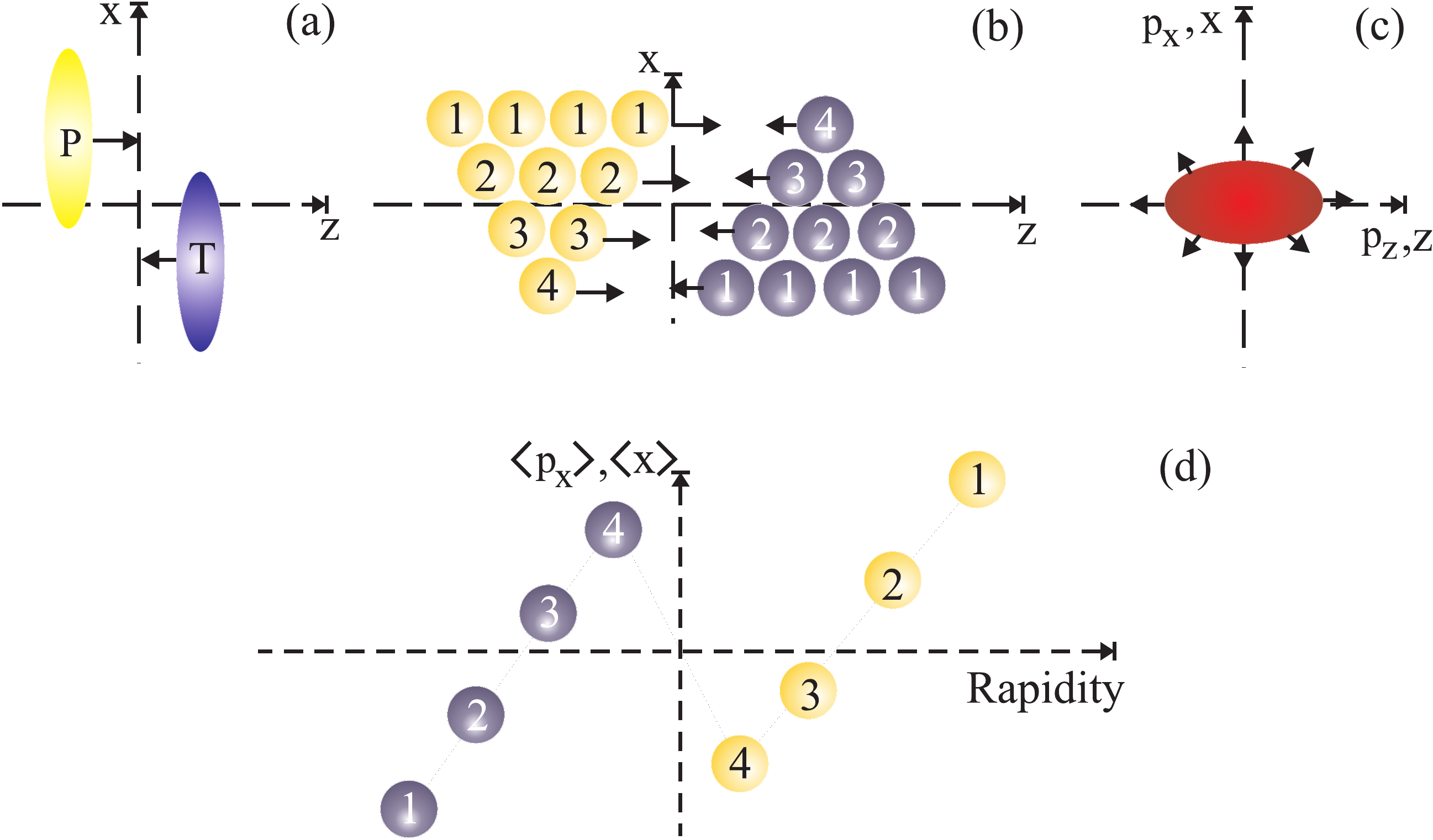}
     \caption{A sketch of a medium central symmetric heavy-ion collision
	progressing in time (a) and (c), and the rapidity distribution
	of $\langle p_x \rangle$ and $\langle x \rangle$ in (d). 
	In (b) the overlap region is magnified and the ``spectators'' 
	are not shown. In these figures $x$ is the
	coordinate along the impact parameter direction and $z$ is 
	the coordinate along the projectile direction~\cite{Snellings:1999bt}. }
     \label{fig:wiggle}
   \end{center}
\end{figure}

Where the colliding nuclei start to overlap, dense matter is created which 
deflects the remaining 
incoming nuclear matter (see Figs.~\ref{fig:DirectedElliptic}
and~\ref{fig:antiflow}). 
The deflection of the remnants of the incoming nucleus at positive rapidity is
in the $+x$ direction leading to $\langle p_x \rangle > 0$, and the remnants of 
the nucleus at negative rapidity 
are deflected in the $-x$ direction thus having a  $\langle p_x \rangle < 0$.
The magnitude of the deflection probes the compressibility of the created 
dense
matter. It probes the system at early time because the deflection happens 
during the passing time of the colliding heavy-ions.  
At AGS energies this is considered the dominant mechanism for generating 
directed flow. The observed directed flow at AGS and SPS energies is an
almost linear function of rapidity so that at these energies the slope
$\dd v_1(y)/\dd y$ at midrapidity was often used to quantify the strength 
of directed flow.

At higher energies the linear dependence of directed flow is expected 
to break down;
at midrapidity the directed flow is predicted to be very small and it
is possible that the slope at midrapidity has a sign opposite to that
in the beam rapidity region. This so-called `wiggle', whereby the
directed flow changes sign three times outside the beam fragmentation
region as illustrated in Fig.~\ref{fig:wiggle}(d), is very sensitive to the equation of
state~\cite{Snellings:1999bt,Brachmann:1999xt,Csernai:1999nf,Csernai:2004gk}. 
The wiggle can have different physical origins. 
Using a hydrodynamic approach it is observed in
Refs.~\cite{Brachmann:1999xt,Csernai:1999nf} that this wiggle
structure only appears under the assumption of a QGP equation of
state, thus becoming a signature of the QGP phase transition.
For another class of models, e.g. cascade models, 
the initial conditions are defined as illustrated in Fig.~\ref{fig:wiggle}. 
Here the rapidity loss of the incoming nucleons at positive rapidity 
is larger at negative $x$ than at positive $x$ while for the incoming nucleons
at negative rapidity the rapidity loss 
is larger at positive $x$ than at negative $x$. This, in addition to 
a positive space momentum correlation,
can also cause a wiggle structure in the directed
flow~\cite{Snellings:1999bt}. In this scenario the measured directed flow 
is sensitive to the magnitude of the rapidity loss and the strength of 
the space-momentum correlation, e.g. radial flow.
More recently, taking into account this initial velocity gradient along 
the $x$-direction (see Fig.~\ref{fig:becat}), it is
argued that one also has to consider the collective motion due to angular 
momentum conservation. 
It has been shown in Ref.~\cite{Becattini:2007sr} that this contributes to 
the in-plane expansion rate and as argued in 
Ref.~\cite{Troshin:2007cp} this should also lead to directed flow.

\subsubsection{System size and energy dependence; extended longitudinal scaling}

\begin{figure}[t]
\begin{center}
 \includegraphics[width=0.6\textwidth]{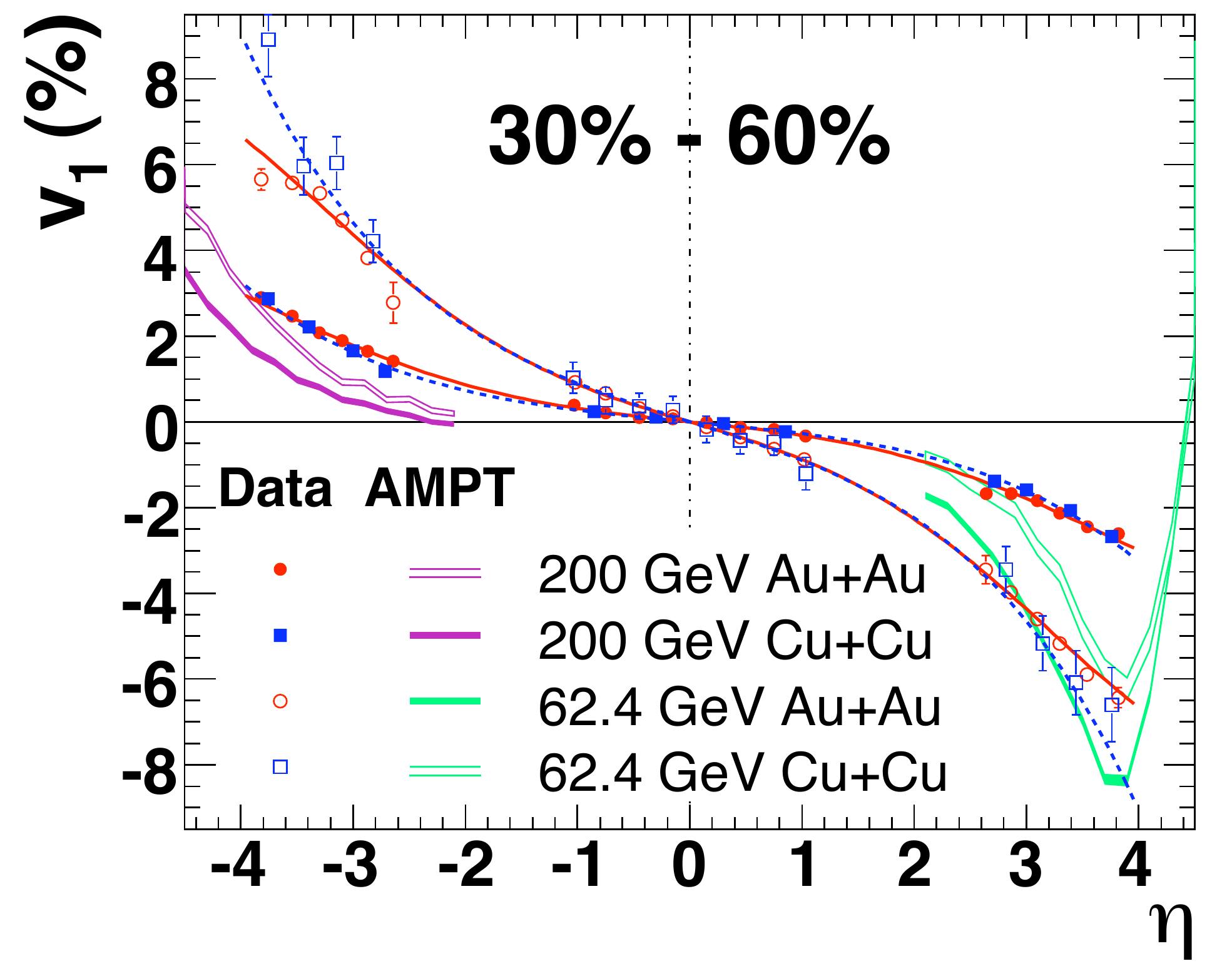}
  \caption{Directed flow of charged hadrons for two beam energies 
  and two colliding species as a function of pseudorapidity for 30--60\% 
  centrality. The event plane was determined from spectator neutrons in the shower maximumm detecors of the zero-degree calorimeters. The solid and dashed curves are odd-order polynomial fits 
  to demonstrate the forward-backward symmetry of the measurements. 
  The AMPT model calculations are plotted for the same conditions as the data 
  (see legend, plotted only on one side of $\eta = 0$ to reduce clutter). From 
  Ref.~\cite{Abelev:2008jga}. }
  \label{fig:v1_3060}
\end{center}
\end{figure}

Figure~\ref{fig:v1_3060} shows the measured directed flow of charged hadrons 
at RHIC for two beam energies and two colliding species.
Clearly the magnitude of the directed flow is very small at midrapidity, as 
was expected, and the directed flow is not a linear function of rapidity as 
it was at lower beam energies.
However there is also no sign of a wiggle structure in the observed charged 
particle directed flow. 
To rule out the models predicting such a wiggle, directed flow of identified 
particles has to be measured because in the case of charged hadrons 
the wiggle could be masked due to the opposite sign of the wiggle for nucleons 
and pions as predicted in Ref.~\cite{Snellings:1999bt}. 
Figure~\ref{fig:v1_3060} also clearly shows that at the same colliding energy and at the same fraction of cross section, the magnitude of directed flow is 
the same for Au+Au and Cu+Cu collisions over the whole measured 
pseudorapidity range.
This is remarkable because the Au+Au system is three times more massive 
and indeed most model calculations, as shown for AMPT~\cite{{Lin:2001zk}} in the figure, predict a stronger directed flow for the more massive system. 
In addition the figure shows that the directed flow decreases with 
collision energy at fixed rapidity.
\begin{figure}[t]
\begin{center}
  \includegraphics[width=0.65\textwidth]{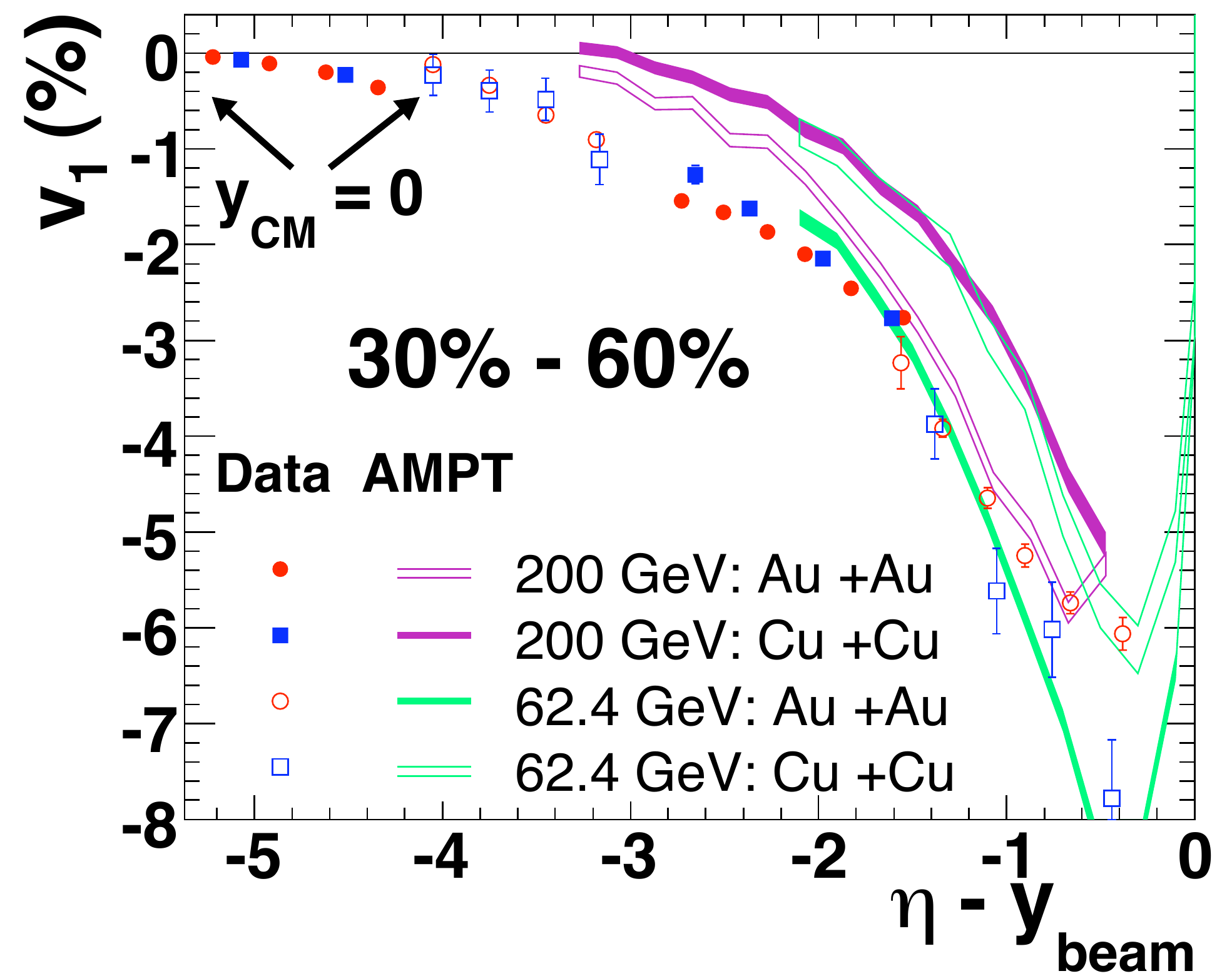}
  \caption{Directed flow of charged hadrons as a function 
  of $\eta -  y_{\mathrm{beam}}$, for 30--60\%  centrality
  Au+Au and Cu+Cu collisions, each at two energies~\cite{Abelev:2008jga}. }
 \label{fig:longscaling}
\end{center}
\end{figure}

To investigate the collision energy dependence in Fig.~\ref{fig:longscaling} 
the directed flow is plotted as function 
of $\eta -  y_{\mathrm{beam}}$, i.e. in the frame of incoming nucleons. 
In this frame the directed flow for different species and collision energies 
is shown to collapse into a unified curve.
The AMPT calculations clearly do not predict such a scaling; in AMPT 
the scaling only approximately holds for each 
colliding species separately.
This remarkable scaling as a function of collision energy and for 
the two different colliding systems was to our knowledge
not predicted in any model calculation. However, the recent work in 
Refs.~\cite{Becattini:2007sr,Troshin:2007cp}
taking into account the initial flow fields might provide a plausible
explanation.
 
\subsection{Elliptic flow}

\subsubsection{In-plane elliptic flow}
\label{EllipticFlow}

\begin{figure}[hbt]
\begin{center}
 \centerline{ \includegraphics[width=.6\textwidth]{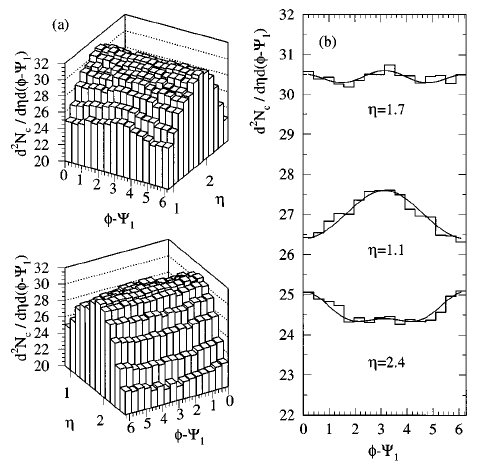} }
  \caption{First observation of in-plane elliptic flow of charged particles
at 11.8 A GeV by the E877 Collaboration~\cite{Barrette:1996rs}. 
The first harmonic event plane was determined by (mostly nucleon)
directed flow detected by the Target Calorimeter.
Mid-rapidity is $\eta \approx 1.7$.}
  \label{fig:E877inplane}
\end{center}
\end{figure}

\begin{figure}[hbt]
\begin{center}
  \includegraphics[width=.6\textwidth]{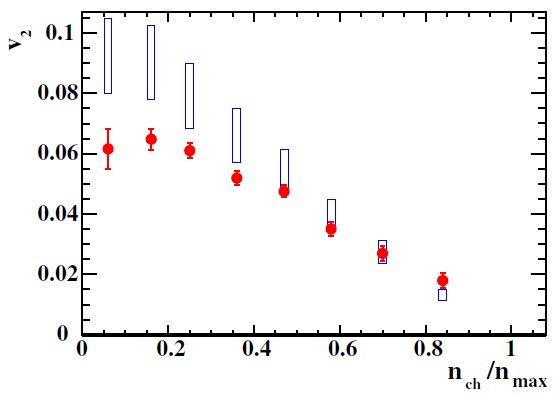}
  \caption{First RHIC results~\cite{Adler:2002pu} of elliptic flow plotted vs the charged hadron multiplicity divided by the maximum observed multiplicity. Blue boxes show estimates  based on hydrodynamic calculations.}
  \label{fig:v2_STAR_first}
\end{center}
\end{figure}

\begin{figure}[hbt]
\begin{center}
\centerline{  \includegraphics[width=.6\textwidth]{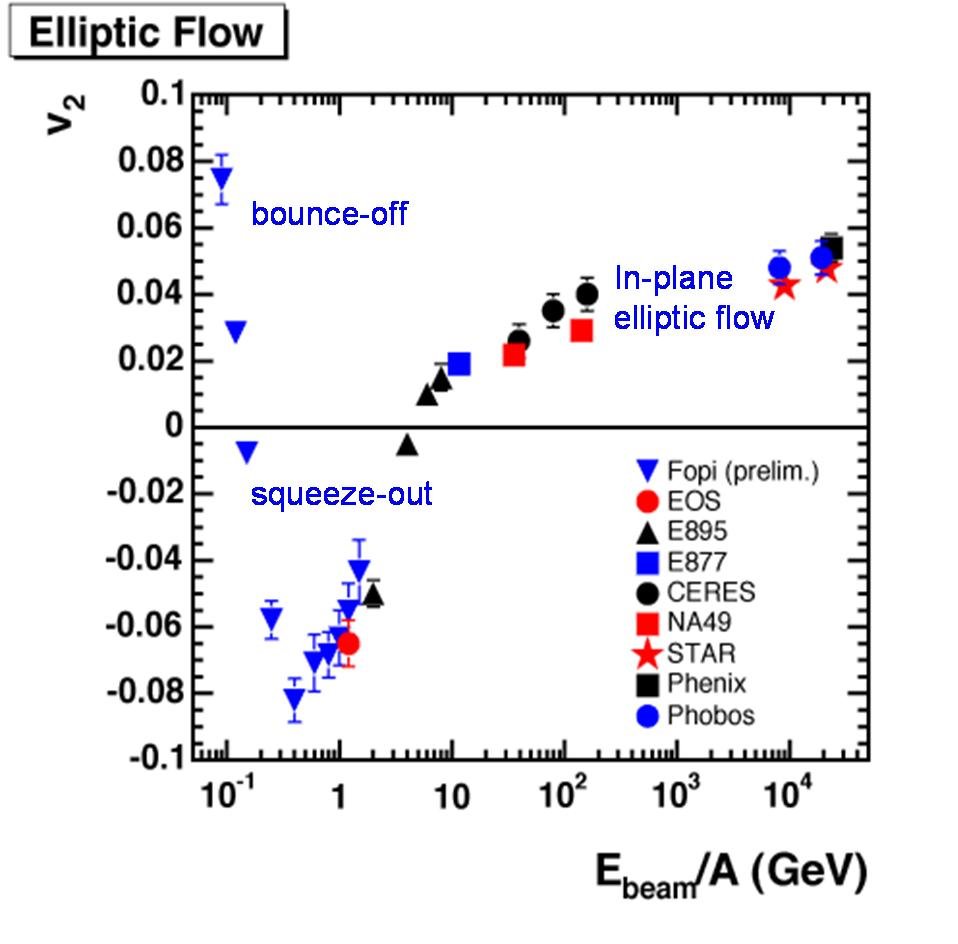}}
  \caption{Elliptic flow for midrapidity 25\% most central
  collisions as a function of beam energy~\cite{Wetzler}.}
  \label{fig:v2excitation}
\end{center}
\end{figure}

Elliptic flow has attracted the most attention in recent years. 
Based on hydrodynamic calculations, in-plane elliptic flow was
suggested~\cite{Ollitrault:1992bk} as a signature of collective
expansion in ultrarelativistic nuclear collisions. Note that the
situation is different at lower energies, where the elliptical shape of the
particle transverse momentum distribution at midrapidity is elongated
in the direction
perpendicular to the reaction plane and interpreted as due to shadowing by
spectator nucleons, a phenomenon called {\em squeeze-out}.
Only at high energies, when the longitudinal size of the
Lorentz contracted nuclei become negligible compared to the transverse 
size and, correspondingly,  the passing time too small compared to the
characteristic time for the development of elliptic flow, of the size
of the nuclei radius, 
the shadowing goes away and elliptic flow fully develops in-plane.
In-plane elliptic flow was observed at the AGS by the
E877 Collaboration~\cite{Barrette:1996rs} (Fig.~\ref{fig:E877inplane}) 
and later, during the AGS low
energy scan, the E895 Collaboration performed measurements of
elliptic flow in the transition region from out-of-plane to in-plane.
The first measurements of elliptic flow at RHIC~\cite{Adler:2002pu}
(Fig.~\ref{fig:v2_STAR_first}) showed that $v_2$ approached the
predictions of ideal hydrodynamic models; this observation
was taken as a signature of the rapid thermalization of the system.
Figure~\ref{fig:v2excitation} shows the excitation function 
of $v_2$ including other lower beam energies and
later measurements from SPS and RHIC.

\subsubsection{Low density and ideal hydro limits, $v_2/\eps$ plot}
\label{secLv2_limits}

As elliptic flow should be zero in the absence of rescattering, 
in the low density limit (LDL) it
is directly proportional to the particle density in the
transverse plane~\cite{Heiselberg:1998es,Voloshin:1999gs}. 
As particle density increases, the anisotropy is expected also to
increase, saturating at some level, assuming no modification to the
constituent interactions. 
In addition, as the elliptic flow should be zero for a totally 
azimuthally symmetric system, for small anisotropies in the initial geometry,
elliptic flow must be proportional to this spatial anisotropy. 
Usually, for this purpose one uses the eccentricity defined by
\be
\eps=\frac{\mean{y^2-x^2}}{\mean{x^2+y^2}},
\label{eq:eps}
\ee  
where the average is taken  over the initial geometry with some
weight. For the weight one can use the nuclear profile density
(participant nucleons), entropy or
energy densities, or something else (e.g. number of binary collisions).
Note that as long as the eccentricity is small, elliptic flow should be
directly proportional to eccentricity calculated with any weight, but
it is important to keep weight the same when comparing results
obtained in different calculations.
For numerically large eccentricities the direct proportionality could break in principle, but as was shown in the very first
hydrodynamic calculation by Ollitrault~\cite{Ollitrault:1992bk}
(though he used somewhat different measure of eccentricity) the
proportionality holds well even for rather large values of $\eps$.
It was pointed out by Sorge~\cite{Sorge:1998mk}, who tried to study 
the effect of the QGP phase transition
based on RQMD calculations, that the centrality dependence of the
scaled elliptic flow $v_2/\eps$ (he used a different notation) would
exhibit non-monotonic dependence in response to the softening of the
equation of state. 

Based on all the above observations Voloshin and 
Poskanzer~\cite{Voloshin:1999gs}
proposed to plot all the experimental data as
$v_2/\eps$ vs particle density in the transverse
plane, $1/S(dN_{ch}/dy)$, where the initial overlap area $S$ and
eccentricity are taken from Glauber model calculations.
The idea of the plot is to compare the results obtained at different
collision energies, with different projectiles, and at different centralities.
Non-smooth dependence would be indicative of new physics (for
example, deconfinement) and saturation could signal an approach to
ideal hydrodynamical evolution.
\begin{figure}[htb]
\begin{center}
\centerline{  \includegraphics[width=0.6\textwidth]{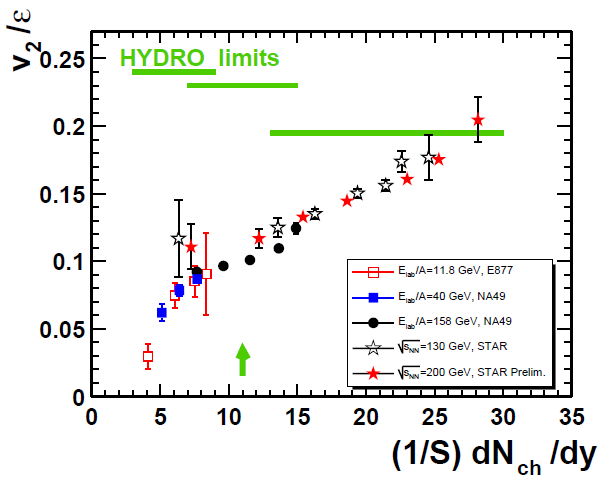}}
\caption{
 Compilation of $v_2/\eps$ data~\cite{Alt:2003ab} vs particle density at midrapidity. Green lines
 indicate ideal hydrodynamic predictions for AGS, SPS and RHIC
 collisions energies~\cite{Kolb:2000sd}.
}
\label{fig:na49_v2/eps}
\end{center}
\end{figure}
Figure~\ref{fig:na49_v2/eps} shows NA49 results~\cite{Alt:2003ab}
together with results obtained at the AGS and preliminary results from
RHIC.  
The SPS and RHIC flow values were obtained with higher order
cumulants and the eccentricities were taken from an optical Glauber model.
As will be discussed in detail in section~\ref{sec:v2_fluc}, this
combination represents the best for comparison to theoretical calculations, as
it is most free from both nonflow and fluctuations in the initial 
geometry of the overlap region. 
For this plot, elliptic flow values were integrated over the
entire \pt region; the data, if measured in a limited \pt window were extrapolated to correct for this. Also, when rapidity
density was not measured, the pseudorapidity density has been
used with rescaling based on model calculations.
For discussions of other systematic uncertainties see 
the original paper~\cite{Alt:2003ab}. 

Figure~\ref{fig:na49_v2/eps} attracted a lot of attention as the data show a continuous
rise reaching the ideal hydrodynamic expectations (shown by green
lines) in the most central collisions at RHIC energies.
The green arrow indicates the position of the color
percolation phase transition predicted by Satz~\cite{Satz:2002ku};
unfortunately the systematic uncertainties in the experimental data
(see discussion in Ref.~\cite{Alt:2003ab}) do not allow a definite
statement whether the data exhibit any non-smooth behavior at this place. 
Future beam energy scans at RHIC should clarify this point.

\begin{figure}[htb]
\begin{minipage}[b]{0.48\textwidth}
  \includegraphics[width=0.95\textwidth]{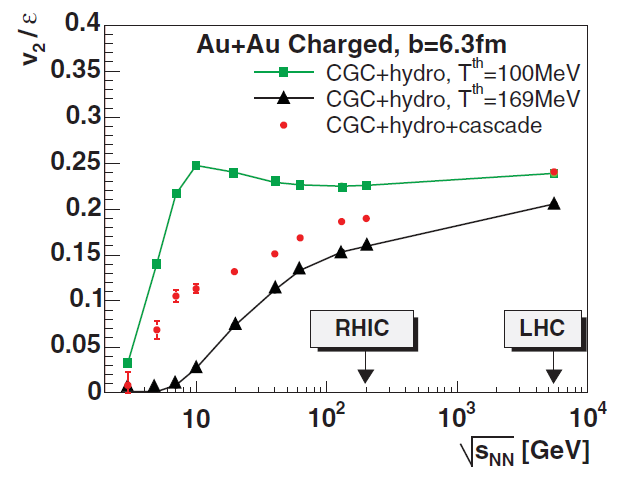}
   \caption{Hydro+cascade calculations for $v_2/\eps$ vs beam energy 
	with CGC initial conditions and different freeze-out conditions.
	The points without a line through them were calculated with a cascade afterburner. Predictions for LHC are on the right side of the graph~\cite{Hirano:2007gc}.} 
\label{fig:HydroCascadeLHC}
\end{minipage}
\hspace{\fill}
\begin{minipage}[b]{0.48\textwidth}
  \centerline{\includegraphics[width=0.9\textwidth]{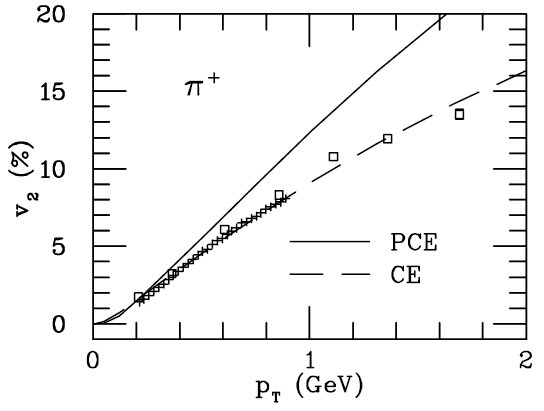}}
\caption{Ideal hydrodynamic calculations of pion $v_2(\pt)$ in two scenarios~\cite{Huovinen:2007xh}. 
Solid line (Partial Chemical Equilibrium) indicates the
results using parameters best fit to spectra. Dashed line assuming chemical equilibrium. }
\label{fig:pasi}
\end{minipage}
\end{figure}

Strong elliptic flow observed in central Au+Au collisions at the highest RHIC
energies, consistent with prediction of ideal hydrodynamics, lead to
the picture of strongly coupled quark-gluon plasma, 
sQGP~\cite{Gyulassy:2004zy}.
Taking into account the significance of the $v_2/\eps$ plot in 
helping to establish the sQGP picture, 
all components of this plot has been recently reevaluated
and we now discuss in detail recent developments in this area. 
Along with several indirect indications that in central Au+Au
collisions thermal equilibrium is not complete, it was found that 
very small viscous effects can lead to 
a significant reduction in the predicted elliptic
flow compared to the ideal hydro case.
There are significant effects even for viscosity equal to the conjectured lower limit of shear viscosity to entropy ratio ($\eta/s$)~\cite{Kovtun:2004de}.
Continuous freeze-out as implemented by a cascade afterburner also has 
a considerable effect~\cite{Hirano:2007gc}, see
Fig.~{\ref{fig:HydroCascadeLHC}}. Lower values of elliptic flow would 
contradict experimental measurements if other effects,
responsible for an increase of elliptic flow (compared to the
``standard'' hydrodynamic calculation) could not be identified. 
Several such effects have been reported:
\begin{itemize}
\item
First, it was demonstrated by Huovinen~\cite{Huovinen:2007xh}
that ideal hydro calculations, if tuned to describe spectra, 
yield larger elliptic flow than thought previously,
which emphasizes the need to describe the spectra and elliptic flow
simultaneously.  
Figure~\ref{fig:pasi} shows calculations of elliptic flow in two
scenarios; one (Partial Chemical Equilibrium) shown by the solid line is a fit to the spectra and significantly over-predicts elliptic flow.
\item
It was shown, that in some models, e.g. CGC, 
the initial eccentricity  can take significantly larger values 
than in the optical Glauber model
that is usually used to set initial conditions in hydro calculations. 
The larger eccentricities inevitably lead to larger elliptic 
flow~\cite{Hirano:2005xf,Drescher:2006pi,Drescher:2007ax,Lappi:2006xc}. 
\item
Flow fluctuations, the nature of which is much better understood
in recent years, lead to an increase of {\em apparent} flow 
relative to the participant plane~\cite{Voloshin:2007pc}. 
\item
Finally, it was noticed that gradients in the initial velocity
field (Fig.~\ref{fig:becat}) also increase the final values of 
elliptic flow~\cite{Becattini:2007sr}.
\end{itemize}

Any of the above mentioned effects can be quite significant, each leading to
20--30\% or even larger changes in the values of $v_2$. 
The inclusion of these effects into one reliable model is still under way.

\subsubsection{Viscous effects}
\label{sec:viscous}
\begin{figure}[htb]
\begin{center}
  \includegraphics[width=0.6\textwidth]{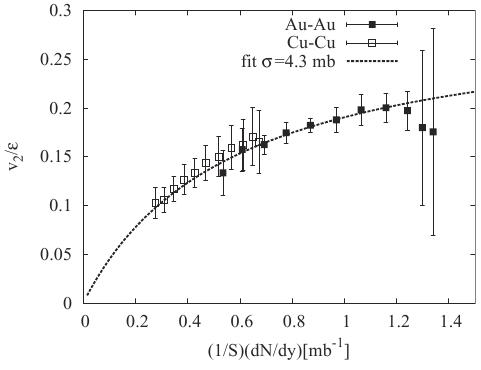}
\caption{Fit to $v_2/\eps$ vs particle density in terms of Knudsen number for Cu+Cu and Au+Au~\cite{Drescher:2007cd}. The data are from PHOBOS using the eta subevent method.}
\label{fig:knudsen}
\end{center}
\end{figure}

An attempt of model independent 
analysis of $v_2/\eps$ dependence on particle density based on a
parameterization in terms of {\em Knudsen number} has been developed 
in Refs.~\cite{Bhalerao:2005mm,Drescher:2007cd}. By definition, $1/K$ is the
mean number of collisions per particle, and ideal hydrodynamics
corresponds to the limit $1/K \rightarrow \infty$. The authors use the expression
$v_2/\eps =(v_2/\eps)_{\mathrm{hydro}} (1+K/K_0)^{-1}$, where the parameter
$K_0\approx 0.7$ is independently estimated from a fit of
model calculations to the data (see Fig.~\ref{fig:knudsen}). 
The authors conclude that at 
RHIC we might be up to 30\% below the ideal ``hydro limit'' even for
the most central collisions. 
Their estimate of the viscosity yields values of
$\eta/s=0.11-0.19$ depending on the CGC or Glauber initial conditions.
(These initial conditions are discussed more in Sec.~\ref{sec:v2_fluc}.)
Similar fits to STAR data~\cite{Tang:2008if} lead to similar conclusions.

\begin{figure}[htb]
\begin{center}
\centerline{\includegraphics[width=0.6\textwidth]{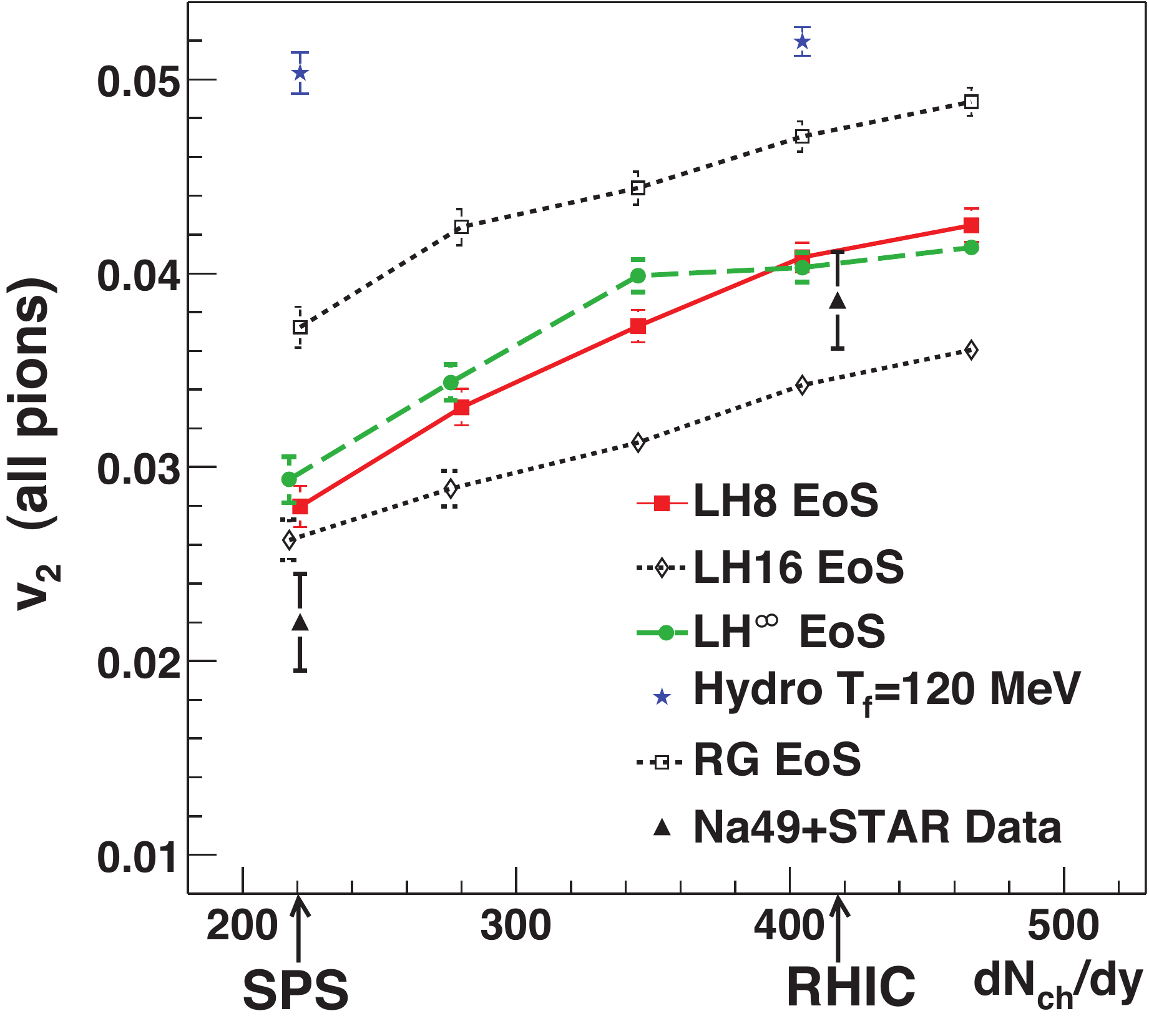}}
\caption{
Hydro+RQMD results~\cite{Teaney:2000cw} for elliptic flow vs rapidity density. Experimental data are the two points at the top.
LH8 denotes the results obtained for an EoS with latent heat 0.8~GeV/fm$^3$.
} 
\label{fig:TeaneyShuryak}
\end{center}
\end{figure}

The magnitude of the {\em viscous effects} could be judged already from the
early calculations~\cite{Teaney:2001av}
where the hydrodynamical evolution at some intermediate stage was joined to
the transport model RQMDv2.4 to simulate late (viscous) evolution of the
system and differential freeze-out. 
Using the transport code to describe the late hadronic stages of the
system evolution and freeze-out, allowed a satisfactory
description of elliptic flow for SPS and RHIC data using 
the same EoS (QGP, including first order phase transition
with a latent heat of 0.8~GeV/fm$^3$), see Fig.~\ref{fig:TeaneyShuryak}.
At SPS,  $v_2$ values have been found about a factor of 2 lower compared
to ideal hydrodynamic predictions. 
The importance of continuous freeze-out and late hadronic viscosity
have been also demonstrated in similar calculations by Hirano,
Fig.~\ref{fig:HydroCascadeLHC}. 
Somewhat smaller in magnitude, but again a similar effect, was obtained in a
hydro+uRQMD hybrid calculation in Ref.~\cite{Nonaka:2006yn}.

Taken together, hybrid model results show that in this approach one achieves
very reasonable, and at the moment probably the best description of the data. 
The collision energy and centrality dependence in the hybrid models are mostly
due to changes in the relative time the system spends in the sQGP state compared
to the hadronic gas and the (continuous) freeze-out. 
Note that these models employ ideal hydrodynamics, 
once again arguing for very small viscosity of the early QGP stage. 
Viscose effects in this approach are totally due to the hadronic cascade
phase. 

\begin{figure}[htb]
\begin{center}
\centerline{\includegraphics[width=0.6\textwidth]{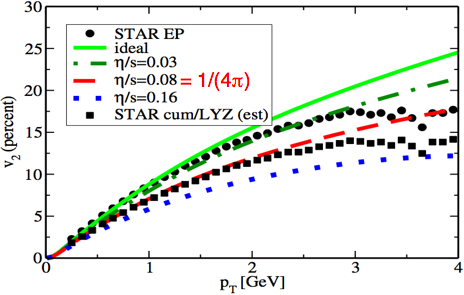}}
\caption{Viscous hydro calculations~\cite{Romatschke:2007mq, Luzum:2008cw}
compared to minimum bias STAR data~\cite{:2008ed}. 
The lower STAR points have an estimated correction for nonflow and fluctuation
effects based on the four-particle cumulant and Lee-Yang Zeros methods shown in Fig.~\ref{fig:nonflow}.}
\label{fig:romat}
\end{center}
\end{figure}

\begin{figure}[htb]
\begin{center}
\includegraphics[width=0.6\textwidth]{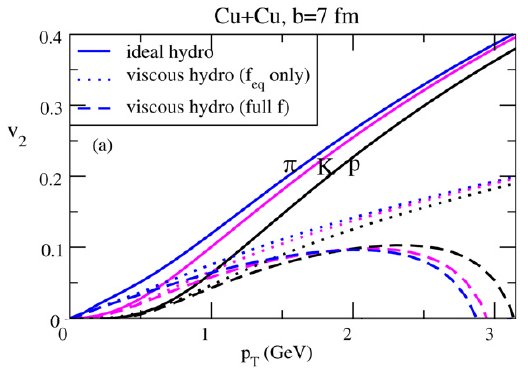}
\caption{
$v_2(\pt)$ for $\pi$, $\K$, and p particles from ideal and viscous ($\eta/s=1/(4\pi)$) 
hydrodynamics for Cu+Cu mid-central collisions~\cite{Song:2007ux}.} 
\label{fig:song}
\end{center}
\end{figure}

Later, viscosity was attempted to be introduced directly into
hydrodynamic calculations~\cite{Teaney:2003kp}. 
Recently there have been performed
calculations~\cite{Romatschke:2007mq, Song:2007fn, Song:2007ux, Dusling:2007gi}
of the hydrodynamical expansion with viscous terms explicitly included 
into the equations. 
Now everyone agrees~\cite{Song:2008si} on the significance of the viscous 
effects even for the conjectured minimal value of shear viscosity 
to entropy ratio ($\eta/s=1/(4\pi)$). 
The results presented in Figs.~\ref{fig:romat} and~\ref{fig:song} show 
that even minimal viscosity leads to $\sim$25--30\% reduction in flow 
values in Au+Au collisions and probably more than 50\% in Cu+Cu.

Note that viscosity coefficients calculated in pQCD are usually much larger than 
would be allowed by the data. 
In this sense, noteworthy are the recent calculations~\cite{Xu:2007jv} 
which emphasize the importance of taking into account  
$2 \leftrightarrow 3$ processes.  
With these effects included, the viscosity coefficient appears 
to be about an order of magnitude smaller compared to previous calculations
and falls into the ``allowable'' range of the data.

\subsubsection{Initial eccentricity and $v_2$ fluctuations}
\label{sec:v2_fluc}

\begin{figure}[htb]
\begin{center}
  \includegraphics[width=0.5\textwidth]{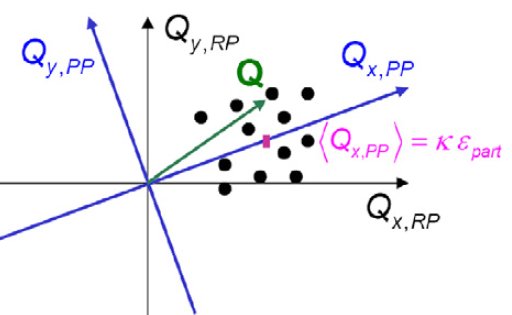}
\caption{Flow vector distribution at 
fixed $(\eps_x,\eps_y)$ showing $\mean{Q}$ along the participant plane $x$-axis~\cite{Voloshin:2007pc}.}
\label{fig:planesQ}
\end{center}
\end{figure}

\begin{figure}[htb]
  \includegraphics[width=0.49\textwidth]{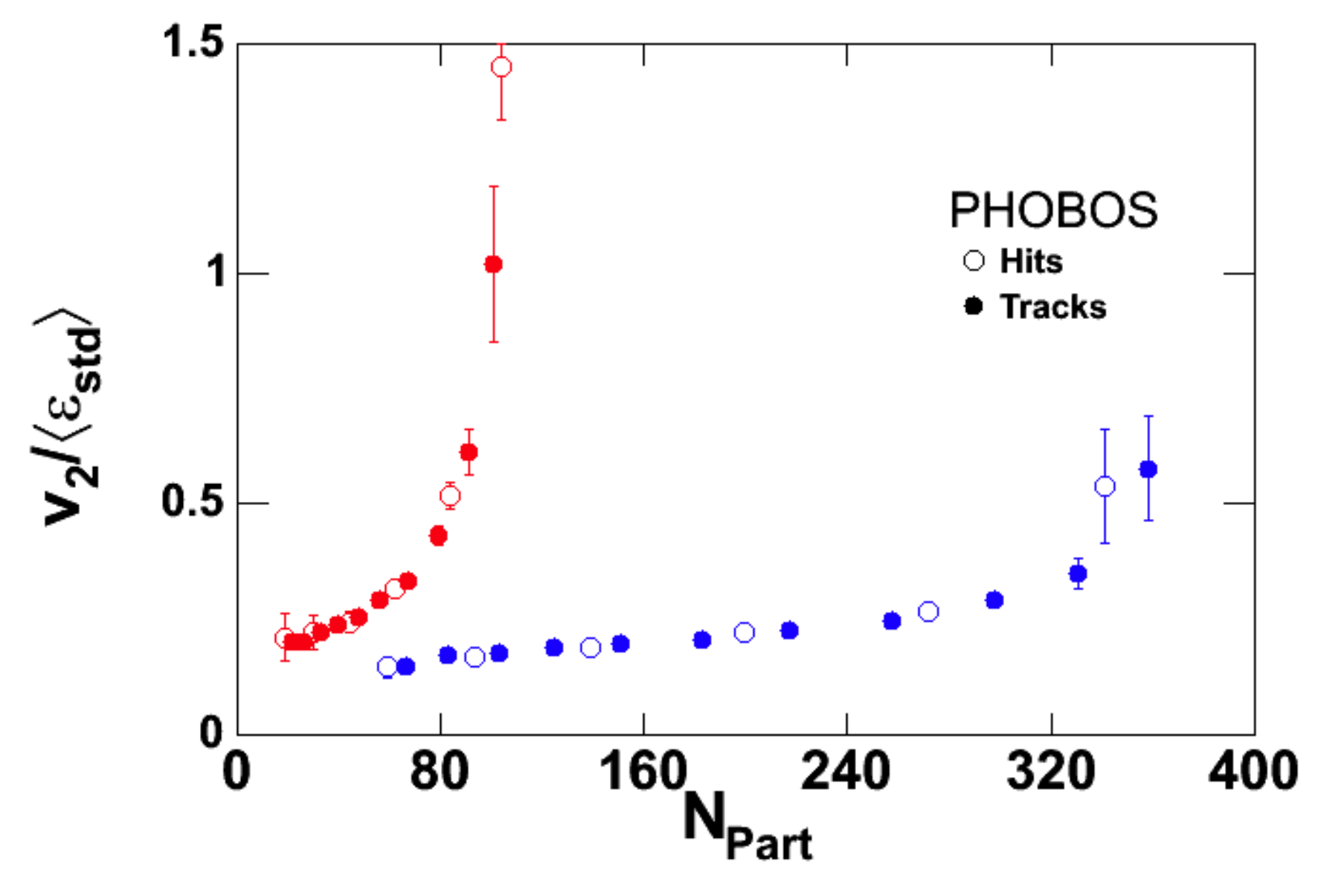}
  \includegraphics[width=0.49\textwidth]{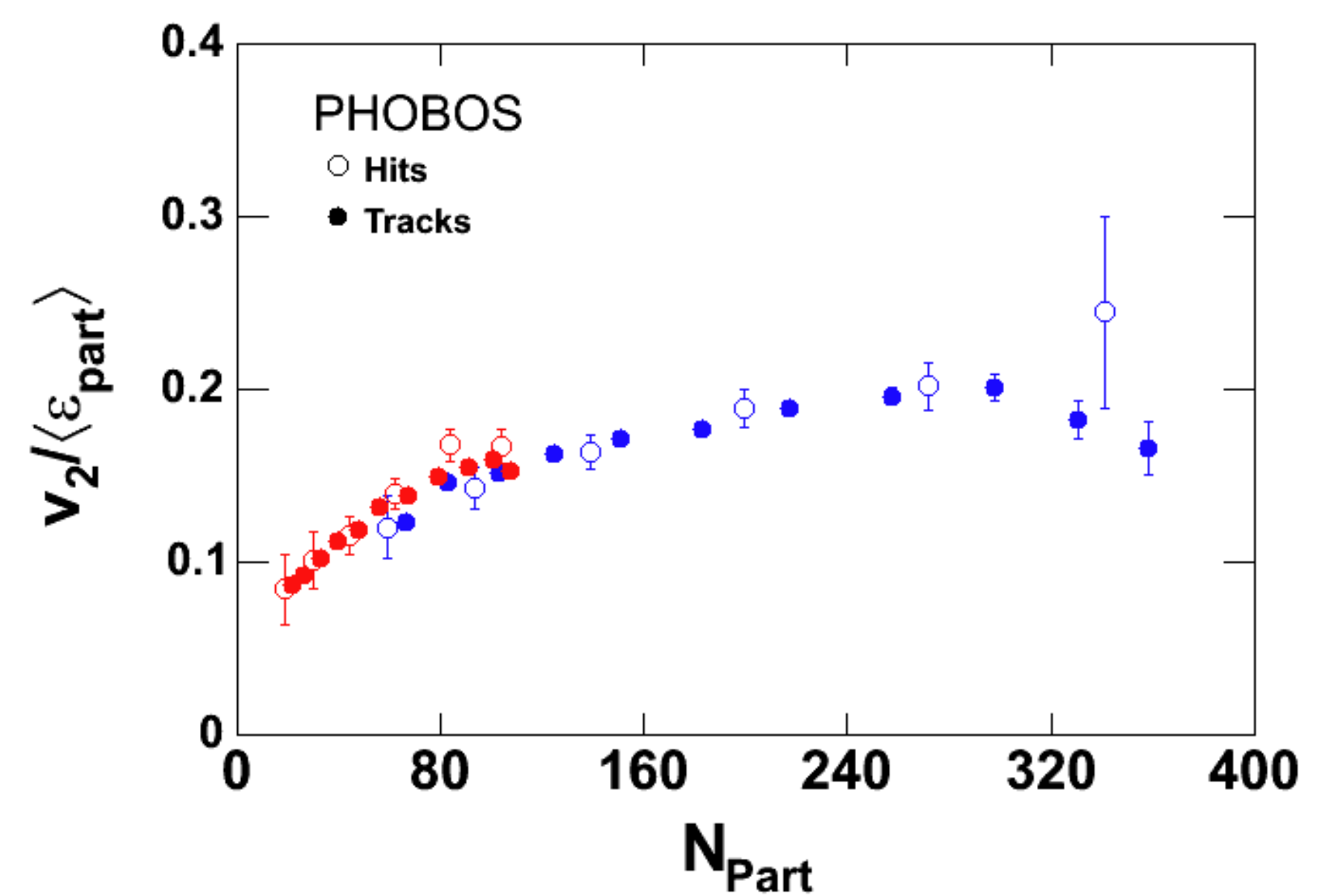}
\caption{Left: Measured $v_2$ of charged hadrons scaled by \epsstd \ from 
Cu+Cu (upper curve) and Au+Au (lower curve) vs the number of participants. 
Right: Rescaled with \epspart~\cite{Alver:2006wh}. The data were obtained with the eta subevent method.}
\label{fig:PHOBOS_eps}
\end{figure}

The role of fluctuations in  the initial system geometry~\cite{Miller:2003kd}
defined by nuclear {\em participants}~\cite{Manly:2005zy} 
(interacting nucleons or quarks) has been greatly clarified 
in recent years~\cite{Voloshin:2007pc, Bhalerao:2006tp,Voloshin:2006gz,Alver:2008zz,Broniowski:2007ft,Andrade:2006yh}.
The following picture emerges: 
at fixed impact parameter, the geometry of the {\em participant zone}
(see Fig.~\ref{fig:planes}) fluctuates, both, in terms of the value of the
eccentricity as well as the orientation of the major axes. 
Then the anisotropy develops along the plane spanned by the minor
axis of the participant zone and the beam direction, 
the so called {\em participant plane}. As the true
reaction plane is not known and the event plane is estimated from
the particle azimuthal distribution, 
the apparent (participant plane) flow appears to be always
bigger (and always in-plane with $\vtpp>0$) compared to the
flow as projected onto the reaction plane $\vtrp$ in Fig.~\ref{fig:planesQ}.
The importance of using the proper values of the initial
eccentricity is illustrated in Fig.~\ref{fig:PHOBOS_eps}. 
The $v_2$ values from Cu+Cu and Au+Au collisions are vastly different when
scaled by \epsstd \ (Eq.~(\ref{eq:eps})) but agree nicely when scaled by
\epspart \ (Eq.~(\ref{eq:epsxy}))~\cite{Alver:2006wh}.

It was noticed in Ref.~\cite{Voloshin:2007pc} that in collisions 
of heavy nuclei the fluctuations in the eccentricity can be well 
described by a two-dimensional Gaussian with the same width in both directions:
\begin{equation}
(\eps_x,\eps_y) = \left( \left< \frac{\sigma_y^2-\sigma_x^2}{\sigma_y^2+\sigma_x^2} \right>,
\left< \frac{2\sigma_{xy}}{\sigma_y^2+\sigma_x^2} \right> \right) ,
\label{eq:epsxy}
\end{equation}
where $\sigma_x^2=\mean{x^2} - \mean{x}^2$, $\sigma_y^2=\mean{y^2} - \mean{y}^2$, and $\sigma_{xy}=\mean{xy} - \mean{x}\mean{y}$. 
What is not trivial is that for such Gaussian
fluctuations the higher order cumulant flow values ($v\{n\}, n\ge 4$) are not only insensitive to nonflow but also to eccentricity fluctuations. 
All of the higher cumulants are exactly equal to $\vtrp$, the flow 
projected onto the reaction plane. At the same time, the participant plane
flow become unmeasurable because flow fluctuations can not
be separated from nonflow contributions by means of correlation
measurements. See Ref.~\cite{opv}.    

The fact that higher order cumulants of eccentricity are very close
to the ``standard'' values \epsstd~was initially observed in numerical 
calculations~\cite{Voloshin:2006gz}, and as an approximate result
in~Ref.~\cite{Bhalerao:2006tp}. Numerically, deviations were observed only for
lighter (Cu+Cu) systems, which were traced to a break in the Gaussian
approximation~\cite{Alver:2008zz}.
Experimentally, this question can be addressed by comparing the values of
higher order cumulant flow with results obtained with 
the first harmonic reaction
plane determined by spectator neutrons (at RHIC with the help of Zero Degree
Calorimeters). 
Note also that using the first harmonic reaction plane one can also 
address the validity of the Gaussian approximation by direct 
fluctuation measurements~\cite{Wang:2006xz}. 

An important conclusion from that study of eccentricity fluctuations
was  that in most cases the measurements of higher order cumulant
flow values provide elliptic flow relative to the reaction plane
and not the participant plane. Similarly, the same value
is given by several other methods, such as Lee-Yang Zeros, Bessel
Transform, and the fit to q-distribution. 
This greatly simplifies the comparison of hydrodynamic calculations
to data, as it says that in such calculations one should not
worry about how to take into account fluctuations in the initial
eccentricity (which is a non-trivial task) but just compare to the
``right'' measurement, $\vtrp$, e.g. $v_2\{4\}$.  
This understanding also explains some earlier calculations
with the uRQMD model~\cite{Zhu:2005qa,Zhu:2006fb}. 
There, it was shown that using higher cumulants and/or the LYZ method 
one indeed can measure elliptic flow very well, but it was not at all clear 
why there was no trace of the effects of flow fluctuations, 
which were expected in this model.  

Unfortunately this progress in understanding the nature of fluctuations
does not help in resolving the problem of {\em measuring} 
flow fluctuations and nonflow. 
Strictly speaking, to separate those effects one is required to
make assumptions.  Most often, the azimuthal correlations
between particles with large rapidity separation are used to suppress nonflow contributions.     
The problem with this method is that there are no reliable estimates of
how well they suppresses nonflow and also how much the flow
fluctuations (in this case correlations) change after imposing such a cut. 

\begin{figure}[htb]
\begin{minipage}[b]{0.48\textwidth}
  \includegraphics[width=0.95\textwidth]{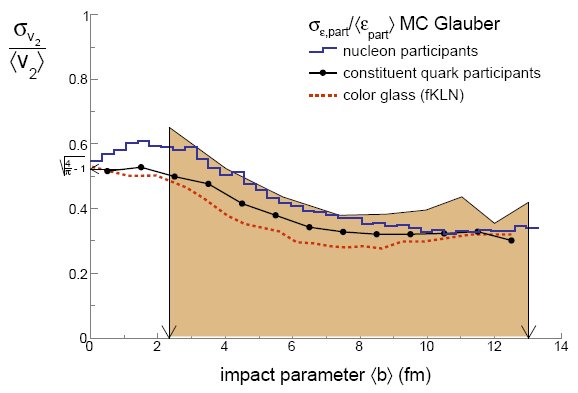}
\caption{The shaded area shows the elliptic flow fluctuations as estimated by STAR~\cite{Sorensen:2008zk}. The curves are MC Glauber calculations for nucleons, quarks, and the CGC. }
\label{fig:flucStar}
\end{minipage}
\hspace{\fill}
\begin{minipage}[b]{0.48\textwidth}
  \includegraphics[width=0.95\textwidth]{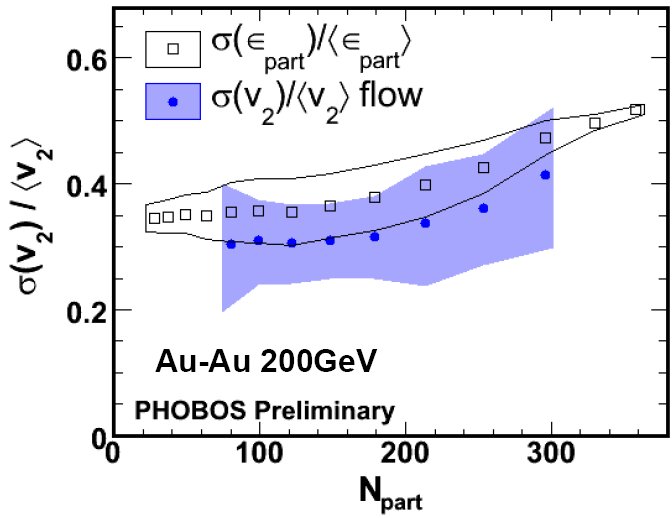}
\caption{The shaded area shows the elliptic flow fluctuations as estimated by PHOBOS~\cite{Alver:2008hu}. The lines are estimates from MC Glauber calculations.}
\label{fig:flucPhobos}
\end{minipage}
\end{figure}

At the QM'08 conference the PHOBOS~\cite{Alver:2008hu} and 
STAR~\cite{Sorensen:2008zk} collaborations have presented their
revised (compared to QM'06) results on flow fluctuations.
These results are in good agreement,
see Figs.~\ref{fig:flucStar},~\ref{fig:flucPhobos}. 
In Ref.~\cite{Sorensen:2008zk} a conservative approach is taken and only upper
limits on fluctuations are reported. The PHOBOS Collaboration 
uses estimates of nonflow effects from correlations
with large rapidity separation and reports a more restrictive range for
fluctuations. Both agree that the current
measurements exhaust the (nucleon) eccentricity values obtained 
in the MC Glauber model and in
this sense somewhat favor models which predict smaller relative
fluctuations, such as the CGC model or a MC Glauber model taking into account
constituent quark substructure. 

\subsubsection{(Pseudo)rapidity dependence}

\begin{figure}[htb]
\begin{minipage}[b]{0.48\textwidth}
  \includegraphics[width=0.95\textwidth]{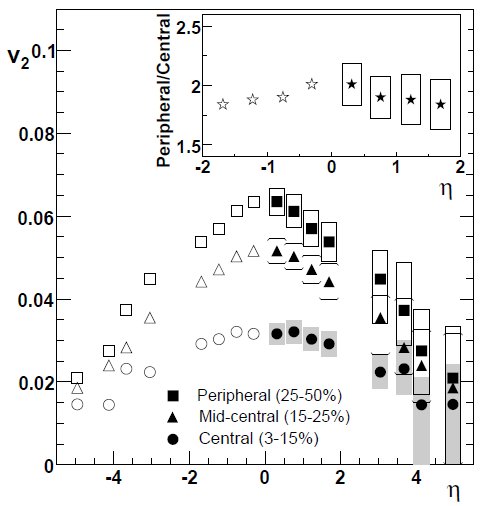}
  \caption{
  Elliptic flow vs pseudorapidity at different centralities~\cite{Back:2004mh}. Insert shows ratio of peripheral to central. The data were obtained with the eta subevent method.}
  \label{fig:v2etaPhobos}
\end{minipage}
\hspace{\fill}
\begin{minipage}[b]{0.49\textwidth}
  \includegraphics[width=.99\textwidth]{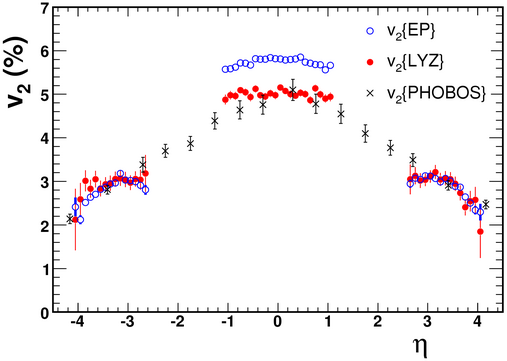}
  \caption{Pseudorapidity dependence of elliptic 
  flow for $\sqrtsNN = 200$~GeV Au+Au from the event plane method (STAR), the
  LYZ method (STAR), and using an $\eta$ gap (PHOBOS). 
  From Ref.~\cite{:2008ed}.}
  \label{fig:v2etaSTAR}
\end{minipage}
\end{figure}

\begin{figure}[htb]
\begin{minipage}[b]{0.48\textwidth}
\begin{center}
  \includegraphics[width=0.99\textwidth]{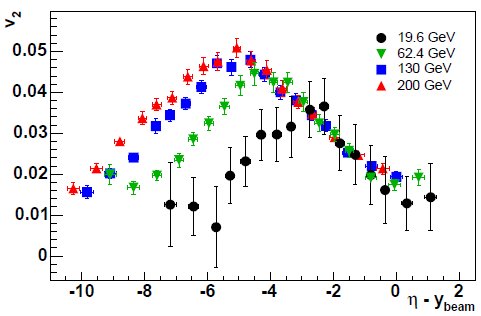}
  \caption{
   $v_2$ as a function of $\eta-y_{\mathrm{beam}}$ at four collisions energies~\cite{Alver:2006wh}. The data were obtained with the eta subevent methiod.
} 
  \label{fig:v2etaLimiting}
\end{center}
\end{minipage}
\hfill
\begin{minipage}[b]{0.48\textwidth}
\begin{center}
  \includegraphics[width=.99\textwidth]{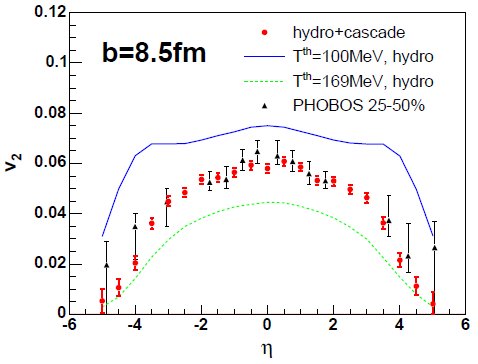}
  \caption{
 Pseudorapidity dependence of elliptic flow from PHOBOS compared to a hydro+cascade model~\cite{Hirano:2007gc}. The lines are for sudden freeze-out at different temperatures. The data were obtained with the eta subevent method.
}
  \label{fig:v2eta_hirano}
\end{center}
\end{minipage}
\end{figure}

The measured pseudorapidity dependence of elliptic flow is shown in
Figs.~\ref{fig:v2etaPhobos}--\ref{fig:v2eta_hirano}. In general $v_2$ exhibits a
$dN_{ch}/d\eta$ dependence, with a maximum at mid-rapidity and falls off in the fragmentation regions.
Such behavior would find a natural explanation in the LDL.
Collision energy and nuclear size dependence of $v_2(\eta)$, 
Fig.~\ref{fig:v2etaPhobos}, would also roughly
agree with scaling proportional to the charged particle density.
For 10--40\% centrality Fig.~\ref{fig:v2etaSTAR} compares the PHOBOS and STAR results. It appears that using the Lee-Yang Zeros method has about the same effect as having a pseudorapidity gap.

Similar to the limiting fragmentation picture of charged
particle density, elliptic flow exhibits the same behavior.
Figure~\ref{fig:v2etaLimiting} shows how the PHOBOS  $v_2(\eta)$ data 
plotted against $\eta-y_{\mathrm{beam}}$, where $y_{\mathrm{beam}}$ is the projectile rapidity, collapses on the universal curve in the fragmentation region. The original hydrodynamic models predictions for $v_2(\eta)$ 
are far from the data (see the solid curve in Fig.~\ref{fig:v2eta_hirano}). 
Note that 3d hydrodynamic calculations 
are computationally difficult and the dependence of the initial
conditions on rapidity is far from being known.
In all calculations so far a boost-invariant initial geometry is
assumed. Adding a cascade hadronic afterburner greatly improves agreement
with data (see Fig.~\ref{fig:v2eta_hirano}) similar to the integral flow discussed above 
(Figs.~\ref{fig:HydroCascadeLHC}
and~\ref{fig:TeaneyShuryak}). 

\subsubsection{Low \pt region: mass splitting}
\label{sec:v2_lowpt}

For charged hadrons, shown in Fig.~\ref{fig:romat},
the elliptic flow increases almost linearly as a function of \pt
reaching values of about 0.15 at large $\pt$.  At low transverse
momenta, the dependence is well described by hydrodynamics. 

\begin{figure}[htb]
\begin{center}
  \includegraphics[width=0.5\textwidth]{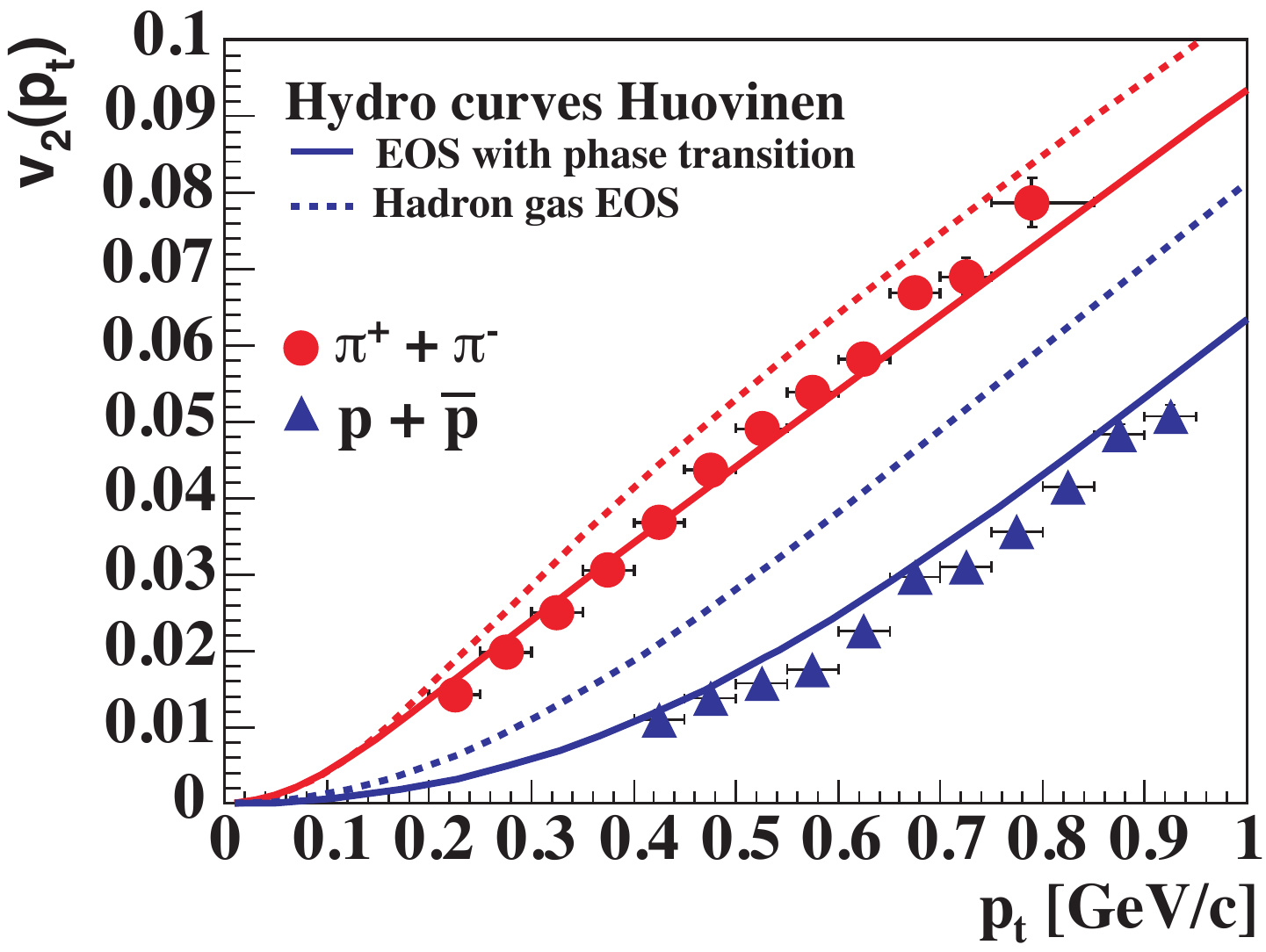}
\caption{Comparison of $v_2(\pt)$ dependence of pions and protons with
hydrodynamic calculations for a hadron gas and also including a phase 
transition~\cite{Huovinen:2001cy,Snellings:2006qw}. }
\label{fig:MassSplit}
\end{center}
\end{figure}
As explained in section~\ref{sec:rad-ani} in a  
(locally) thermalized system, like in hydrodynamics, the interplay of radial expansion and anisotropic flow should lead to a specific
dependence of the differential flow $v_2(\pt)$ on the mass of 
the particle~\cite{Voloshin:1996nv,Huovinen:2001cy,Borghini:2005kd}.
Figure~\ref{fig:MassSplit} shows $v_2$ as function of transverse momentum for two particle species.
As expected, at low \pt the elliptic flow clearly depends on the mass of the particle with $v_2$ at a fixed \pt decreasing with increasing mass.
The hydrodynamic model calculations of $v_2(\pt)$ for pions and
(anti-)protons in Fig.~\ref{fig:MassSplit} are performed for two 
equations of state: the full curves are for an
EoS which incorporates the effect of a phase transition from a QGP to
a hadron gas, the dashed curves are for a hadronic EoS without phase
transition.  The hydro calculations clearly predict the observed
behavior rather well with a better description of the measurements
provided by the EoS incorporating a phase transition.  For the pions
the effect of a phase transition is less pronounced compared to the
protons. The lighter particles are more affected by
the temperature, thus less sensitive to the collective flow
velocity and vice versa for the heavier particles. One should not, however, conclude from the good fit of the ideal hydro calculations to the data
in Fig.~\ref{fig:MassSplit} that the EoS which includes the phase 
transition is the only allowed EoS.
To draw conclusions about the EoS one first has to better understand 
the initial 
conditions~\cite{Hirano:2005xf,Drescher:2006pi,Drescher:2007ax,Lappi:2006xc}, 
has to have a more realistic description of the phase 
transition~\cite{Huovinen:2005gy}, 
and quantify the effects of viscous 
corrections~\cite{Teaney:2003kp,Romatschke:2007mq, Luzum:2008cw, Song:2007fn, Song:2007ux, Dusling:2007gi}.

\begin{figure}[htb]
\begin{center}
  \includegraphics[width=0.65\textwidth]{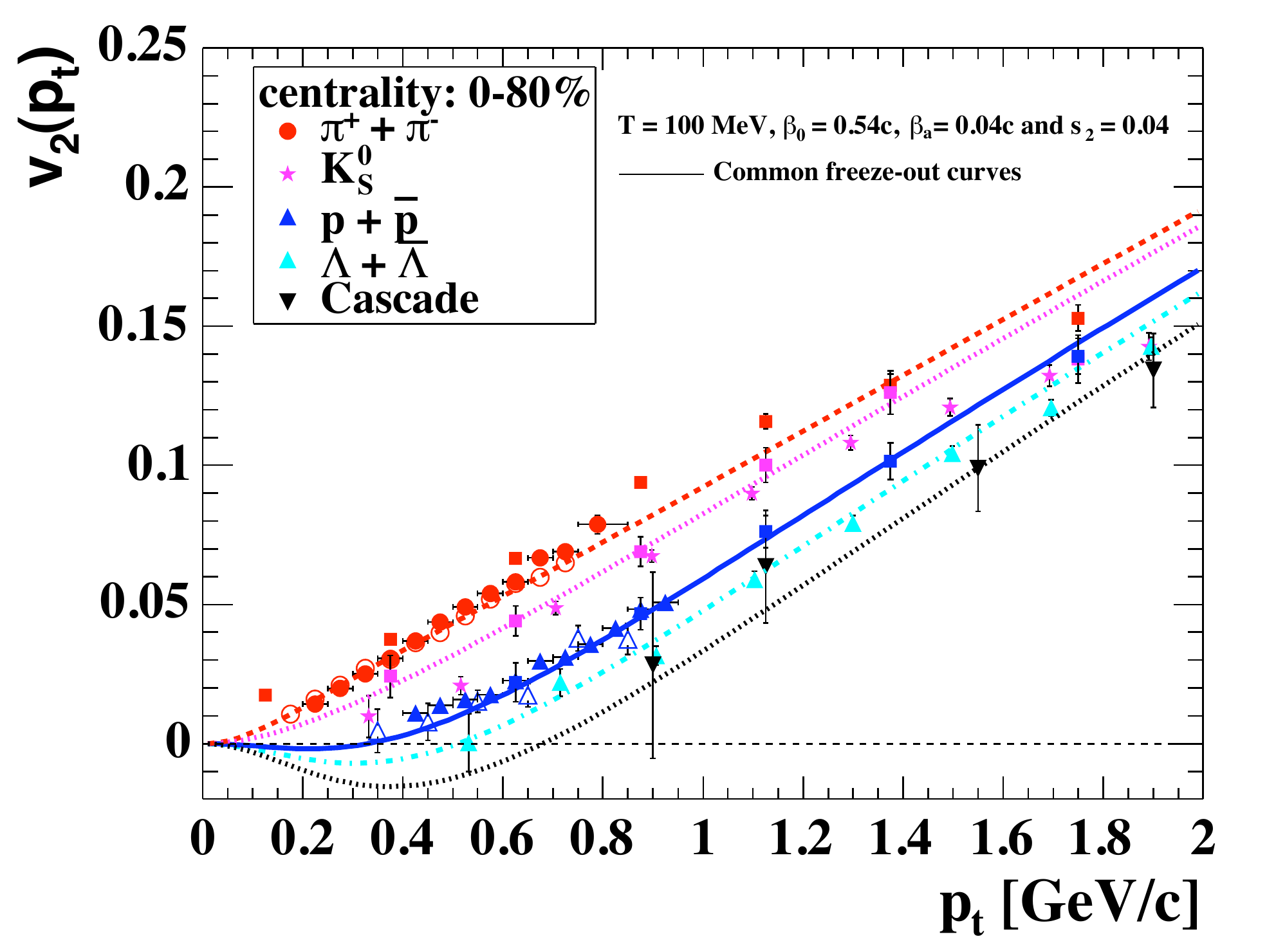}
\caption{Comparison of minimum bias $v_2(\pt)$ dependence on particle mass with
blast-wave model fits taken from Ref.~\cite{Snellings:2003kk}. The data are a compilation of PHENIX and STAR $v_2(\pt)$ measurements and the curves are 
from the fit to pion and proton  $v_2(\pt)$ performed in Ref.~\cite{Adler:2001nb}.}
\label{fig:MassSplitBW}
\end{center}
\end{figure}
Figure~\ref{fig:MassSplitBW} shows that the elliptic flow of 
the different mass particles at low \pt can be
characterized rather well by a common set of four freeze-out parameters: the
temperature, the mean radial flow velocity, the azimuthal dependence
of the radial flow velocity and the source
deformation~\cite{Adler:2001nb}. In hydrodynamics, these
parameters are not independent since they are related by the initial 
conditions and the equation of state. 
This also explains why $v_2\{\pt\}$ differential flow for heavier particles is sensitive to the EoS.  

The four parameter blast-wave model fit, to the extent that it provides 
an accurate description of the
system at freeze-out, has an advantage over hydrodynamic 
and cascade models in that it allows 
extraction of these freeze-out parameters without knowledge of  
the systems history before freeze-out. 
The blast-wave parameters obtained by fitting the pion and 
(anti-)proton $v_2(\pt)$ in Ref.~\cite{Adler:2001nb} 
also described the later obtained measurements of $v_2(\pt)$ for the heavier 
$\Lambda$ and Cascade particles, as can be seen in Fig.~\ref{fig:MassSplitBW}. 
In addition, the obtained $s_2$ parameter which describes
the source deformation was later confirmed also by azimuthally 
sensitive femtoscopy measurements~\cite{Adams:2003ra}.

In ideal hydrodynamics the mass ordering in $v_2$ persists up to large 
$\pt$, although less pronounced because the $v_2$  
of the different particles start to approach each other.
It is seen that at higher \pt the measurements start to deviate
significantly from hydrodynamics for all particle species,
and that the observed $v_2$ of the heavier 
baryons is larger than that of the lighter mesons. This mass
dependence is the reverse of the behavior observed at low $\pt$.
This is not expected for hadrons in hydrodynamics and is also 
not expected if the 
$v_2$ is caused by parton energy loss (in the latter case 
there would be, to first order, no dependence on particle type).
An elegant explanation of the unexpected particle type dependence
and magnitude of $v_2$ at large \pt is provided by the coalescence 
picture~\cite{Voloshin:2002wa,Molnar:2003ff} and is discussed 
in Sec.~\ref{sec:v2_coalescence}.

\subsubsection{Constituent quark number scaling}
\label{sec:v2_coalescence}

\begin{figure}[htb]
\begin{minipage}[b]{0.48\textwidth}
  \includegraphics[width=1.1\textwidth]{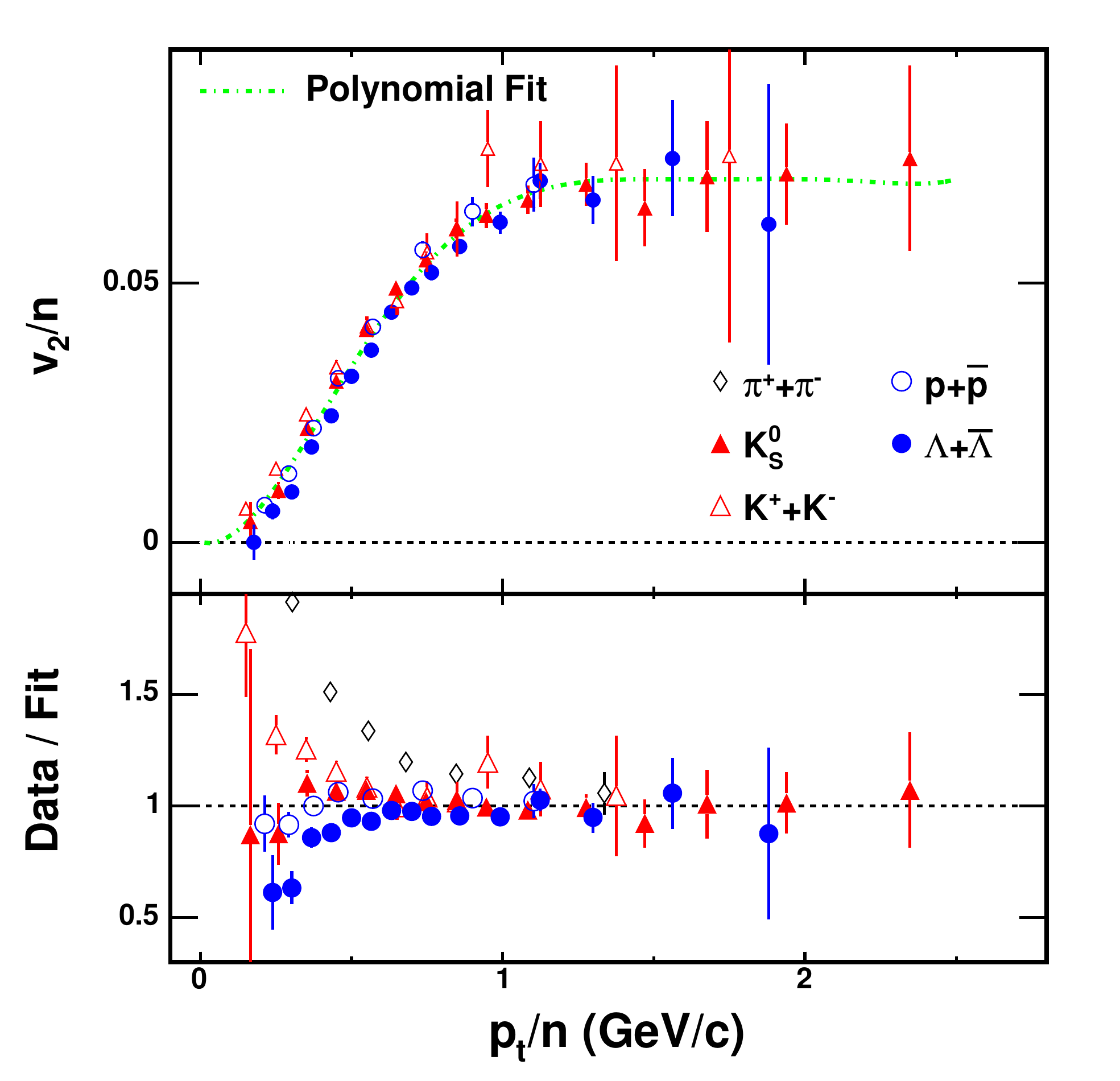}
  \caption{
Test of the constituent quark number scaling of elliptic flow for minimum
bias Au+Au collisions at $\sqrtsNN=200$~GeV. 
The dashed lines are polynomial fits~\cite{Abelev:2007qg}.
}
  \label{fig:coales}
\end{minipage}
\hspace{\fill}
\begin{minipage}[b]{0.48\textwidth}
  \includegraphics[width=0.99\textwidth]{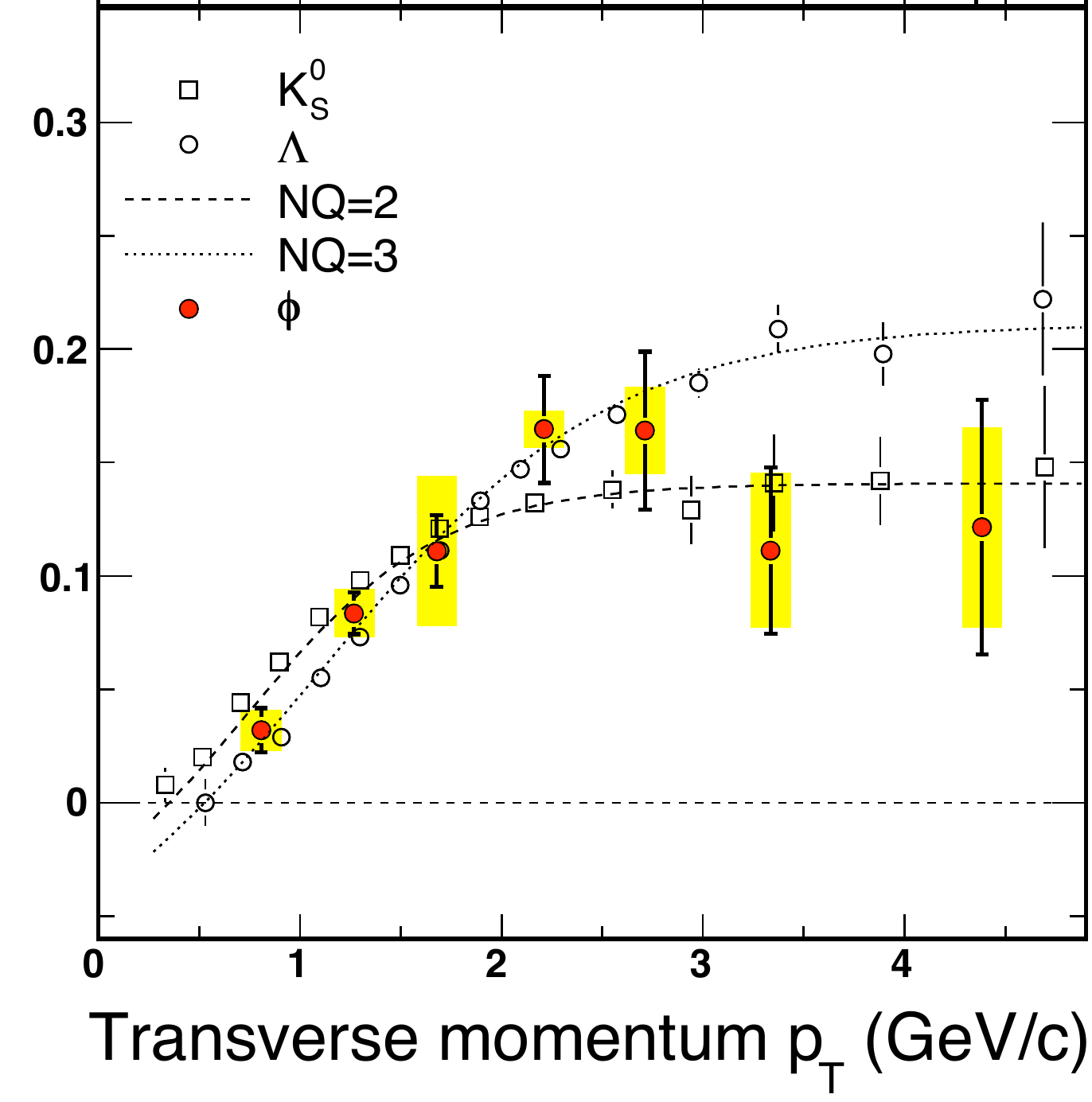}
  \caption{
  Minimum bias $v_2$ as a function of \pt for the $\phi$ meson, 
  \Kzs, and \llam~\cite{Abelev:2007rw}. 
  The dashed and dotted lines represent what is expected from 
   coalescence for two- and three-quark particles.}
  \label{fig:phiScaling}
\end{minipage}
\end{figure}

Elliptic flow of identified particles measured in Au+Au
collisions at RHIC exhibits a remarkable 
{\em scaling} with the {\em number of constituent quarks} --
an apparent dependence 
of hadron elliptic flow at intermediate transverse momenta, 
$p_T \sim 2-4$~GeV/$c$, on the number of constituent quarks in the hadron.
This  observation is of particular interest and importance,
as it indicates that the system is in a deconfined stage.
In a more general sense, it appears that high energy nuclear 
collisions provide a  window of opportunity to prove that hadron
production indeed happens via the constituent quark phase. 
It has been noticed in Ref.~\cite{Voloshin:2002wa} that if hadrons are
formed via coalescence of the constituent quarks then there
should be a region in the transverse momentum space where particle yield
would be proportional to the quark density to the power equal to the
number of constituent quarks in the produced hadron, 
2 for mesons and 3 for baryons. 
Besides other important consequences, such as enhanced relative
production of baryons in this transverse momentum region, 
this picture leads 
to the constituent quark scaling of elliptic flow, 
$v_2(p_T) \approx n\ v_{2}(p_T/n)$, where $n$ is the number of
constituent quarks in the hadron~\cite{Voloshin:2002wa,Molnar:2003ff}.
Figure~\ref{fig:coales} shows this scaling holds with good accuracy.
Note that while the scaling itself is limited to a specific region 
in transverse momentum, the coalescence mechanism can be valid at all smaller
momenta. The reason for the scaling violation at lower momenta 
might be that the equations used to describe coalescence break down.
For identified particles the scaling for mesons 
and baryons is based on the equations
\be
\frac{d^3n_M}{d^3p_M} \propto [\frac{d^3n_q}{d^3p_q}(p_q\approx p_M/2)]^2
\; \; \; \; \; \; \; \; 
{\rm and}\;\; \;\;
\frac{d^3n_B}{d^3p_B} \propto [\frac{d^3n_q}{d^3p_q}(p_q\approx p_B/3)]^3,
\label{eq:coales}
\ee
which are valid only if the probability of coalescence is relatively low.
At low transverse momentum, where {\em most} of the quarks hadronize via
coalescence these equations break unitarity. Note that according
to these equations, the hadron yield scales with power 2 or 3 of the quark
density. For a discussion of the particle spectra calculations 
and further development of the coalescence picture we refer to the review in Ref.~\cite{Fries:2008hs}.

In this picture the quantity $v_{2}(p_T/n)$ is interpreted as the elliptic
flow of constituent quarks. 
This means deconfinement -- as the constituent quarks must be in
a deconfined phase in order to be freely ``reshuffled'' into final hadrons. 
This could be the first, and very strong
argument for an observation of deconfined matter at RHIC.  
The $\phi$ meson is an interesting test of constituent quark scaling. 
It has the mass of a proton but contains only two quarks. 
Figure~\ref{fig:phiScaling} shows that its $v_2$ values are consistent 
with the two-quark curve thereby supporting the quark picture.

\begin{figure}[htb]
\begin{center}
  \includegraphics[width=1.\textwidth]{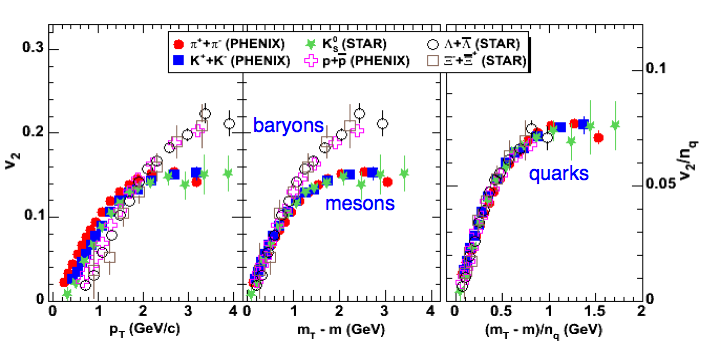}
  \caption{$v_2$ for $\sqrtsNN= 200$~GeV Au+Au as a function 
  of \pt and \ket, and also $v_2/n_q$ vs $\ket/n_q$ in the last 
  panel~\cite{Adare:2006ti}. }
  \label{fig:v2scaling}
\end{center}
\end{figure}
Figure~\ref{fig:v2scaling} shows that plotting $v_2$ vs
transverse kinetic energy $\ket = m_T - m$, where $m_T = \sqrt{\pt^2 + m^2}$, results in the formation of distinct branches for mesons and baryons. 
Scaling with the number of constituent quarks $n_q$ then coalesces the
two branches into one curve.
At the moment there is no agreement on the reason for such a universal
scaling except that re-plotting the data vs transverse kinetic energy
to some extent compensates for the effect of radial flow (see the
discussion in section~\ref{sec:rad-ani}). But we mention that
transverse kinetic energy scaling is an approximation to a more
general scaling found in the Buda-Lund model~\cite{Csanad:2005gv}.
Also, a quark recombination model based on q-qbar resonance interactions may  account for this phenomenon\cite{Ravagli:2008rt}.

\begin{figure}[htb]
\begin{minipage}[b]{0.48\textwidth}
  \includegraphics[width=0.99\textwidth]{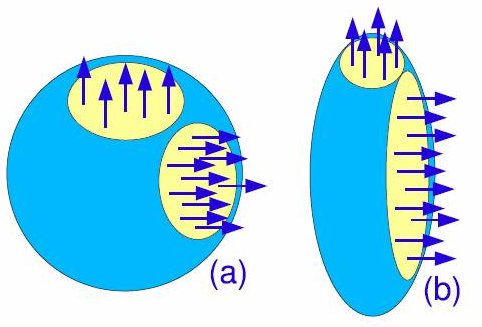}
  \caption{
Two possible source configurations, corresponding to the same quark
elliptic flow, but only case (a) leads to quark-number 
scaling~\cite{Pratt:2004zq}. 
}
  \label{fig:TwoSources}
\end{minipage}
\hspace{\fill}
\begin{minipage}[b]{0.48\textwidth}
  \includegraphics[width=0.99\textwidth]{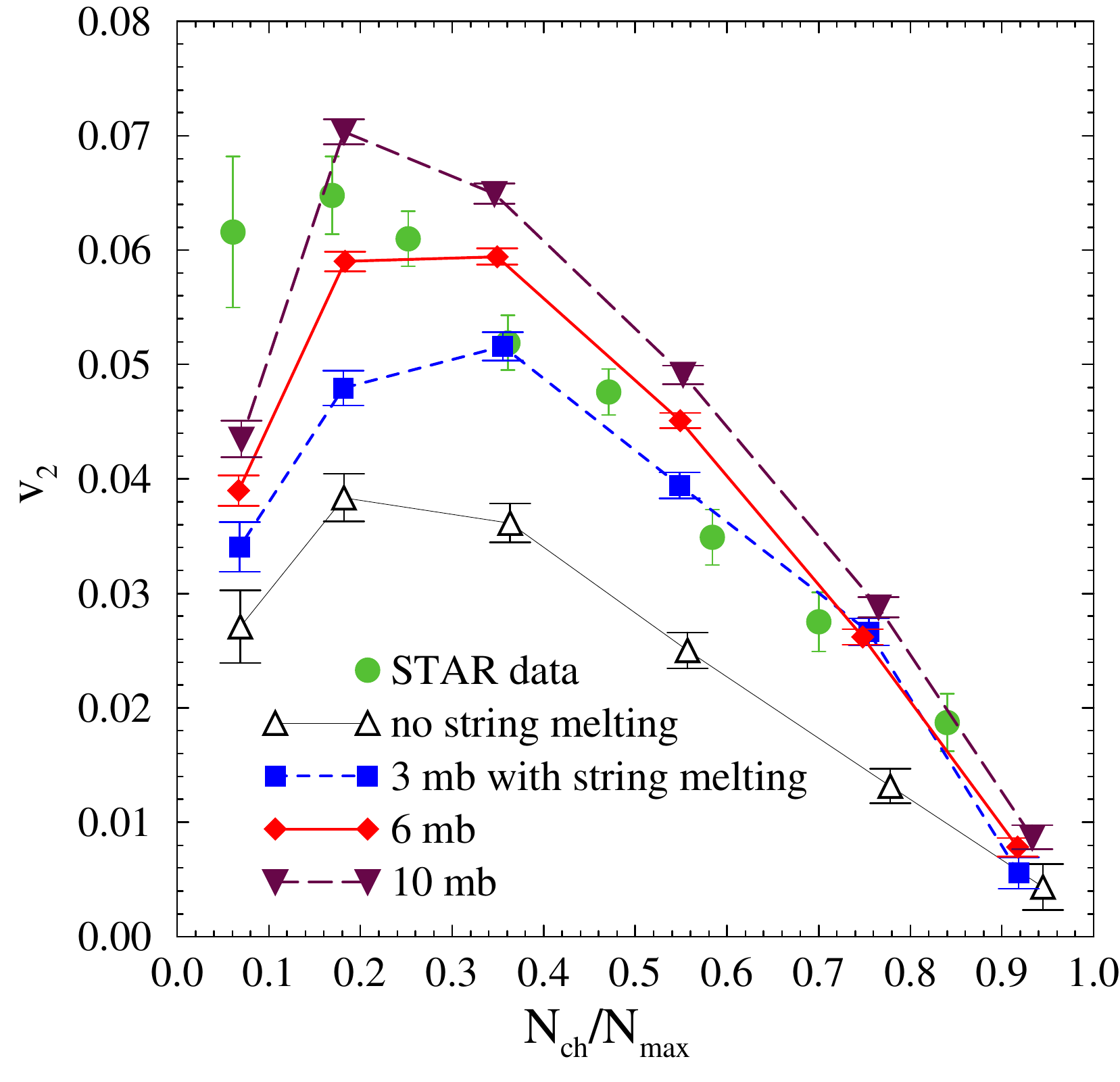}
  \caption{
Centrality dependence of elliptic flow in Au+Au collisions at
$\sqrtsNN=130$~GeV~\cite{Adams:2004bi} compared to results 
of the AMPT model with string melting and different 
cross-sections~\cite{Lin:2001zk}. } 
  \label{fig:v2ampt}
\end{minipage}
\end{figure}

A detailed study of necessary conditions 
for the quark-number scaling to be valid
has been performed by Pratt and Pal~\cite{Pratt:2004zq}.
Two limiting possibilities are illustrated in Fig.~\ref{fig:TwoSources},
where the particles (quarks)  of a given velocity are represented by arrows. 
In case (a) the effective volumes of
the right-moving and upward-moving particles are the same but the
right-moving particles have a larger phase space density. In case (b) the
densities are the same but the volumes differ.
Both cases correspond to the same quark $v_2$, but only case (a) would lead
to the quark-number scaling. 
Based on this analysis one might conclude that constituent quark scaling 
contradicts local thermalization 
and freeze-out at a constant phase-space density. Even if true, this
does not diminish the validity of the conclusion on
deconfinement, but points to a very interesting possibility 
that the system created in the heavy ion collision 
can be in a deconfined but not completely thermalized state.
It also does not exclude the possibility that thermalization happens
only for lower transverse momenta.

\begin{figure}[htb]
\begin{center}
  \includegraphics[width=0.7\textwidth]{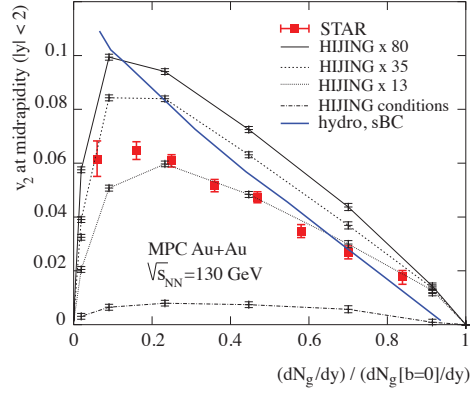}
  \caption{
Elliptic flow from MPC calculations compared to data 
for $\sqrtsNN= 130$~GeV Au+Au~\cite{Molnar:2001ux}.
}
  \label{fig:mpc_v2}
\end{center}
\end{figure}

The success of quark coalescence as a hadronization mechanism
in describing particle production at intermediate transverse
momenta hints on the importance of the constituent quark dynamics in the
evolution of the system in general. Recall that
the transport models~\cite{Lin:2001zk,Molnar:2001ux} 
in their standard configuration fail to describe the strong
increase of elliptic flow with energy. 
They have to significantly increase the parton transport cross 
section or the matter density in order to reach the experimental values.
On the other hand if one looks carefully into what particular 
parameters are required in order to describe the data, an interesting picture
emerges: the density and partonic cross sections are just what one would 
expect for a system of constituent quarks. 
For example, Fig.~\ref{fig:v2ampt} shows 
a comparison of the experimental data to the 
AMPT~\cite{Lin:2001zk} model calculations in the so-called 
melted string scenario. 
The main assumption of this scenario is that the total number of partons 
in the system equals the number of constituent quarks in the produced 
hadrons, exactly what one would use for a model based on 
the picture of a system of constituent quarks. 
Note that the model describes the data best with a cross
section of about 5~mb, which is what one expects for constituent quarks.
Similar conclusion can be drawn from Fig.~\ref{fig:mpc_v2} which shows
the results of calculations in the MPC~\cite{Molnar:2001ux} model, using as a reference the
standard HIJING gluon density ($dN_g/dy = 1200$ for central Au+Au
collisions) and a cross section of about 2~mb. Again, rescaling the parton density
to that corresponding to the total number of quarks in the final hadrons and
increasing the cross section to 5--6~mb one would achieve a reasonable
description of the data.

\begin{figure}[htb]
\begin{center}
  \includegraphics[width=0.5\textwidth]{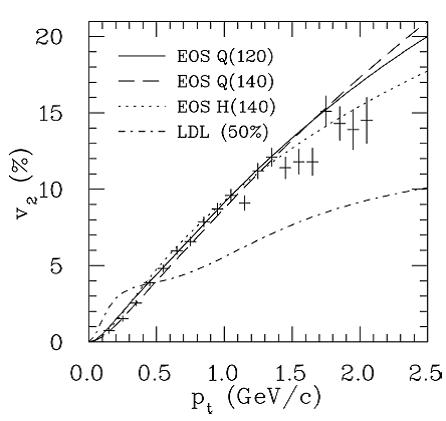}
  \caption{
	Calculations~\cite{Kolb:2000fha} compared to
  experimental data. The LDL model results are shown by the dash-dot
  line. Other lines are hydrodynamic calculations with different
  equations of state.
} 
  \label{fig:v2pt_ldl}
\end{center}
\end{figure}

Figure~\ref{fig:v2pt_ldl} shows the comparison of the experimental
data on $v_2(\pt)$ to calculations in the low density
limit~\cite{Kolb:2000fha}. 
According to the equations of the LDL, $v_2(\pt)$ saturates at
transverse momenta about a few times the particle mass. 
The calculations, shown by a dashed curve, exhibits this very behavior
as first exhausting flow of pions and then at higher \pt that of
protons. Indeed the results do not resemble the data. But it is
clear that similar calculations performed with masses of the constituent
quarks (a few times that of the pion) taken together with coalescence
enhancement at intermediate transverse momenta could well be
consistent with the data. (See also the upper panel in
Fig.~\ref{fig:coales} and the right panel of Fig.~\ref{fig:v2scaling}, which can be considered as $v_2(\pt)$ of constituent quarks.) 
This again strongly advocates for a major role of  constituent quark
dynamics in the system evolution.

Note that, in principle, hadronization always, even in small systems and at
lower energies, occurs via constituent quark coalescence. 
The importance of the observation of quark number scaling in elliptic flow, is that it
constitutes deconfinement -- ``free'' kinetics of quarks in the system. In this
sense this scaling is a signature of deconfinement and should be violated  
under conditions when deconfinement does not happen.

\subsubsection{High \pt region}

Elliptic anisotropy at high transverse momenta is an 
interesting observable as it is believed that it reflects the path length
dependence of high \pt parton energy loss~\cite{Snellings:1999gq}.
At sufficiently high transverse momentum in Au+Au collisions, 
hadron yields are thought to contain a significant fraction originating 
from the fragmentation of high energy partons, 
resulting from initial hard scatterings. 
Calculations based on perturbative QCD (pQCD) predict
that high energy partons traversing nuclear matter lose energy 
through induced gluon radiation~\cite{Gyulassy:1990ye,Wang:1991xy}. 
The energy loss (jet quenching) is expected to depend strongly on
the color charge density of the created system and the traversed path length
of the propagating parton. 
In non-central heavy-ion collisions, the geometrical
overlap region has an almond shape in the transverse plane, 
with its short axis in the reaction plane. 
Depending on the azimuthal emission angle, partons traversing such a system, 
on average, experience different path lengths and therefore different energy loss. This mechanism introduces an azimuthal anisotropy in particle production 
at high transverse momenta~\cite{Snellings:1999gq,Wang:2000fq,Gyulassy:2001kr}.

\begin{figure}[htb]
\begin{center}
    \includegraphics[width=0.7\textwidth]{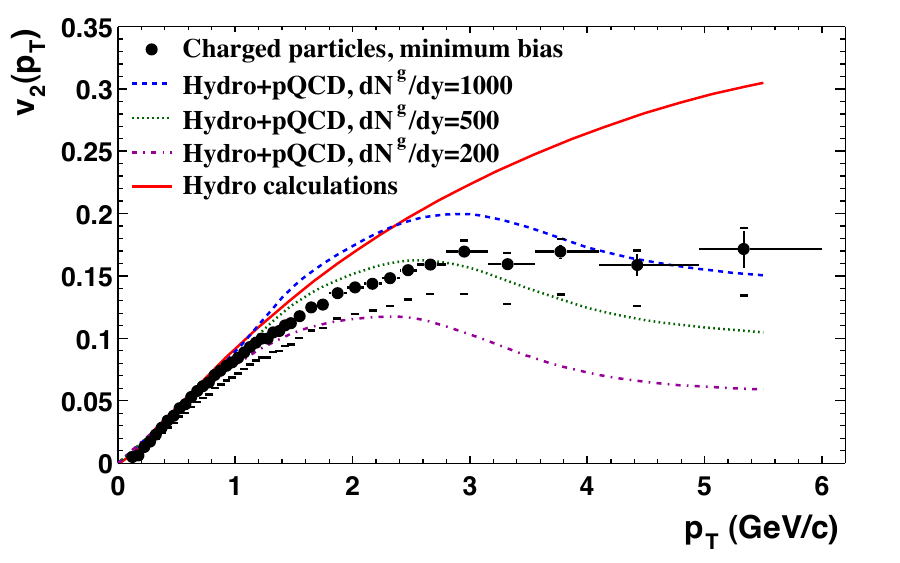}
    \caption{The predicted magnitude of $v_2$ at higher \pt based on energy 
    loss expectations for different values of the gluon density in a static 
    medium~\cite{Gyulassy:2000gk}. The data were obtained with a modified event plane method where particles with $\Delta \eta < 0.5$ were excluded from the calculation of the event plane~\cite{Adler:2002ct}. }
    \label{fig:predict_highpt}
\end{center}
\end{figure}

The dashed lines in Fig.~\ref{fig:predict_highpt} show the first 
quantitative theory predictions based on energy-loss 
calculations in a static medium~\cite{Gyulassy:2000gk} and are compared 
to STAR measurements~\cite{Adler:2002ct}. 
However, as was discussed in the previous
section, it was realized that in the \pt range of $2-6$ GeV/$c$ hadron 
yields might not dominantly originate from the fragmentation of high energy 
partons but are instead mostly produced by quark coalescence. 
Therefore in order to compare to predictions of parton energy loss $v_2$ has 
to be measured above $\pt=6$ GeV/$c$. 
Unfortunately nonflow contributions which at low \pt are modest are 
significant at high $\pt$. 
In fact nonflow might even start to dominate the measured two-particle 
azimuthal correlations at high \pt which 
complicates making reliable $v_2$ measurements.

\begin{figure}[htb]
\begin{center}
    \includegraphics[width=0.8\textwidth]{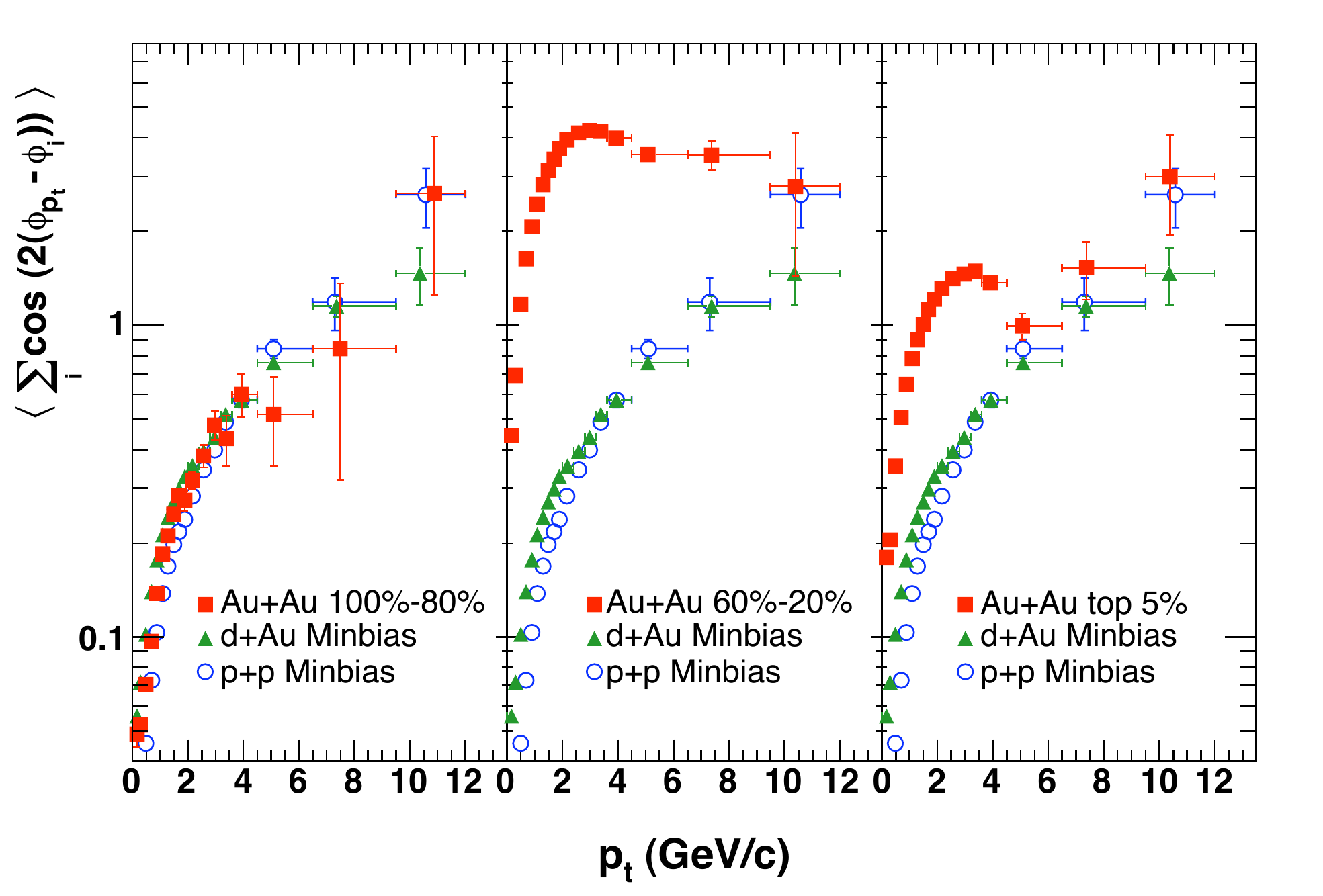}
    \caption{Azimuthal correlations in p+p, d+Au, and Au+Au collisions of different centralities~\cite{Adams:2004bi,Adams:2004wz}.}
    \label{fig:AApp}
\end{center}
\end{figure}
Assuming that there is no anisotropic flow in p$+$p and d$+$Au collisions
one can estimate the contribution from nonflow by comparing the two- 
particle azimuthal correlation 
in these systems with the azimuthal correlation in Au$+$Au.
Figure~\ref{fig:AApp} shows the measured azimuthal correlation as a
function of transverse momentum in p$+$p and d$+$Au collisions as well 
as for three different centralities in Au$+$Au.
The two-particle azimuthal correlation shown is defined in Eq.~(\ref{eq:QAApp}). 
It is seen that the magnitude of the azimuthal correlations in p$+$p,
d$+$Au and  very peripheral Au$+$Au is comparable over the complete measured 
transverse momentum range.
This indicates that in peripheral Au$+$Au collisions nonflow correlations 
are the dominant contribution. 
On the other hand, for mid-central and central Au$+$Au collisions at low 
and intermediate \pt the observed azimuthal correlation is much stronger 
than in p$+$p and d$+$Au as is expected in the presence of strong anisotropic 
flow. However, at high $\pt$, even for these mid-central and central Au$+$Au 
collisions, the two-particle azimuthal correlation 
are again dominated by the correlations already seen in p$+$p and d$+$Au. 
The remaining difference in the azimuthal correlation at high \pt can be used 
to estimate the elliptic flow, which is done in Refs.~\cite{Adams:2004bi,Adams:2004wz}.

\begin{figure}[htb]
\begin{center}
  \includegraphics[width=0.6\textwidth]{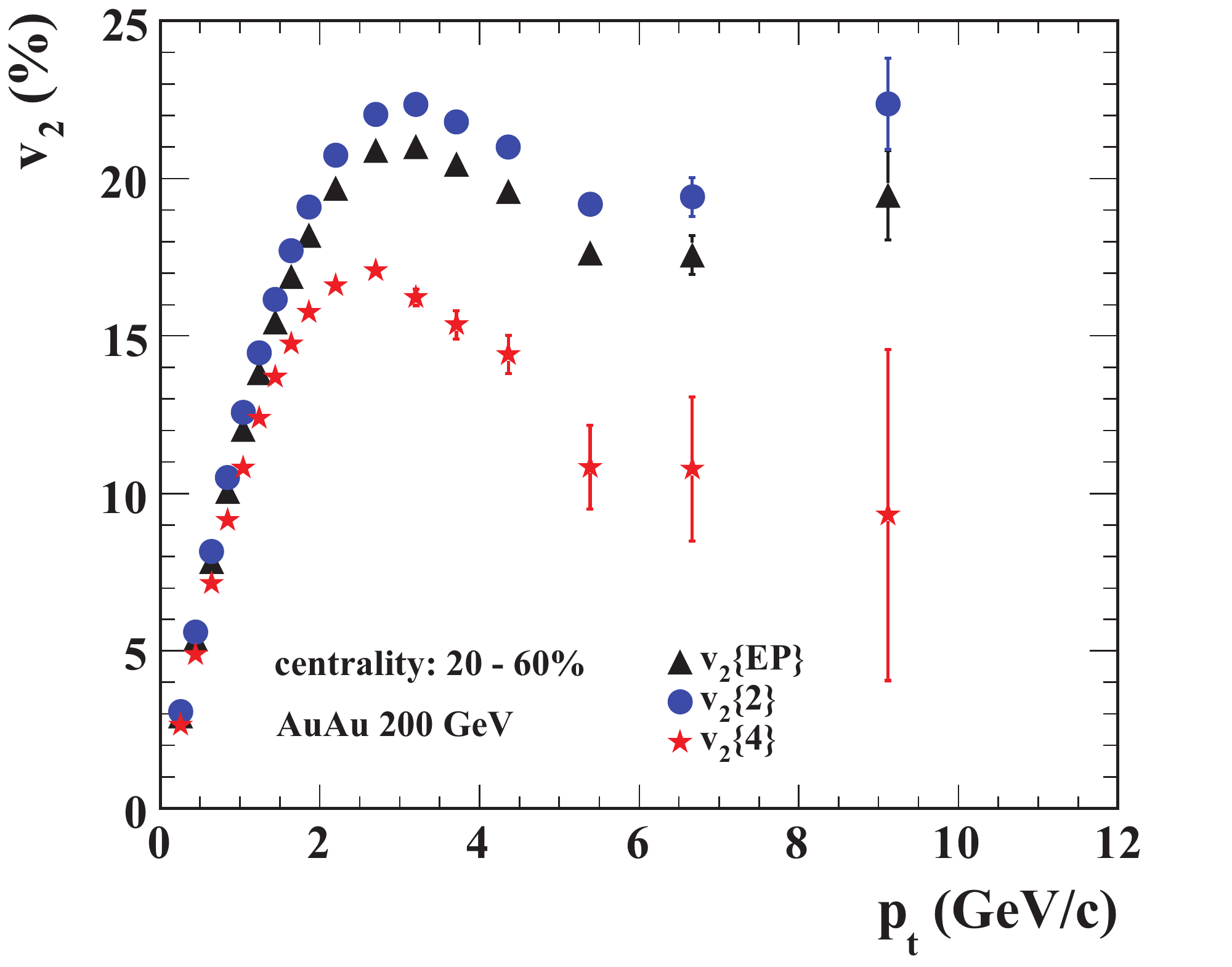}
  \caption{Comparison of $v_2(\pt)$ measured using 2- and 4- particle 
  correlations compared to event plane results. The centrality was 20--60\% for $\sqrtsNN=200$~GeV Au+Au~\cite{ThesisYuting:2007}. }
  \label{fig:v2_highpt}
\end{center}
\end{figure}
Instead of using p$+$p and d$+$Au collisions as a reference for nonflow, 
the large data sample obtained at RHIC in Au+Au collisions 
at $\sqrtsNN = 200$~GeV now also allows for calculating 
the anisotropic flow coefficients  up to $10$~GeV/$c$ using higher 
order cumulants. 
Figure~\ref{fig:v2_highpt} shows the $v_2(\pt)$ obtained using 
the $4^{th}$ order cumulant, and the values are compared 
to the $v_2(\pt)$ obtained using the event plane method and the 2-particle 
cumulant. The $v_2$ values from the event plane method and  
2-particle cumulant exhibit the same large values at high \pt as seen in 
Fig.~\ref{fig:AApp}. The $v_2$ obtained from
the $4^{th}$ order cumulant method at high \pt is at least a factor of two smaller, however still significant. 
These $v_2\{4\}(\pt)$ values can be compared to model 
calculations to better constrain the mechanism responsible for parton energy loss.

\begin{figure}[htb]
  \includegraphics[width=0.49\textwidth]{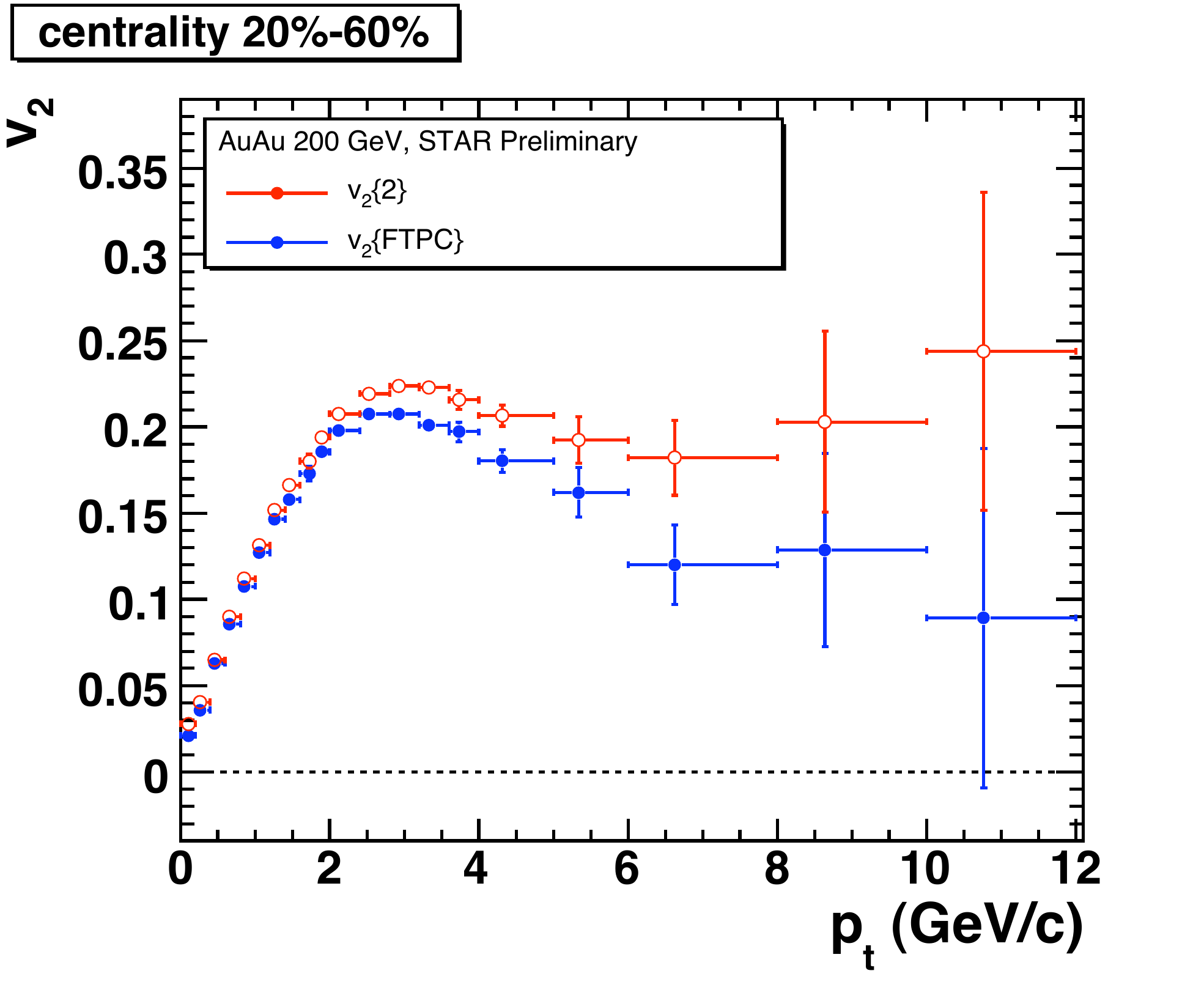}
  \includegraphics[width=0.49\textwidth]{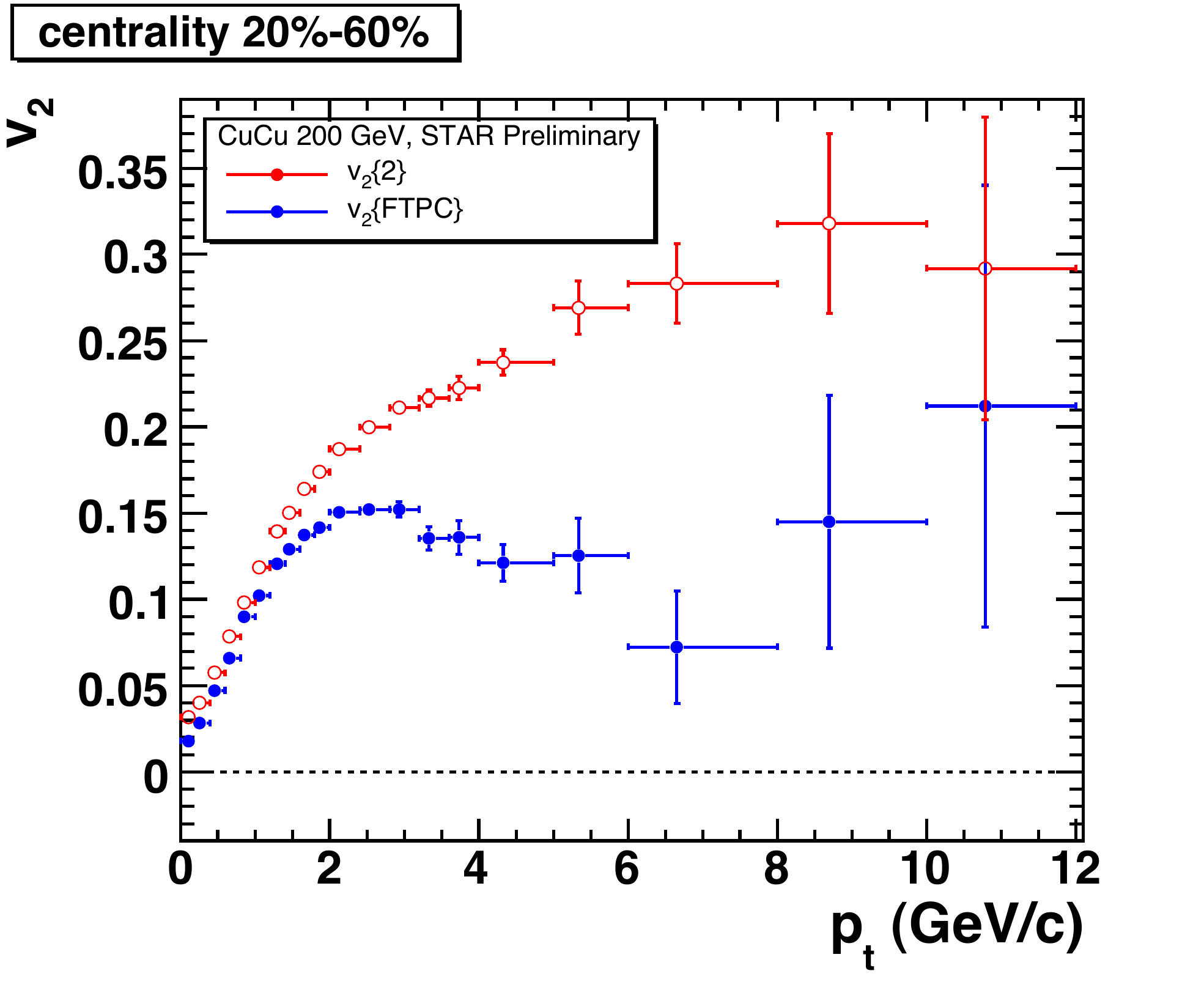}
  \caption{
Comparison of $v_2(\pt)$ measured from the 2-particle cumulant method and from the event plane method using the STAR FTPCs. 
The centrality was 20--60\% for $\sqrtsNN=200$~GeV~\cite{Voloshin:2007af}. Au$+$Au is on the left and Cu$+$Cu on the right.
}
  \label{fig:v2_highpt_ftpc}
\end{figure}

Another approach to reduce nonflow is to make sure that there is a large gap in rapidity between the particles used to determine the event plane and the particles used to calculate the elliptic flow. 
In STAR, in addition to the ZDCs, the Forward TPCs (FTPCs) which are located at  $2.9 \lt \eta \lt 3.9$ can be used for this~\cite{Voloshin:2007af}. 
Figure~\ref{fig:v2_highpt_ftpc}~left compares the $v_2(\pt)$ measured at midrapidity ($\eta \lt |0.9|$) from the 2-particle cumulant and from the event plane method using the STAR FTPCs. 
As was already observed, the nonflow (here the difference 
between $v_2\{2\}$ and $v_2\{\mathrm{FTPC}\}$) is significant above 2 GeV/$c$ and becomes large above 6 GeV/$c$. 
Assuming that the nonflow is completely removed with $\Delta \eta \sim 2$ the difference between $v_2\{4\}$ 
from Fig.~\ref{fig:v2_highpt} and $v_2\{\mathrm{FTPC}\}$ in Fig.~\ref{fig:v2_highpt_ftpc}~left shows the difference between $v_2$ in the reaction plane and $v_2$ in the participant plane with fluctuations. However, because the curves still turn up at very high $\pt$, there is probably still some nonflow even with the event plane determined in the FTPCs.

Clearly for mid-central collisions in Au$+$Au the nonflow contribution at high $\pt$ is already significant. In case
of Cu$+$Cu, i.e. a smaller system, the elliptic flow for the same midcentral collisions is expected to be smaller 
while the relative nonflow contribution is expected to be larger.
Figure~\ref{fig:v2_highpt_ftpc}~right shows the $v_2(\pt)$ measured in Cu$+$Cu with the $2^{nd}$ order cumulant and 
with the event plane method using the STAR FTPCs~\cite{Voloshin:2007af}. 
It is seen that the $v_2$ in Cu$+$Cu obtained from the two-particle cumulant method at intermediate \pt (4--12 GeV/$c$) 
is even larger than in Au$+$Au. 
In contrast the $v_2(\pt)$ in Cu$+$Cu obtained from the event plane method using the FTPCs is smaller than in Au$+$Au over almost the whole $\pt$ range. 
This confirms that in Cu$+$Cu indeed the $v_2(\pt)$ is smaller than in Au$+$Au while the nonflow effects are larger.
Even though statistical uncertainties are large for $v_2\{\mathrm{FTPC}\}$ above 6 GeV/$c$ the transition to again 
increasing values of $v_2$ might indicate that a rapidity gap of $\Delta \eta \sim 2$ is still not large enough to remove all nonflow in Cu$+$Cu. 

\subsubsection{Rare probes}

Elliptic flow of heavy flavor and direct photons,
often called by the common name rare probes, attract significant 
attention as they 
provide important additional information about the dynamics and
properties of the sQGP created in heavy ion collisions. 
Heavy flavor elliptic flow is studied either directly or via single electrons from semi-leptonic decays.

\begin{figure}[htb]
\begin{center}
  \includegraphics[width=0.6\textwidth]{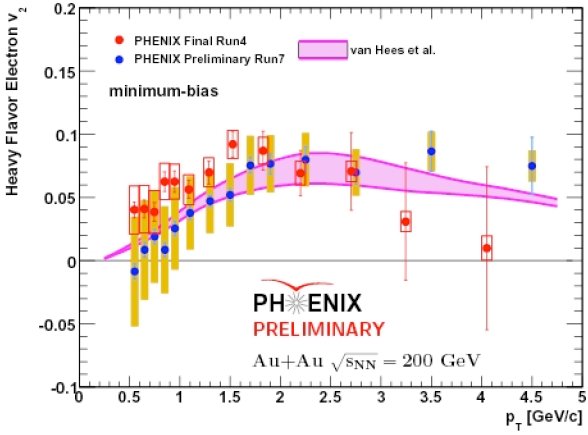}
  \caption{
Heavy flavor electron elliptic flow~\cite{Awes:2008qi}. 
The shaded area shows the model prediction based on a relativistic Langevin simulation of heavy quarks~\cite{vanHees:2005wb}.
}
  \label{fig:single_e}
\end{center}
\end{figure}

Elliptic flow of hadrons containing charm or bottom quark(s) provide
direct information on the collectivity of heavy flavor quarks in the
deconfined stage. So far there exist data from PHENIX on flow of
single electrons and very limited data on flow of $J/\Psi$.
Simulations show that for $\pt >$ 2--3~GeV/$c$ most electrons are from
semi-leptonic decays of charm and bottom, and at $\pt >$ 5--8~GeV/$c$ it could
be dominated by bottom decays. At such transverse momentum, 
electrons to a large degree preserve
the direction of heavy flavor mesons and thus can be studied to evaluate the elliptic flow of charm or bottom quarks. 
In a simple coalescence picture~\cite{Lin:2003jy} the elliptic flow
of charm hadrons is a combination of the elliptic flow of
corresponding quarks taken at appropriate transverse momentum (roughly
shared in proportion to the quark masses). It appearers that, for
example, charm meson elliptic flow at transverse momentum of about
3~GeV/$c$ may differ more than a factor of 2 depending on whether charm
quarks participate in collective flow~\cite{Lin:2003jy}.

\begin{figure}[htb]
\begin{center}
  \includegraphics[width=0.6\textwidth]{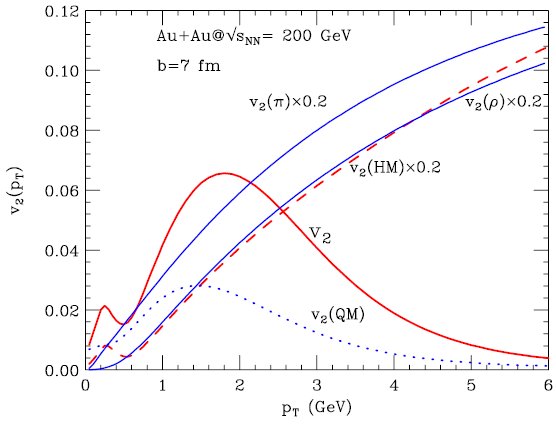}
  \caption{ $v_2(\pt)$ of thermal photons~\cite{Chatterjee:2005de} in
    Au+Au collisions at $\sqrtsNN=200$~GeV at $b=0$. 
	Quark and hadronic matter contributions are shown separately. 
	Elliptic flow of pions and $\rho$ mesons obtained with the pair-wise method are shown for comparison. }
  \label{fig:v2_thermalPhotons}
\end{center}
\end{figure}

\begin{figure}[htb]
\begin{center}
  \includegraphics[width=0.8\textwidth]{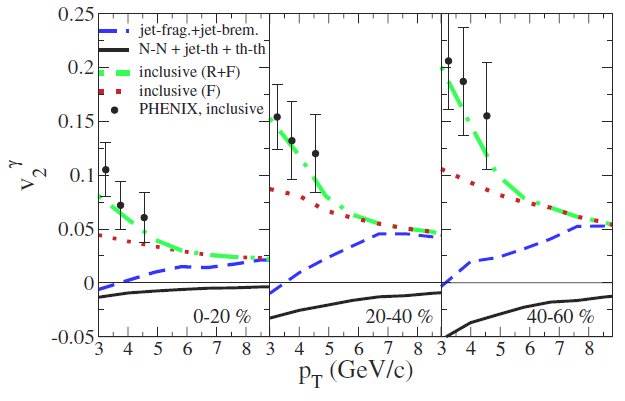}
  \caption{
	Photon $v_2(\pt )$ for Au+Au collisions at RHIC~\cite{Turbide:2005bz} obtained with the event plane determined from the beam-beam counters at $\mid \eta \mid$ from 3.1 to 3.9. 
	The dashed line shows jet-fragmentation and induced
  bremsstrahlung only, while the solid lines give jet-photon
  conversion, primary hard and thermal photons. The dotted line shows
  direct photons from decay of neutral mesons from jet
  fragmentation. The dot-dashed line also includes photons from pion
  decays. 
}
\label{fig:v2_allPhotons}
\end{center}
\end{figure}

PHENIX measurements~\cite{Adler:2005ab,Awes:2008qi}, shown in
Fig.~\ref{fig:single_e}, indicate that heavy quarks flow at the same
level as light ones. From this measurement (combined with heavy flavor
electron spectra) PHENIX derives an upper limit on viscosity $\eta/s
\le (1.3-2)/(4\pi)$, confirming that the system behaves as a near
perfect liquid. 
Results on elliptic flow of $J/\Psi$ have just started to
appear~\cite{Awes:2008qi}, and in spite of large errors they immediately
attract attention as they are largely consistent with out-of-plane elliptic
flow, thus possibly indicating strong effects of radial 
flow~\cite{Krieg:2007bc}, as discussed in Sec.~\ref{sec:rad-ani}.

Elliptic flow of direct photons has a very rich physics as there have been
identified several mechanisms which lead to significant elliptic flow.
The most important are: (i) thermal photons, (ii) photons from high
\pt parton fragmentation (in vacuum), (iii) bremsstrahlung
and ``jet conversion'' photons
from fast partons propagating inside the medium.
Thermal photons originate both in QGP and hadronic phases. 
Calculations~\cite{Chatterjee:2005de} show that even though the QGP phase is
short due to higher temperature, it gives the largest contribution to
photons at intermediate transverse momenta 
(see Fig.~\ref{fig:v2_thermalPhotons}). 
High \pt photons come from hard parton-parton scattering with
subsequent usual fragmentation (leading to positive \vt) or to
bremsstrahlung, which is stronger out-of-plane leading
to negative $v_2$. The combination of all these effects results in
a rather complicated picture (see Fig.~\ref{fig:v2_allPhotons}). 
There is hope that with large statistics different mechanisms could be separated with different isolation cuts.
Finally, note that the mechanism, similar to that of ``jet conversion''
could also lead to an enrichment of jet fragments with some rare
particles (including strangeness content). It would lead to the
corresponding negative contribution to the elliptic flow of that 
particle~\cite{Liu:2008kj}.

\subsection{Higher harmonics}

\begin{figure}[htb]
\begin{center}
\includegraphics[width=.48\textwidth]{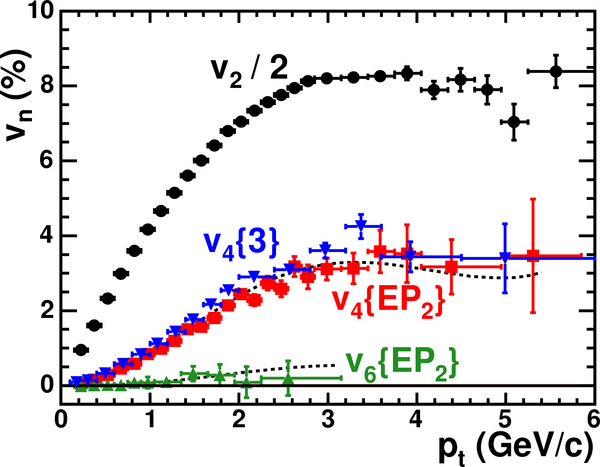} 
\caption{$v_2$, $v_4$, and $v_6$ vs \pt with respect to the second harmonic event plane, for minimum bias 200 GeV Au+Au 
collisions. Also shown is $v_4$ from three-particle cumulants.
The dotted curves are $1.2\ v_2^2$ and $1.2\ v_2^3$~\cite{Adams:2003zg}.}
\label{fig:v4v6}
\end{center}
\end{figure}

\begin{figure}[htb]
\begin{center}
\includegraphics[width=1.0\textwidth]{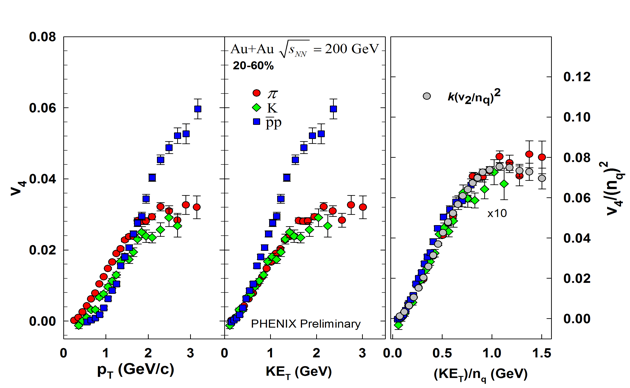} 
\caption{$v_4$ for 20--60\% centrality $\sqrtsNN = 200$~GeV Au+Au as a function of \pt and $\ket$, and also 
$v_4/(n_q)^2$ vs $\ket/n_q$ in the last panel~\cite{Huang:2008vd}. The event plane was determined with the RxNP detector at $\mid \eta \mid$ from 1.0 to 2.8.}
\label{fig:v4scaling}
\end{center}
\end{figure}

\begin{figure}[htb]
\begin{center}
\includegraphics[width=.6\textwidth]{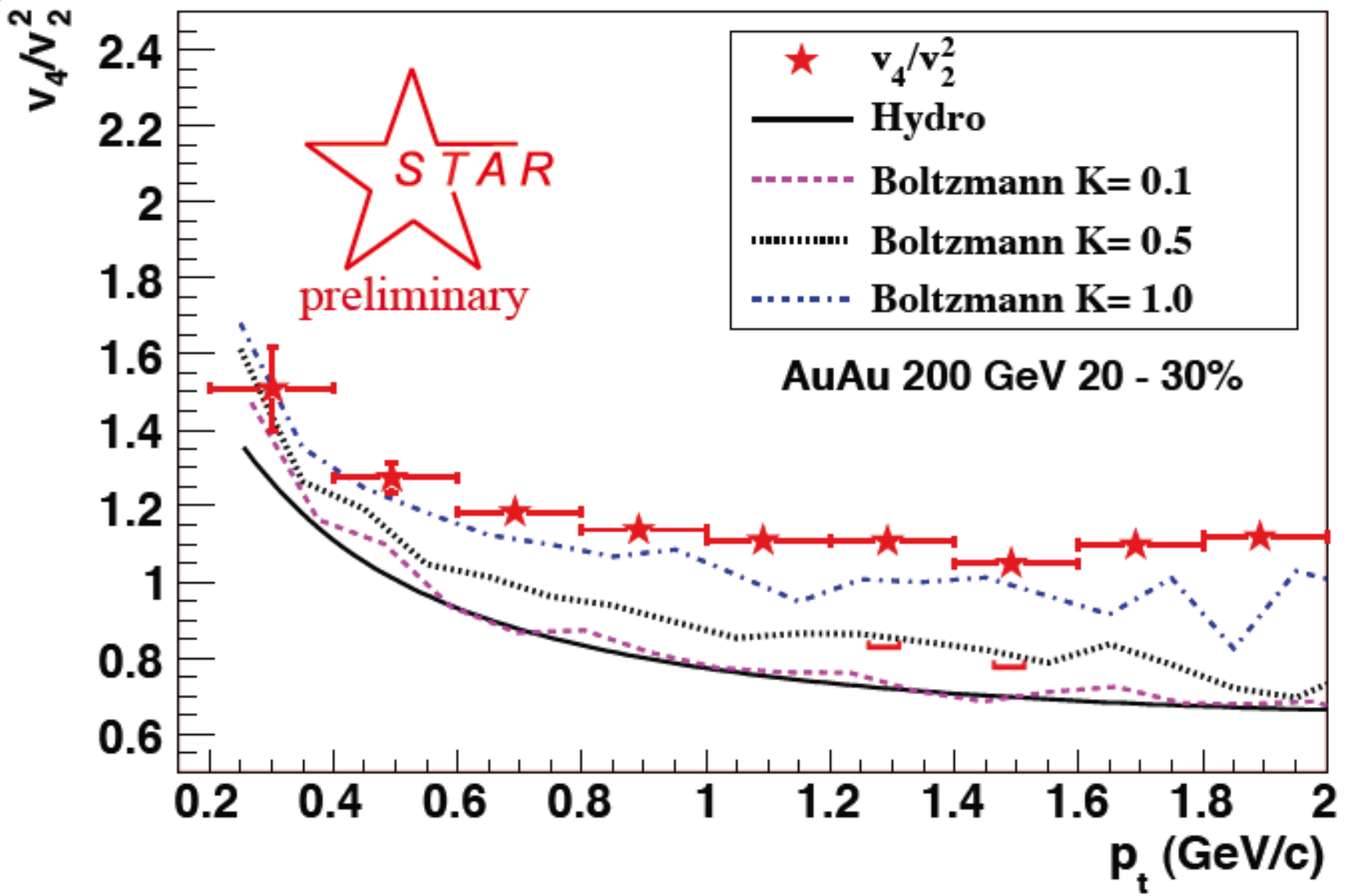} 
\caption{$v_4/v_2^2$ vs \pt for 20--30\% centrality for $\sqrtsNN = 200$~GeV Au+Au. Also shown are ideal hydro and Boltzmann calculations 
with various Knudsen numbers~\cite{Tang:2008if, ThesisYuting:2007}. }
\label{fig:v4STAR}
\end{center}
\end{figure}

The higher harmonics give more detail about the shape of the azimuthal 
anisotropy. Values for $v_4$ and $v_6$ are shown in
Fig.~\ref{fig:v4v6}~\cite{Adams:2003zg}.
The values appear to scale as $v_n \propto v_2^{n/2}$ as shown 
in Fig.~\ref{fig:v4v6}. 
For identified particles in Fig.~\ref{fig:v4scaling}, $v_4$ appears 
to have the same scaling as $v_2$ with transverse kinetic 
energy $\ket$, but vs $\ket/n_q$ it seems to scale as the square 
of the number of constituent quarks~\cite{Huang:2008vd} as one would 
guess because $v_4 \propto v_2^2$~\cite{Adams:2003zg}. 
The proportionality constant can be seen in Fig.~\ref{fig:v4STAR}. 
For ideal hydro this ratio should be $1/2$~\cite{Borghini:2005kd} 
at high $\pt$. Boltzmann calculations indicate that a finite Knudsen 
number~\cite{Gombeaud:2007ub} is needed to describe the deviation of 
the data from non-viscous hydro.

In simple coalescence models~\cite{Kolb:2004gi}, the ratio $v_4/v_2^2$ for
hadrons is related to $v_4/v_2^2$ for quarks:
\begin{equation}
\left[ v_4/v_2^2\right]^{\mathrm{Meson}}_{2 \pt} \approx 1/4 + (1/2)\left[
  v_4/v_2^2\right]^{\mathrm{Quark}}_{\pt} ,
\end{equation}
\begin{equation}
\left[ v_4/v_2^2\right]^{\mathrm{Baryon}}_{3 \pt} \approx 1/3 + (1/3)\left[
  v_4/v_2^2\right]^{\mathrm{Quark}}_{\pt} ,
\end{equation}
where here $\pt$ is the quark $\pt$.  The $v_4/v_2^2$ for mesons can
also be related to $v_4/v_2^2$ for baryons:
\begin{equation}
\left[ v_4/v_2^2\right]^{\mathrm{Baryon}}_{3 \pt} \approx 1/6 + (2/3)\left[
  v_4/v_2^2\right]^{\mathrm{Meson}}_{2\pt} .
\end{equation}
From the results for identified particle $v_4$ it appears that the quark $v_4/v_2^2$ is approximately 2~\cite{Abelev:2007qg}.

\section{Conclusion and outlook}
\label{sec:conclusion}

The wealth of information obtained recently in the area of anisotropic
flow is far from being fully explored. The theoretical understanding
of the experimental data is rapidly developing and greatly contributes
to the overall picture of heavy ion collisions dynamics as well as the
properties of the new state of matter, sQGP.
Unfortunately, we have not been able to discuss
many other important measurements related to the event anisotropy,
such as azimuthally sensitive identical and non-identical particle 
femtoscopic measurements~\cite{Voloshin:1995mc1996ch,Voloshin:1997jh} 
(for a recent review, see Ref.~\cite{Lisa:2005dd}) 
or two particle correlations relative to the
reaction plane~\cite{Adams:2004wz,Bielcikova:2003ku}.

The next years promise to bring new very important data which will shed
more light on quite a few remaining  questions.
The most important developments we expect are in the following directions:
(i) upgrade of RHIC detectors will give more precise information on
identified particle anisotropic flow, (ii) beam energy scan at RHIC
will allow looking very carefully in the region where critical
phenomena are expected (e.g. color percolation, critical point), in
particular to look for any non-monotonic behavior or scaling violation
in the $v_2/\eps$ plot, (iii) U+U collisions at RHIC,
(iv) and, of course, the first data at LHC is
eagerly expected to answer what happens at higher beam energies.

\begin{figure}[htb]
\begin{minipage}[b]{0.48\textwidth}
  \includegraphics[width=0.95\textwidth]{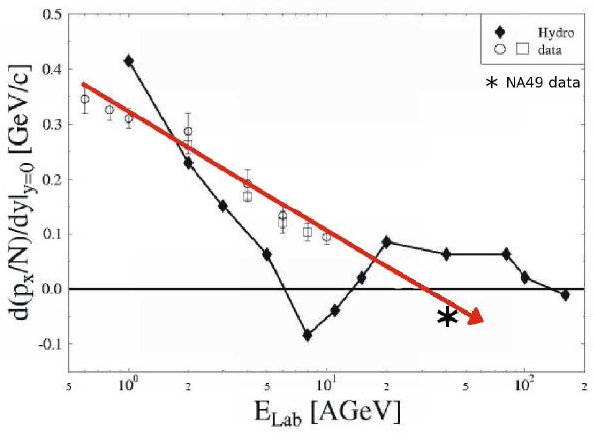}
  \caption{``Collapse'' of directed flow as discussed in 
   Ref.~\cite{Stocker:2007pd}. 
   The old measure of flow, the slope at mid-rapidity, is shown as a function of beam energy with an arrow line through the data. 
   }
\label{fig:collapse}
\end{minipage}
\hspace{\fill}
\begin{minipage}[b]{0.48\textwidth}
  \includegraphics[width=0.95\textwidth]{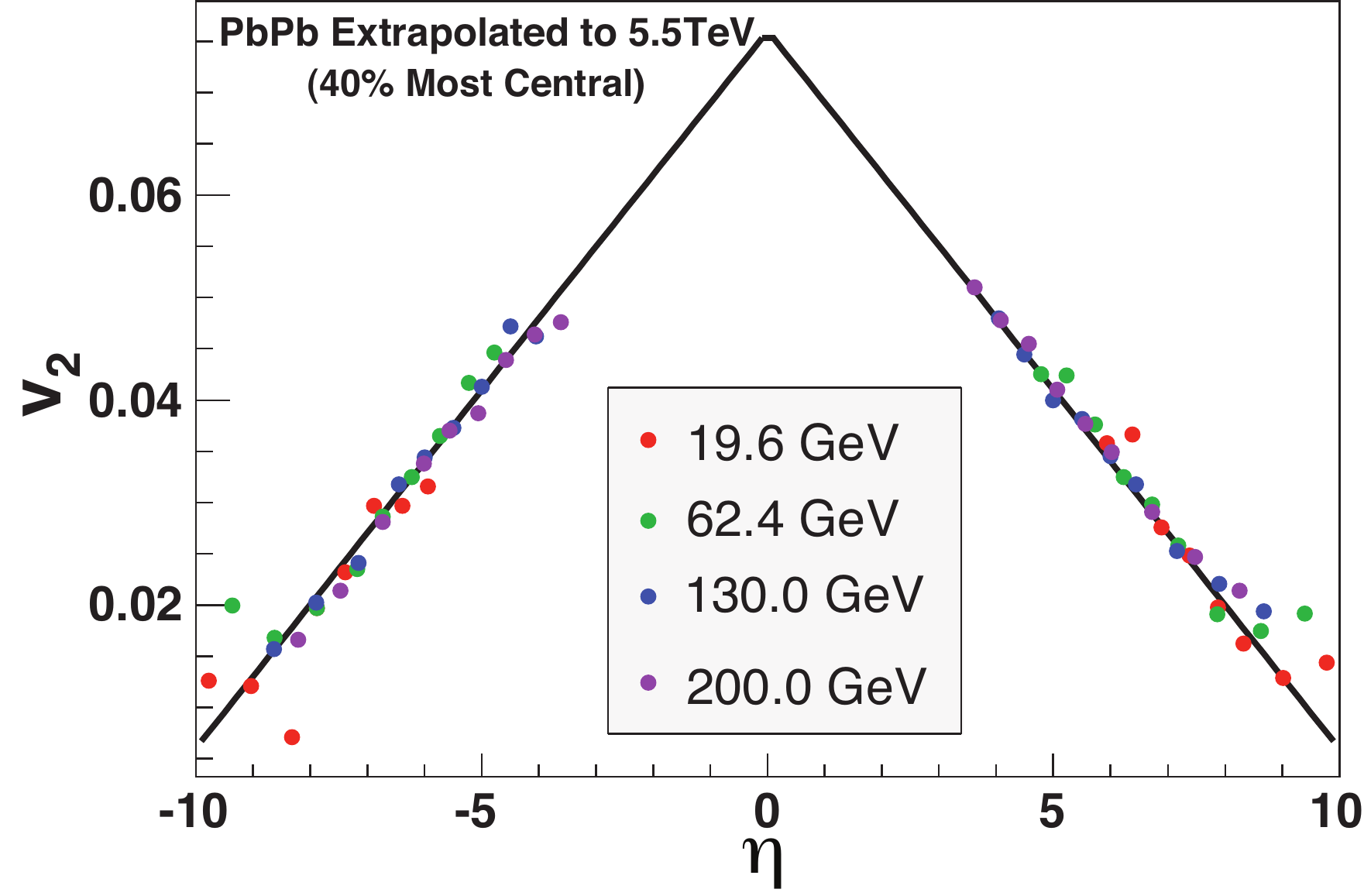}
\caption{Extrapolation of $v_2\{\eta\}$ values to LHC
energies~\cite{Busza:2007ke} based on limiting fragmentation
hypothesis and assuming linear dependence of elliptic flow on $|\eta - y_{\mathrm{beam}}|$. The x-axis is $\eta - y_{\rm{beam}}$ shifted by the $y$ of the LHC. The measurements are from the eta subevent method.}
\label{fig:busza}
\end{minipage}
\end{figure}

The main interest in the RHIC beam energy scan is the search for the
QCD critical point. The scan would cover the energy region from top
AGS energies, over the CERN SPS energies, and higher. In terms of
anisotropic flow the major observables to watch would be a possible
non-smooth behavior in the $v_2/\eps$ dependence on 
particle density~\cite{Voloshin:1999gs}, 
a possible disappearance of constituent quark number scaling, and the
``collapse'' of directed flow~\cite{Stocker:2007pd} (see
Fig.~\ref{fig:collapse}). RHIC also has plans to extend its reach
in terms of energy density using uranium beams. From the first
estimates and ideas of using uranium beams we now have detailed
simulations~\cite{Nepali:2007an} of such collisions 
with methods developed for selection of the
desired geometry of the collision. 

The predictions for the LHC are rather uncertain, 
though most researchers agree
that elliptic flow will continue to increase~\cite{Abreu:2007kv},
partially due to the relatively smaller contribution of viscous effects.  
Simple extrapolations~\cite{Busza:2007ke,Borghini:2007ub} in
Fig.~\ref{fig:busza} of
the collision energy dependence of $v_2$ look rather reliable.
Theoretical calculations predict significantly smaller, thought still
finite, viscous effect at LHC energies.  
One can see it in Fig.~\ref{fig:HydroCascadeLHC} which shows 
$v_2/\eps$ evolution with particle density up to LHC energies 
in 3d-hydro+cascade approach.
Note that there exist some calculations predicting a {\em decrease} of the
elliptic flow~\cite{Krieg:2007sx}. 
Another important observation is an increase in mass dependence
(splitting) of $v_2(\pt)$ due to a strong increase of radial flow.

In summary, we have had very exciting years of anisotropic flow study,
which greatly enriched our understanding of ultra-relativistic nuclear
collisions and multi-particle production in general. We are looking
forward to more new physics with RHIC and LHC. 

\section{Acknowledgments}

We thank all our colleagues for numerous  discussions on many of the
topics discussed in this review. 
We thank our colleagues in STAR for permission to use some of our 
text from published STAR papers. We appreciate comments on the paper 
from Ante Bilandzic, Pasi Huovinen, Claude Pruneau, Ralf Rapp, and Gang Wang.
This work was supported in part by FOM and NWO of the Netherlands, and the HENP Divisions of the Office of Science of the US Department of Energy under Contract Numbers DE-AC02-05CH11231 and DE-FG02-92ER40713.

\end{document}